\pdfoutput=1
\documentclass{article}

\usepackage[utf8]{inputenc}
\usepackage{amsfonts}
\usepackage{amssymb}
\usepackage{multirow}
\usepackage{amsmath}
\usepackage{hyperref}
\usepackage{mathrsfs}
\usepackage{lscape}
\usepackage{caption}
\usepackage{graphicx}
\usepackage{subcaption}
\usepackage{setspace}
\usepackage{multibib}
\usepackage{lipsum}
\usepackage{tcolorbox}
\usepackage{xcolor}
\usepackage[section]{placeins}
\usepackage{chngcntr}
\usepackage{newfloat}
\usepackage{booktabs}
\usepackage{tabularx}
\usepackage{ltablex}
\usepackage{multicol}
\usepackage{authblk}
\usepackage{lineno}
\usepackage{har2nat}
\usepackage{setspace}
\usepackage[onelanguage,ruled,vlined,linesnumbered,resetcount]{algorithm2e}

\newcites{main}{References}
\newcites{app}{References}

\usepackage[section]{placeins}

\DeclareMathOperator{\sign}{sign}

\usepackage{natbib}
	\setcitestyle{authoryear,round,aysep={}}
\title{The Formation of Production Networks: How Supply Chains Arise from Simple Learning with Minimal Information}

\author[1]{Tuong Manh Vu}
\author[2]{Ernesto Carrella}
\author[3]{Robert Axtell}
\author[1]{Omar A. Guerrero}

\affil[1]{The Alan Turing Institute, London}
\affil[2]{Tesco}
\affil[3]{George Mason University}

\begin{document}

\maketitle

\begin{abstract}
We develop a model where firms determine the price at which they sell their differentiable goods, the volume that they produce, and the inputs (types and amounts) that they purchase from other firms. A steady-state production network emerges endogenously without resorting to assumptions such as equilibrium or perfect knowledge about production technologies. Through a simple version of reinforcement learning, firms with heterogeneous technologies cope with uncertainty and maximize profits. Due to this learning process, firms can adapt to shocks such as demand shifts, suppliers/clients closure, productivity changes, and production technology modifications; effectively reshaping the production network. To demonstrate the potential of this model, we analyze the upstream and downstream impact of demand and productivity shocks.
\end{abstract}


\section{Introduction}

Recent events such as Brexit, the COVID-19 pandemic, and the Russia-Ukraine war have shown the importance of production networks to the broadest and most general population, as these disruptive events percolated through different industries and geographies, impacting a heterogeneous set of economic actors.
While researchers in supply chains, operations research, input-output (IO) models, and industrial organization have studied production networks for a long time, quantitative models about the formation and reorientation of production networks are significantly less than qualitative analyses.
Economists and IO researchers typically assume production networks as fixed exogenous structures, and focus on solving models for a price vector in equilibrium.
This approach severely restricts the ability to model realistic responses to shocks, for example, in terms of network reorientation or the duration of transients.
Recent economic models on the endogenous formation of production networks also require strong assumptions such as rational equilibrium and perfect knowledge about network topology and technologies, which have similar implications on how realism such models can reproduce.
Needless to say, such assumptions are poorly supported by empirical evidence from the supply-chain literature in terms of how `far ahead in the network' firms can see and how rational they are in this context \citep{choi2006supply}.
Hence, developing models of the endogenous formation of production networks under uncertainty and with behaviorally plausible foundations remains an area that needs further work.

In this paper, we develop a agent-based model where firms do not know the production functions of the economy nor the productive structure of the economy.
Importantly, uncertainty about production functions extends to the firm's own technology, meaning that an industry does not possess knowledge about the functional form of its own production function other than the information it could gather by trying out different input combinations.
Not needing to assume knowledge about production functions means that, instead of requiring top-down optimization, firms can aim to maximize profits through reinforcement learning by adjusting prices and output volume in a procedural fashion as they experience success and failures in sales outcomes. Using reinforcement learning enables agent to act in uncertain environment, without exploration or mapping, but using immediate reward or penalty received from the environment \citep{ale_ebrahim_dehkordi2023}.
This specification leads to steady-state dynamics where stable buyer-seller relationships emerge between firms.
Moreover, because firms do not need to know specific production functions, the model can accommodate different types of technologies, allowing for a broader set of potential outcomes and experiments that are relevant to understand the implications of technological change.

The rest of the paper is structured in the following way. 
First, we discuss the different strands of literature that have developed production network models.
Second, we present the model.
Third, we analyze the model's behavior through a few theoretical examples with a small hypothetical economy.
Fourth, we illustrate the potential use of the model in the context of the propagation of demand and productivity shocks using a larger hypothetical economy.
Finally, we provide a discussion and conclusion.

\section{Literature review}

Recent quantitative literature on production networks focuses heavily on dynamics taking place on exogenous production networks.
Some notable examples study topics such as the propagation of idiosyncratic shocks \citep{Acemoglu2012}, the amplification of shocks \citep{contreras2014propagation}, offshoring and re-shoring choices \citep{Konig2016}, cascading dynamics of firms' shutdowns \citep{Taschereau-Dumouchel2020}, propagation of supply and demand shocks during COVID \citep{Pichler2021, pichler2022simultaneous}, the impact of labor productivity shocks \citep{baqaee2018cascading}, excess volatility from slow convergence speed to equilibrium \citep{Dessertaine2022}, the amplification of economic growth \citep{Mcnerney2022}, and global supply chains \citep{grazzini2022empirical} (see \citep{carvalho2014micro} for a comprehensive survey).\footnote{\citet{Taschereau-Dumouchel2020} proposes a model where the production network is exogenous, and firms can choose to produce or not; activating and deactivating the presupposed links endogenously.}
These and other examples provide insights on the potential economic dynamics that could happen on a production network that is considered exogenous.

A different and more recent strand of literature focuses on models that explain the endogenous formation of production networks (the existence or not of a link) from economic principles, i.e., from firms choices.
In some cases, these models analyze the formation and destruction of links (inputs choice), while in others, they place more emphasis on the weights of such input-output relations.
While there have been models looking at the endogenous formation of production networks from a purely mechanistic point of view \citep{Atalay2011, Chaney2014, Konig2016, Arata2019, Mcnerney2022}, the most relevant studies are those that try to go beyond the use of stochastic processes by providing economic microfoundations to the network formation mechanisms.
Overall, we can split this literature into rational-equilibrium models and procedural models.
The former class is prevalent in economics, while the latter has been prominent in operations research for supply chain management.

An early example of a rational-equilibrium model is \citet{Carvalho2014}, who specifies the process through which firms choose certain inputs, placing special emphasis on the mechanism that leads production networks to make certain inputs predominant.
Subsequent models consider firms choosing a single intermediate good to form production networks that increase productivity and place them in more central positions \citep{Oberfield2018}; firms forming domestic and foreign customer-supplier relationships to replicate empirical regularities in US IO tables \citep{Zou2018}; industries choosing suppliers and purchasing volumes to reproduce stylized facts such as density and degree distribution \citep{Acemoglu2020}; firms selecting intermediate suppliers to analyze non-linear monetary transmission \citep{Ghassibe2020}; firms in two sectors forming upstream and downstream links to explain stylized facts in the Turkish economy \citep{Bilgin2020}; firms filtering out suppliers to mitigate the uncertainty caused by high volatility \citep{Kopytov2021}; firms choosing suppliers according to the costs implied by spatial distance; and network formation processes shaped by the presence of fixed costs when forming links \citep{Dhyne2023}.
To make these models solvable, the authors focus on equilibrium outcomes and rely heavily on the assumption of specific production functions and the common knowledge (by the agents/firms/industries) about their functional forms.
\citet{Dessertaine2022} criticizes this approach, arguing that real-world economies operate out-of-equilibrium.

Studies using procedural models adopt an algorithmic view by making the dynamics of the network-formation process more explicit--without necessarily focusing on equilibrium outcomes.
This literature has flourished among operations research scholars, and very few studies of this type have crossed over to the economics literature.
An early model of this type by \citet{schieritz2003emergent} combines system dynamics with agent-based modeling to emerge downstream supply structures from income-optimization choices.
\citet{paolucci2008agent} discuss the empirical viability of such approach for the management of supply chains.
One of the first procedural models to cross-over to the economics literature is provided by \citet{Gualdi2016}, who start with a rational-equilibrium model, and add a procedural component where firms can take opportunities to change suppliers in a setting with constant entry and exit of firms.
Their aim is to demonstrate that out-of-equilibrium dynamics can give place to scale-free production networks while preserving several well-known stylized facts about firm dynamics (e,g, firm size and growth-rate distributions).
This model has been extended to study the role of technological change in the evolution of production networks \citep{Gualdi2019}, and the transient dynamics introduced by lock-downs during the Covid-19 pandemic in India \citep{Mandel2024}.
\citet{Lengnick2013} developed a baseline agent-based model that does not depend on the assumption of rationality and can reproduce many stylized facts of business cycle, contrasting against common dynamic stochastic general equilibrium approach.
Note that procedural models still assume that firms have substantial knowledge about the nature of production technologies, as their strategic choices result from a rational optimization process.
We take a different approach through learning.

Having a small number of procedural models in the literature suggests an important knowledge gap in understanding how real-world firms--with limited information and bounded rationality--give place to production networks through their choices of suppliers, clients, volumes, and prices, while remaining profitable.
Our model departs from existing approaches in significant ways, and provides a previously unexplored avenue to investigate important questions about the formation and reorientation of production networks.
Next, we introduce such model.

\section{Model}

Our aim is to focus on the production network problem, not on the dynamics of the entire economy.
Hence, for simplicity, we do not model households and labor, but assume exogenous aggregate demands for each good.
Following the naming convention introduced by \citet{Acemoglu2020}, we use the terms `firm' and `industry' interchangeably in the remainder of the paper.
The logic behind the model is rather simple. 
Firms choose output volumes and prices to try to maximize profits in a dynamical setting; leading them to select and purchase inputs from potential suppliers.
Firms do not know the details of their production technologies, so they learn from experience through trial and error.
When learning is consistent (we explain consistent learning in \autoref{sec:dynamics}), the model reaches a steady state where inter-firm transactions are stable and a production network emerges.
The model was implemented in Python with Numpy vectorization and available at \url{https://github.com/vmtuong/endogenous-production-network}

\subsection{Setup}

There are $N$ industries in the economy, each one producing a differentiated good that is consumed by other firms as well as an end consumer market.
The consumer market of industry $i$ has an aggregate demand

\begin{equation}
    Q^d_i(P_{i}) = f_i(P_{i}),
\end{equation}
where price $P_{i}$ is set by industry $i$.

Firms sell their products to their respective final consumption markets as well as to other firms that use them as inputs.
To generate output, an industry uses a certain production technology described by

\begin{equation}
    Q_{i} = g_i(\mathbf{q}_{i}),
\end{equation}
where $\mathbf{q}_{i}$ is a vector of inputs coming from other industries (including $i$).

Firms do not know the specific form of their own production function (and neither of the other firms); all they see is their own inputs and output.
Hence, firms try to learn how technology works.
Learning happens through the experience of setting prices and output targets, making choices about the amount to be purchased for each input, and receiving responses in terms of profits.
Effectively, this is a procedural optimization process.
Next, we explain how the profit-seeking mechanism works.

Let us begin with industry $i$ at time $t$, with inputs $\mathbf{q}_{i,t}$, price $p_{i,t}$, and output quantity $Q_{i,t}$.
Once all firms have produced their outputs, they engage in purchasing interactions.
Firms interact with each other in random order.
During an interaction, firm $i$ buys quantity

\begin{equation}
    q_{i-j,t} = \min( \mathbf{q}^j_{i,t}, Q^{r}_{j,t} )
\end{equation}
from industry $j$.
Here, $\mathbf{q}^j_{i,t}$ is the $j^{th}$ element of $\mathbf{q}_{i,t}$, and $Q^{r}_{j,t}$ is the residual product of industry $j$ (it is residual because it may have already sold to other firms) before transacting with $i$.
Let us assume that firms always have liquidity, so they can access funds to pay for all the purchased inputs, and profits will be reported after sales.

Once all firms have made their purchases, they sell the remaining production $Q^{m}_{i,t}$ (if any) to the consumption market.
Consumers buy at most the quantity dictated by the aggregate demand at the given price set by the industry.
Thus, it is possible that a firm is unable to sell all its residual product by the end of period $t$.\footnote{For simplicity, and for the time being, let us assume all goods are perishable.}

After completing the sale and purchase process, the amount of goods sold by industry $i$ is $Q^*_{i,t}$, which may be different from the produced quantity $Q_{i,t}$.
Hence, the firm profits are

\begin{equation}
    \Pi_{i,t} = P_{i,t}Q^*_{i,t} - \sum^N_{j=1}P_{j,t}\mathbf{q}^j_{i,t}.
\label{eq:profit_n}
\end{equation}

The profits provide information to the industry about its performance after choosing a price, a quantity, and adjusting its inputs according to their marginal product and prices.
Firms use this information to gradually make adjustments that aim to increase profits.
Note, however, that even if an industry plans to produce a certain quantity $Q_{i,t}$, this may not be feasible due to shortages from its suppliers due to miscoordination.
While miscoordination introduces noise in the learning process (like in the real world), firms manage to learn and set prices and quantities that are consistent with their aggregate demands, driving the economy into a steady state with an endogenous production network.

A key assumption in our model is that firms do not close if they reach negative profits during the learning phase (or in a transient resulting from a shock).
While this assumption is not realistic at the level of firms, it holds for industries in the short and medium term, it is consistent with other models in the literature, and it allows for a simpler specification.
Hence, for this paper, we maintain this assumption.
Next, we elaborate on the learning component of the model.

\subsection{Learning}

Firms adjust their price through a learning process that is known as PID controller \citep{Carrella2014} in the engineering literature, and as directed learning in the behavioral literature \citep{Dhami2016}.
The principle of this learning model is rather simple: firms keep increasing (decreasing) prices if such action raised their profits in the previous period.
They choose the opposite action if profits fell.
Formally, the price in period $t+1$ is determined by

\begin{equation}
    P_{i,t} = \max \left\{ \mu^p, P_{i,t-1} + \sign(\Delta P_{i,t-1}  \Delta \Pi_{i,t-1}) \times \delta^p_{i,t} \right\},\label{eq:pid}
\end{equation}
where the sign function is 1 if $\Delta P_{i,t-1} \Delta \Pi_{i,t-1} > 0$ and -1 if this argument is negative.\footnote{If $\Delta P_{i,t-1} \Delta \Pi_{i,t-1}=0$, then the firm explores the two actions by randomly choosing one direction with equal probability.}
$\mu^p$ prevents prices from becoming zero, and $\delta^p_{i,t-1}$ is the size of the adaptation, which is endogenously determined by 

\begin{equation}
    \delta^p_{i,t}  =  \frac{|P^d_{i,t-1}( Q_{i,t-1} ) - P_{i,t-1}|}{P^d_{i,t-1}( Q_{i,t-1} )}, \label{eq:adapt_p}
\end{equation}
meaning that the magnitude of the adaptation correlates to the proportional difference between the demand's price and the price set by the firm.
It is assumed that industry $i$ is capable of obtaining information about the price that the consumer market would pay for $Q_{i,t-1}$, hence the term $P^d_{i,t-1}( Q_{i,t-1} )$ in \autoref{eq:adapt_p}.
Finally, note that prices are explicit control variables, not numeraries.

The specification of the learning component is quite parsimonious.
First, it avoids potential issues related to defining a large action space since there are only two choices: more or less.
Second, the adaptation step is endogenous, so there is no need to calibrate learning parameters in this component.
Third, the link between action and reward (profit) is direct, so the firm's incentives are explicit while their actions are coherent, which aligns with basic economic intuition about a firm's incentives.

Next, to determine its total output volume, firm $i$ needs to make decisions on purchasing specific amounts of each input $j$.
This decision is slightly more complex than the pricing one as it depends not only on the profit history, but also on the new price of each input, as well as on its marginal product.
Nevertheless, it is possible to separate the quantity-adjustment process into individual problems for each input and to use an extended version of the directed-learning model.
First, let us focus on purchasing input $j$, which corresponds to the $j$th entry of $\mathbf{q}^{i,t}$.
This purchasing decision is determined by

\begin{equation}
    \mathbf{q}^j_{i,t} = \max \left\{ 0, q_{i,j,t-1} + \sign(\Delta Q_{i,t-1}  \Delta \Pi_{i,t-1}) \times \delta^{q}_{i,t} \times \frac{1}{P_{j,t}} \times \frac{\Delta Q_{i,t-1}}{\Delta \mathbf{q}^j_{i,t-1}} \right\},\label{eq:adapt_q}
\end{equation}
where $\delta^q_{i,t}$ is endogenous as $\delta^q_{i,t} = \frac{|Q^d_{i,t-1}( P_{i,t-1} ) - Q_{i,t-1}|}{Q^d_{i,t-1}( P_{i,t-1} )}$, meaning that the magnitude of the adaptation correlates to the proportional difference between the demand's quantity and the quantity set by the firm.

Note that $\mathbf{q}^{i,t}$ is a `purchasing plan', not the actual amount of inputs that the firm acquires.
Since firms adjust prices and volumes, it is possible for an industry to experience supply shortages.
Nevertheless, as we will show in section~\ref{sec:analysis}, the model reaches a steady state with moderate fluctuations and a stable endogenous production network.
This is an important element of the model as questions related to supply chain shocks often involve the sudden shortage of goods, which can lead to amplifying dynamics such as the bullwhip effect.

A key difference in the adaptation of input choices with respect to pricing is that they must consider their potential substitutability.
However, how can a firm determine rates of substitutability if it does not know the functional form of its production function?
To solve this problem, we assume that firms can experiment with their technology to calculate marginal changes at specific levels of production.
The intuition is that, while an industry may not have perfect information about its production technology, it can experiment with it to learn about its response to potential input changes.

Formally, let firm $i$ perform a production experiment consisting of increasing one unit of input $j$ and measuring the change in $i$'s output volume.
The marginal product is captured by $\frac{\Delta Q_{i,t-1}}{\Delta \mathbf{q}^j_{i,t-1}}$ in \autoref{eq:adapt_q}.
The intuition is that inputs that generate larger changes will receive higher priority in a purchasing plan.
Finally, the firm also considers the price of input $j$, and uses this information through $\frac{1}{P_{j,t}}$ when adapting.
Thus, when choosing between two potential substitutes with similar characteristics, an industry will tend to prioritize buying the cheaper one.

Together, \autoref{eq:adapt_p} and \autoref{eq:adapt_q} constitute the learning module of the model.
Since firms do not need to know any functional form of the production function to compute optimal volumes and prices, the model allows for various types of production technologies.
This is an important advantage over most existing frameworks as it allows modifying production functions `on the go' to study the impact of technological changes.

\subsection{Dynamics}\label{sec:dynamics}

The model iterates through time, allowing industries to transact, sell to the consumer market, and adjust prices and quantities.
We study the economy when learning yields a state that is consistent with the markets' aggregate demands.
That is, we say that a firm's learning is consistent if its price and quantity lie in a neighborhood of the aggregate final market demand.

As we show in \autoref{sec:robust_dynamics}, for a given parameterization that yields consistent learning, the outcomes are robust across independent realizations.\footnote{From various numerical analyses, we find that the only situations in which consistent learning is not achieved is when the production functions yield extreme values in output changes, as these introduce large amounts of noise that hinder consistent learning.}
The interaction protocol of the firms is summarized in \autoref{algo:model}.

\vspace{.5 cm}
\begin{algorithm}[!ht]
    \caption{Interaction protocol pseudocode}\label{algo:model}
    \ForEach{period $t$}{
        \ForEach{firm $i$}{
            Produce $Q_{i,t}$\;
        }
        \ForEach{firm $i$}{
            \ForEach{firm $j$}{
               Firm $i$ buys $q_{i-j,t}$ from firm $j$\;
            }
        }
        \ForEach{firm $i$}{
            Firm $i$ sells the remaining quantity $Q^{m}_{j,t}$ to the final market $i$\;
        }
        \ForEach{firm $i$}{
            Firm $i$ calculates profit $\Pi_{i,t}$\;
        }
        \ForEach{firm $i$}{
            Firm $i$ adjusts price $P_{i,t}$ \;
            Firm $i$ adjusts purchase plan $\mathbf{q}^j_{i,t}$\;
        }
    }
\end{algorithm}

\section{Analysis}

Let us begin the analysis of the model by using a three-industry example.
As we move to study shocks in section~\ref{sec:shocks}, we increase the number of industries to five.
Later, in section~\ref{sec:propagation}, we investigate a larger system in the context of propagation.
\label{sec:analysis}

\subsection{Numerical example with linear technology}\label{sec:example_linear}

We specify a production structure where every firm has a linear technology.
While the nature of these technologies is the same for the three industries, their parameters are heterogeneous.
More specifically, we define the system

\begin{equation}
    \begin{split}
      Q_1 &= 2q_{1,1} + 5q_{1,2} + 5q_{1,3}\\
      Q_2 &= q_{2,1} + 5q_{2,2} +  q_{2,3}\\
      Q_3 &= 4q_{3,1} + \qquad \quad  +  4q_{3,3}
    \end{split},\label{eq:linear_qs}
\end{equation}

According to \autoref{eq:linear_qs}, industry 1 is most productive with inputs from industries 2 and 3, and less with its own goods.
Industry 2 is the opposite case from 1, as it is most productive when using its own goods. 
Firm 3 does not need to use input 2 at all, while it is equally productive using goods from 1 and 3.
Note that, overall, industry 1 is the most productive, followed by 3 and then 2.
The difference in marginal product between 3 and 2 is only one unit.

\begin{figure}[ht]
\centering
\caption{Implied productive structure with linear technologies}\label{fig:linear_net}
        \includegraphics[angle=0,width=.4\textwidth]{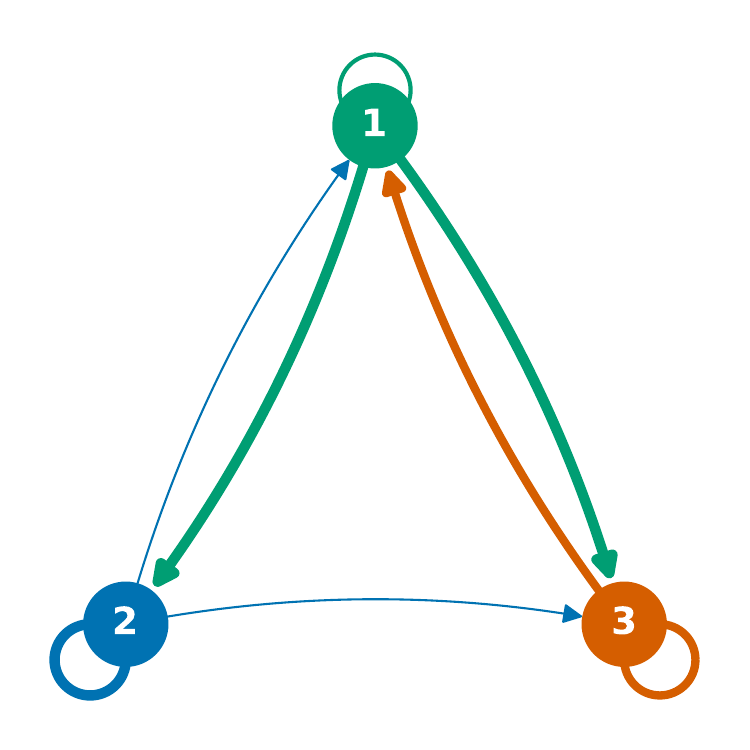}
        \caption*{\footnotesize 
        \textit{\textbf{Notes}}: The arrows indicate potential purchasing relationships.
        For example $3 \rightarrow 1$ means industry 3 could buy inputs from industry 1.
        The circular edges without arrow mean self-loops.\\}
\end{figure}

The implied productive structure is depicted in \autoref{fig:linear_net}, where an arrow $X \rightarrow Y$ indicates that $X$ buys from $Y$.
Note that firm 3 does not use inputs from industry 2.
In this example, firm 3 does not know this, so it will learn that purchasing input 2 does not generate a response in terms of profits.
In general, this zero-knowledge setting allows firms to eventually learn consistently.
However, as the number of industries grow or as production technologies become more coupled (i.e., with more interaction between inputs), learning which inputs are not productive becomes more difficult.
To ameliorate this difficulty, it is possible to introduce a `minimal-knowledge assumption' in which firms are unaware of the functional form of the production technology, but they know which inputs are not used; a reasonable assumption according to how real-world firms operate.
First, we show the example with the zero-knowledge setting and, then, for simplicity, we work with the minimal-knowledge assumption.

The aggregate demands of the final markets to which each industry sells its residual product (after supplying the other industries) is

\begin{equation}
    \begin{split}
      Q^d_1 =&  8000 - 2p_1\\
      Q^d_2 =&  8000 - 0.8p_2\\
      Q^d_3 =& 15000 - 1.5p_3
    \end{split}.\label{eq:linear_demands}
\end{equation}

\begin{figure}[ht]
\centering
\caption{Consistent learning under linear technologies}\label{fig:linear_learn}
        \includegraphics[angle=0,width=.5\textwidth]{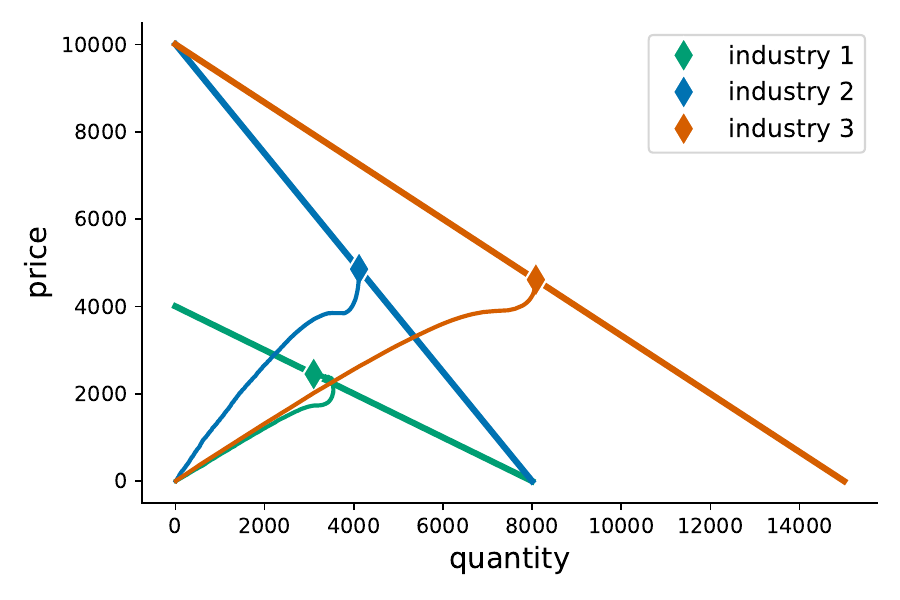}
        \caption*{\footnotesize 
        \textit{\textbf{Notes}}: Diamond markers denote the final price and residual quantity of each firm.\\}
\end{figure}

Firms do not know the functional form of these demands.
Thus, as we have previously explained, consistent learning means that each firm needs to find a residual output and a price that live on its corresponding demand curve.
\autoref{fig:linear_learn} shows an illustrative run of the model, starting at zero-level production.
The plot shows the demand function that each firm faces, as well as the firms' trajectories in the price-quantity space (recall that the quantity in this plot corresponds to residual output volume).
The three trajectories show that all firms are able to learn consistently, as their prices and quantities end up living on the demand curves.\footnote{The diamond markers in \autoref{fig:linear_learn} correspond to the final price and residual quantity of each firm.
While difficult to appreciate in this chart, these points stay in a tight neighborhood, suggesting that the economy reached a steady state.
This is more clearly shown in in \autoref{fig:linear_learn}.}

\begin{figure}[ht]
\centering
\caption{Dynamics under linear technologies}\label{fig:linear_dynamics}
    \begin{minipage}{0.32\textwidth}
        \subcaption{Price dynamics}\label{fig:linear_dynamics.price}
        \includegraphics[angle=0,width=1.\textwidth]{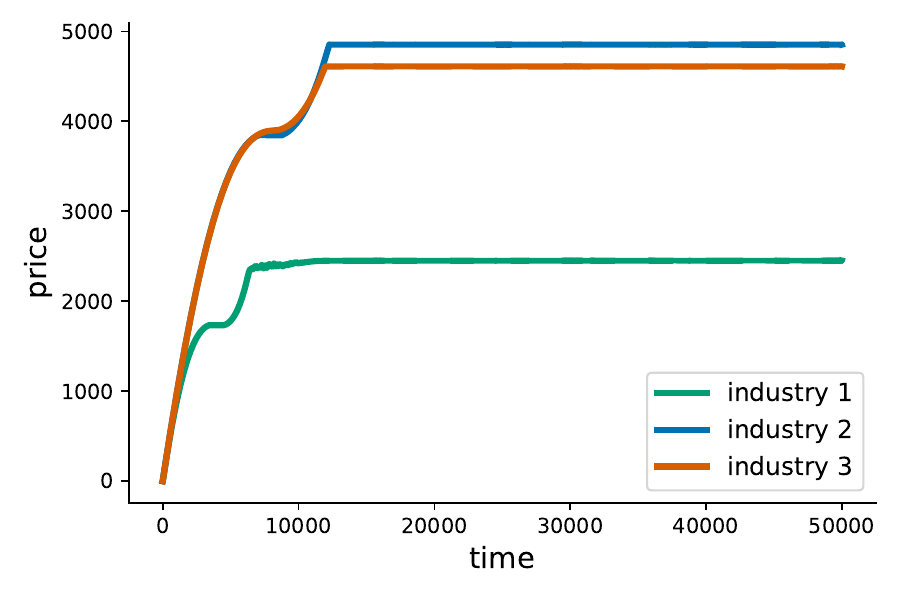}
    \end{minipage}
    \begin{minipage}{0.32\textwidth}
        \subcaption{Output dynamics}\label{fig:linear_dynamics.quantity}
        \includegraphics[angle=0,width=1.\textwidth]{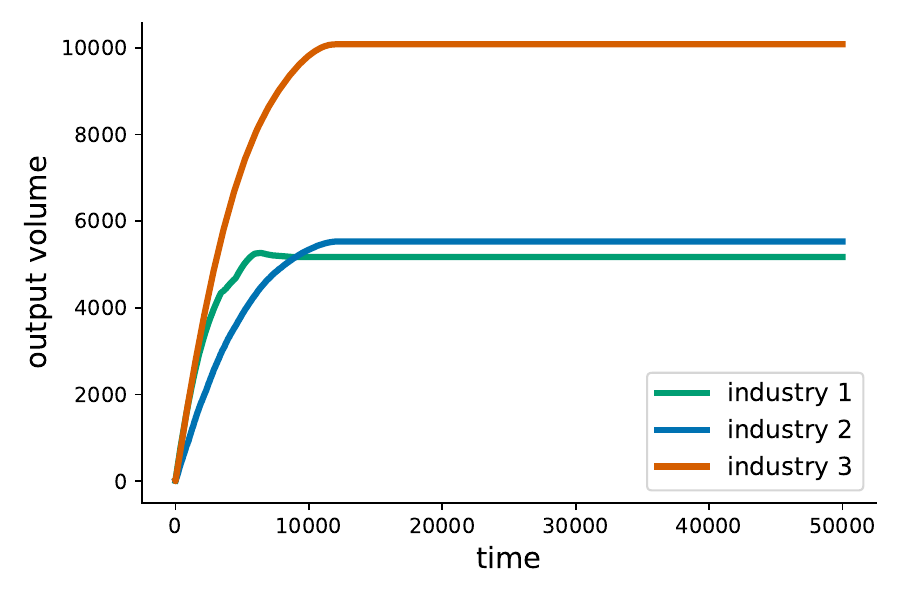}
    \end{minipage}
    \begin{minipage}{0.32\textwidth}
        \subcaption{Profit dynamics}\label{fig:linear_dynamics.profts}
        \includegraphics[angle=0,width=1.\textwidth]{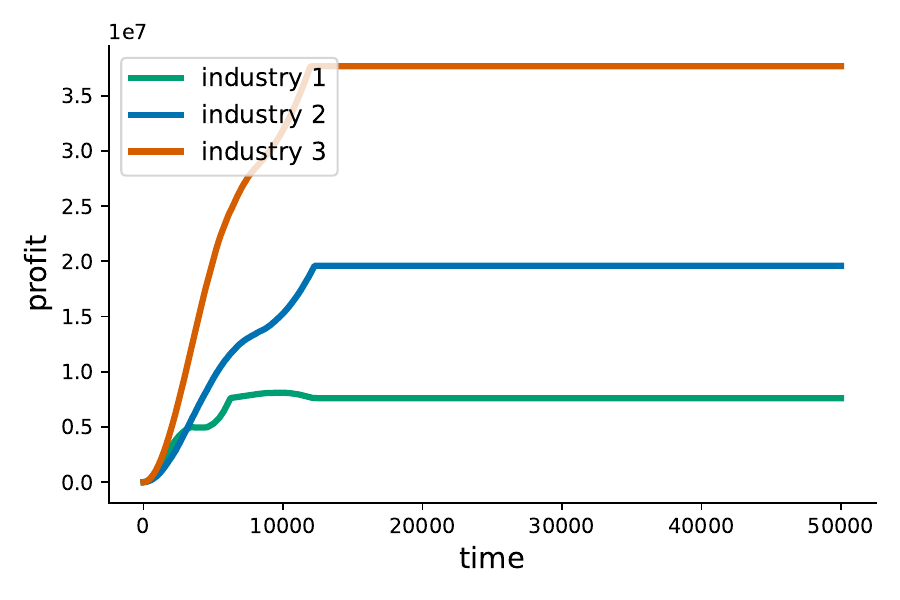}
    \end{minipage}
\end{figure}

Next, let us examine the temporal evolution of the price, output volume, and profit of each industry in \autoref{fig:linear_dynamics}.
This chart shows that, after enough iterations, the three variables reach stable levels.
We show ahead that stability and consistency are insensitive to initial conditions and model stochasticity, providing evidence of robustness.

There are a few interesting highlights in \autoref{fig:linear_dynamics}.
First, firm 1, which has the most productive technology, is not the one making the highest profits.
The most profitable firm is 3, which faces the the largest aggregate demand.
Second, firms 2 and 3 converge to a similar price level, but differ in output volumes, which translates into a substantial difference in profits.
Third, in all three firms, learning price dynamics seem punctuated or step-wise, while output volumes evolve smoothly.

\begin{figure}[ht]
\centering
\caption{Endogenous inter-industry flows under linear technologies}\label{fig:ternary_linear}
    \begin{minipage}{0.32\textwidth}
        \subcaption{Purchased volume}\label{fig:ternary_linear.volumes}
        \includegraphics[angle=0,width=1.\textwidth]{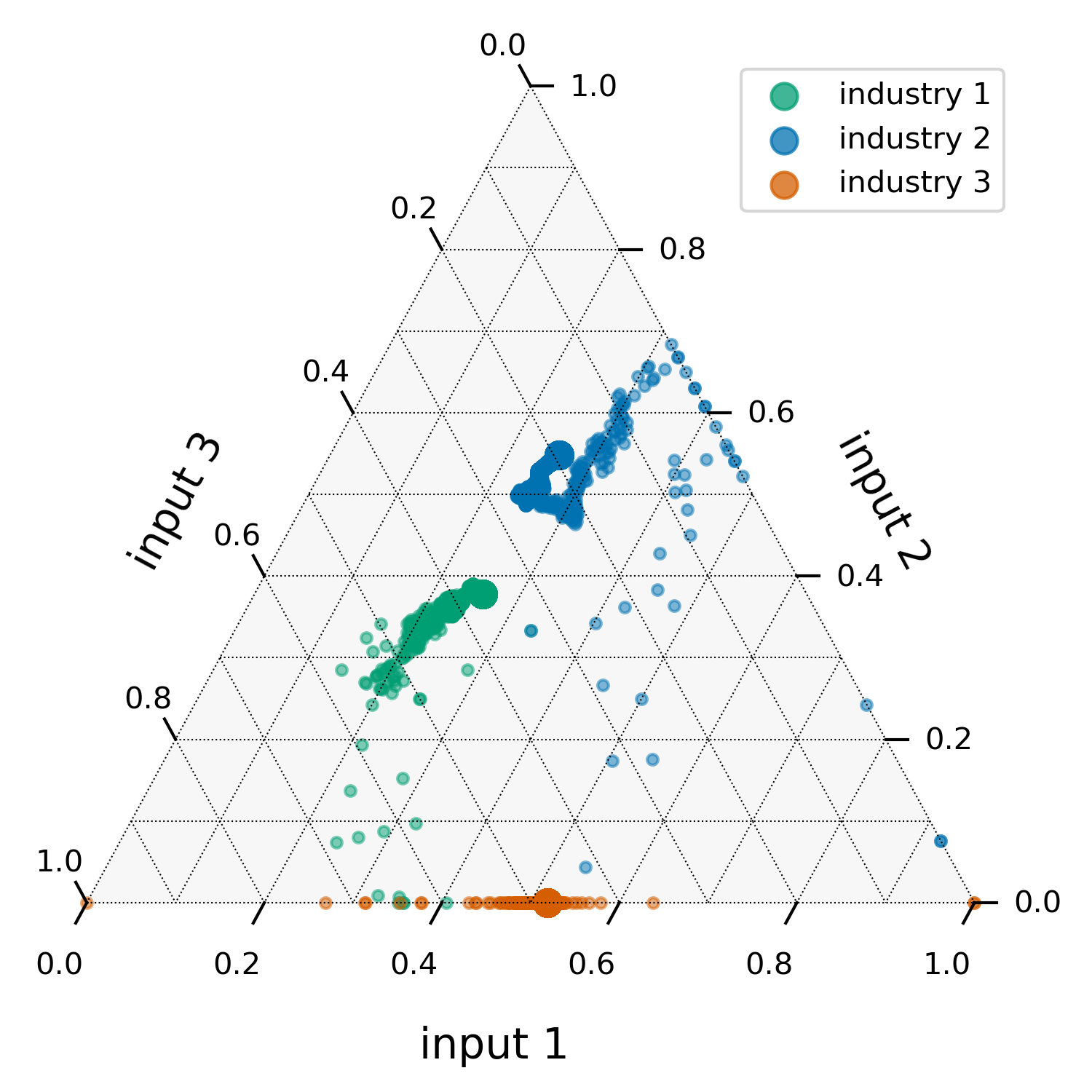}
    \end{minipage}
    \begin{minipage}{0.32\textwidth}
        \subcaption{Costs}\label{fig:ternary_linear.costs}
        \includegraphics[angle=0,width=1.\textwidth]{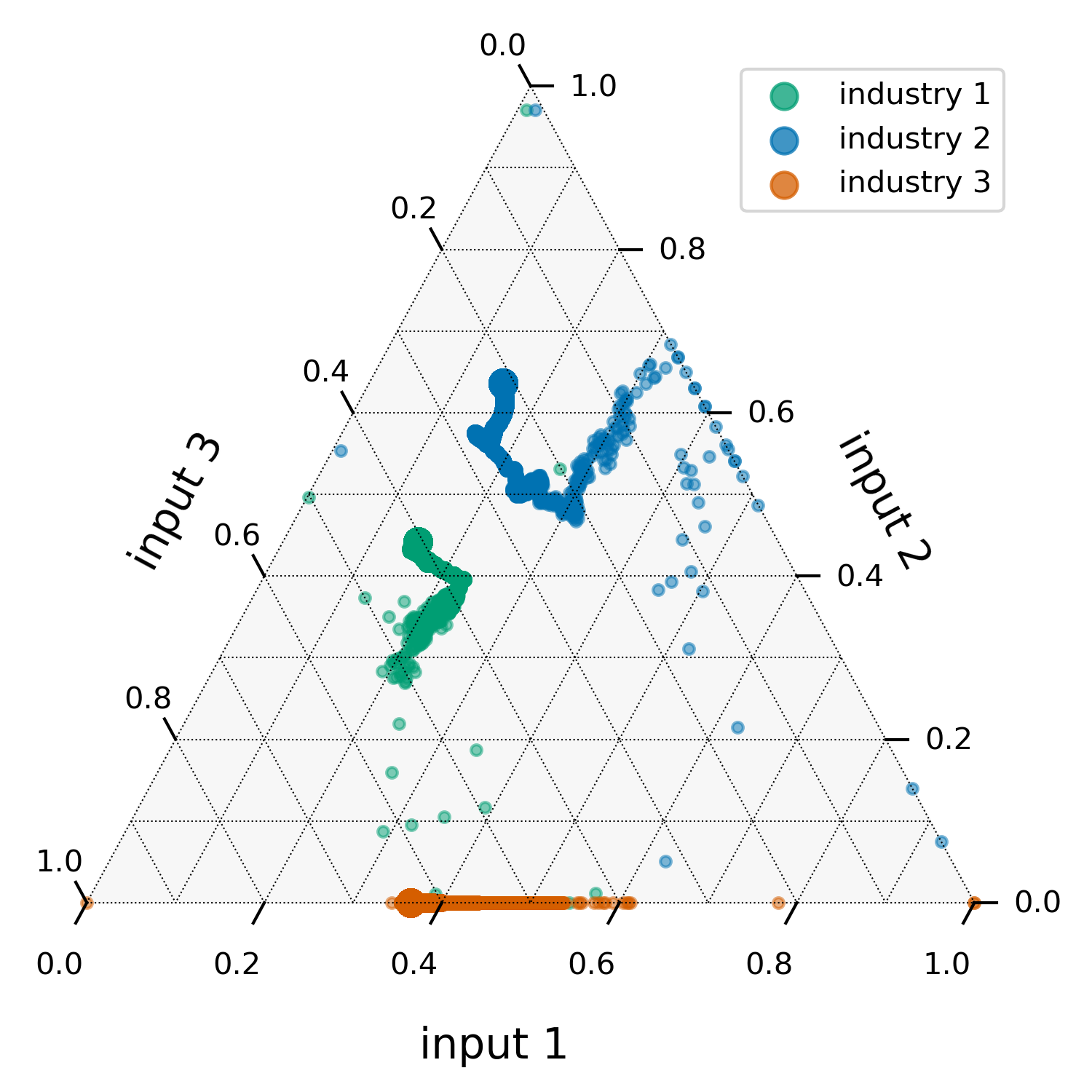}
    \end{minipage}
    \begin{minipage}{0.32\textwidth}
        \subcaption{Sales}\label{fig:ternary_linear.sales}
        \includegraphics[angle=0,width=1.\textwidth]{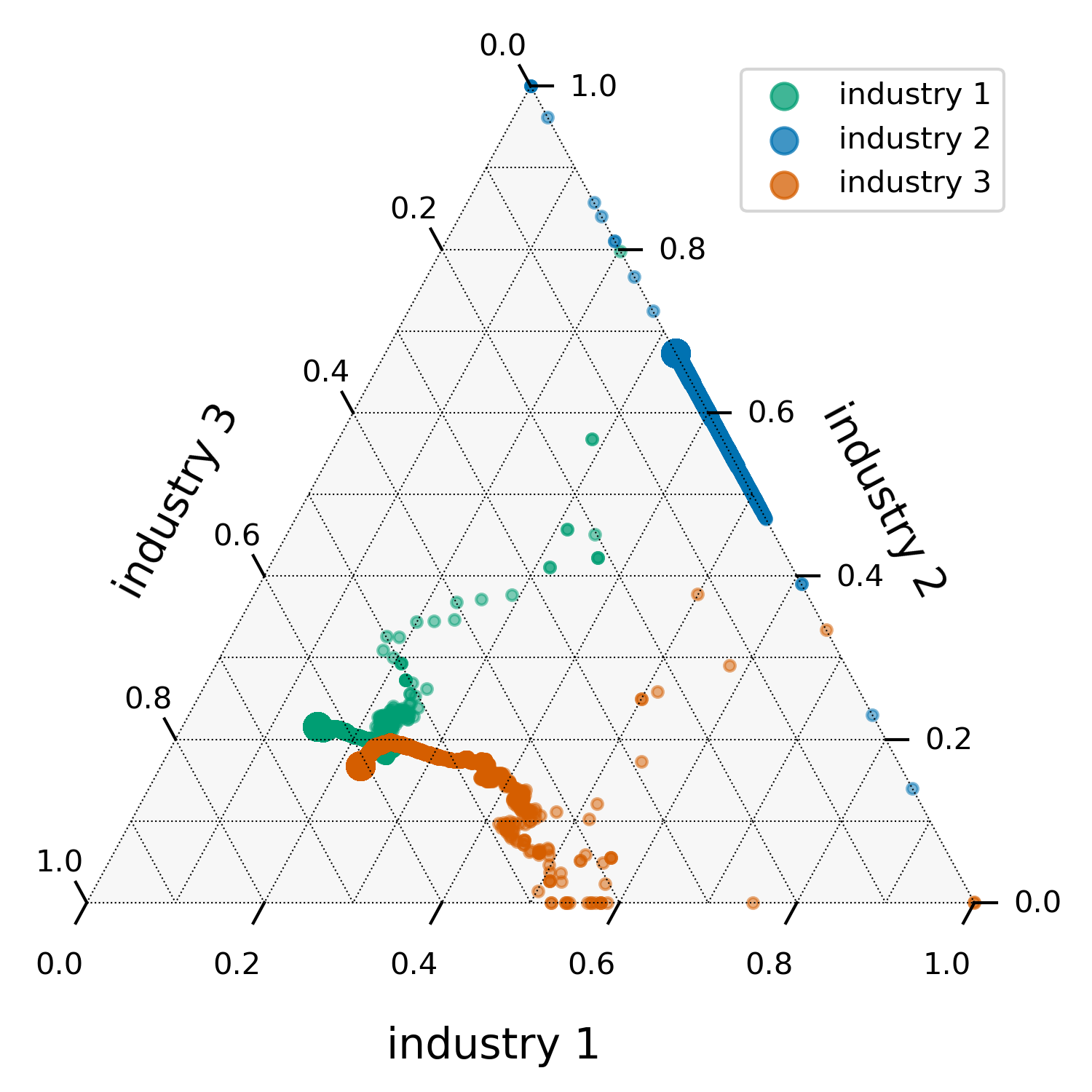}
    \end{minipage}
\caption*{\footnotesize 
\textit{\textbf{Notes}}: Larger and more opaque markers indicate more recent periods.}
\end{figure}

Finally, let us analyze the inter-firm dynamics that give place to the production network endogenously.
More specifically, we track the endogenous flows between industries during the learning process.
To measure such flows, we focus on three types of transactions: (1) the volume that a firm purchases from each industry (including itself), (2) the total cost of such purchases (per input), and (3) the sales that an industry makes to each other.
We analyze each of these measures in terms of their composition, meaning that, for each simulation period, we compute them as fractions of their industry-specific total.
For example, for sales made by industry 1, we calculate the fraction that goes to each of the other firms (including itself).
\autoref{fig:ternary_linear} presents the emergence of the production network through the evolution of these endogenous flows.
These ternary plots capture the distribution of a given variable in terms of the shares coming from each supplier at a point in time.
For example, a dot in the top corner of \autoref{fig:ternary_linear.volumes} would mean that a the corresponding firm purchases 100\% of its inputs from industry 2.
In contrast, a dot in the middle of the triangle would imply that this firm gets one third of its inputs from each industry.\footnote{To read these plots, one can follow the trace of the three straight lines that are closest to the relevant dot.
Each of these lines leads to the axis that indicates the share coming from the industry corresponding to the axis.
The trace belonging to an axis matches the inclination of the axis ticks.}

The first thing to notice is that, according to \autoref{fig:ternary_linear.volumes} and \autoref{fig:ternary_linear.costs}, firm 3 does not purchase inputs from firm 2.
From the perspective of sales, \autoref{fig:ternary_linear.sales} shows that, indeed, industry 2 makes no sales to firm 3.
Hence, without any knowledge about the production technology, industry 3 makes purchasing decisions that are consistent with the implied productive structure of the economy.
The rest of the flows between industries show heterogeneous learning trajectories.
Some of these trajectories explore a bigger space (e.g., those from industry 2), while others move more directly to their steady-state values.
Overall, this depiction of the endogenous formation of a production network shows that simple learning mechanisms and zero knowledge about the productive structure of the economy are sufficient to generate production networks endogenously from basic economic principles.

\subsubsection{Robustness}\label{sec:robust_dynamics}

\begin{figure}[ht]
\centering
\caption{Robust dynamics under different regimes of initial conditions (ICs)}\label{fig:linear_dynamics_sample}
    \begin{minipage}{0.32\textwidth}
        \subcaption{Price under same ICs}\label{fig:linear_dynamics_sample.price}
        \includegraphics[angle=0,width=1.\textwidth]{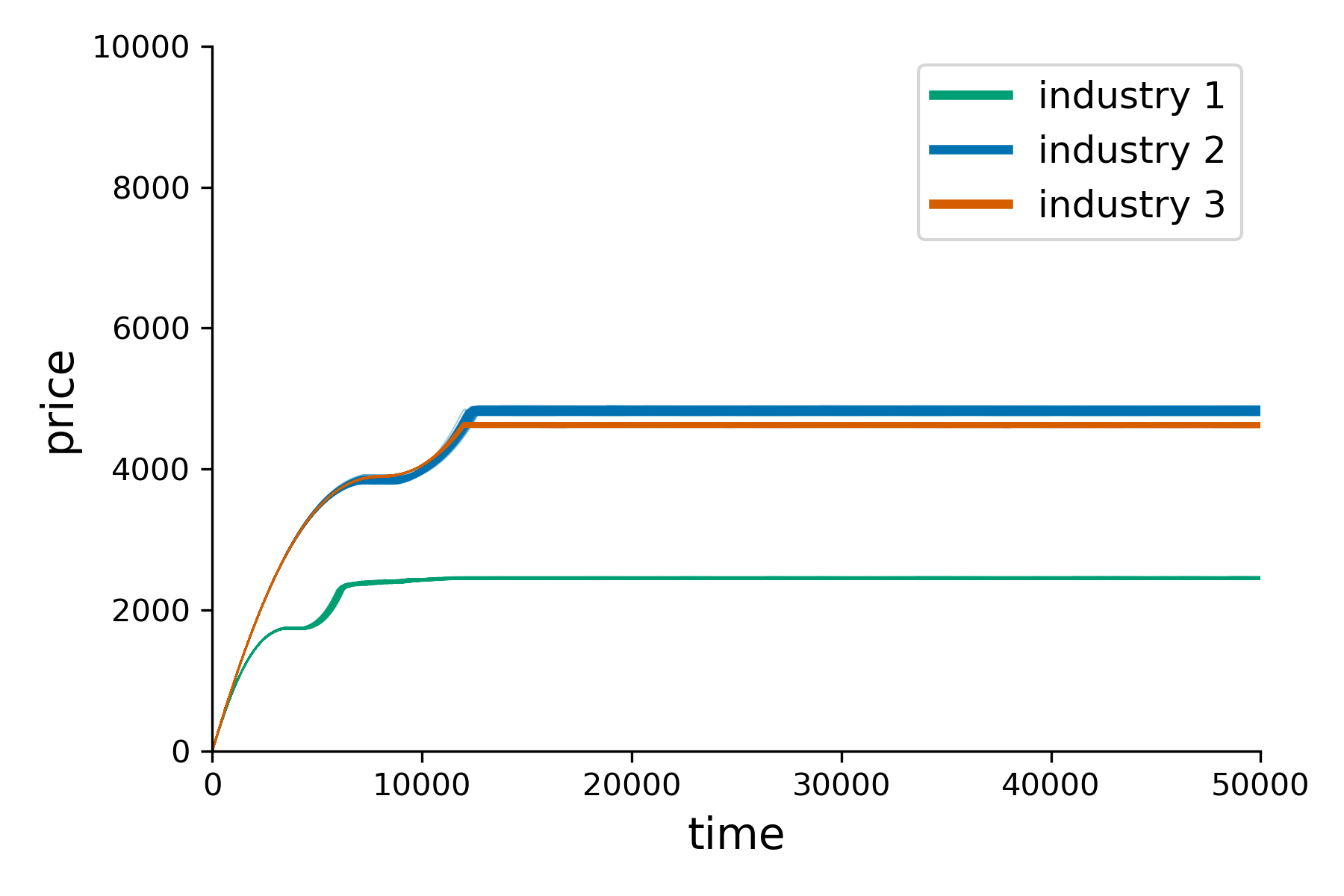}
    \end{minipage}
    \begin{minipage}{0.32\textwidth}
        \subcaption{Output under same ICs}\label{fig:linear_dynamics_sample.quantity}
        \includegraphics[angle=0,width=1.\textwidth]{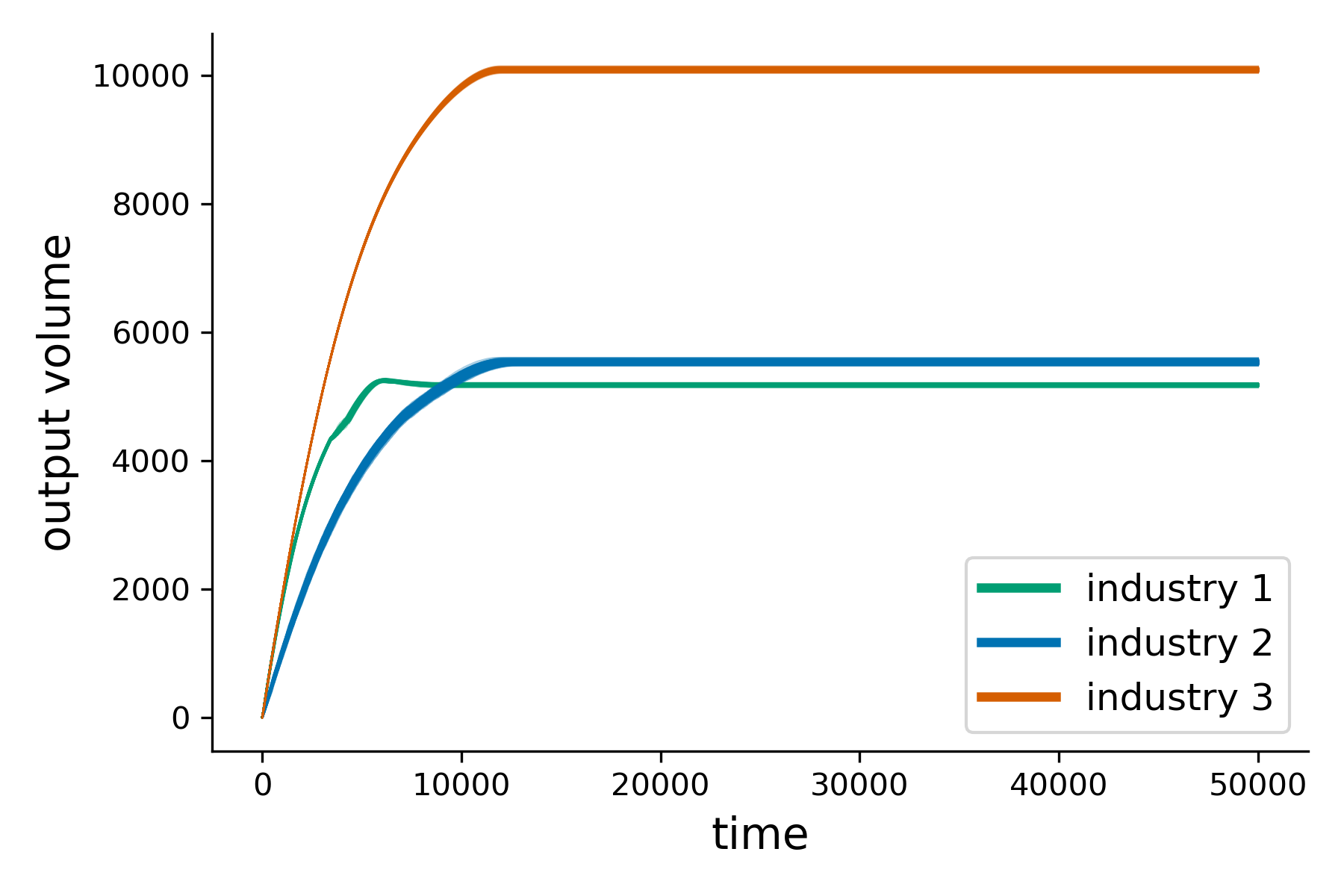}
    \end{minipage}
    \begin{minipage}{0.32\textwidth}
        \subcaption{Profit under same ICs}\label{fig:linear_dynamics_sample.profts}
        \includegraphics[angle=0,width=1.\textwidth]{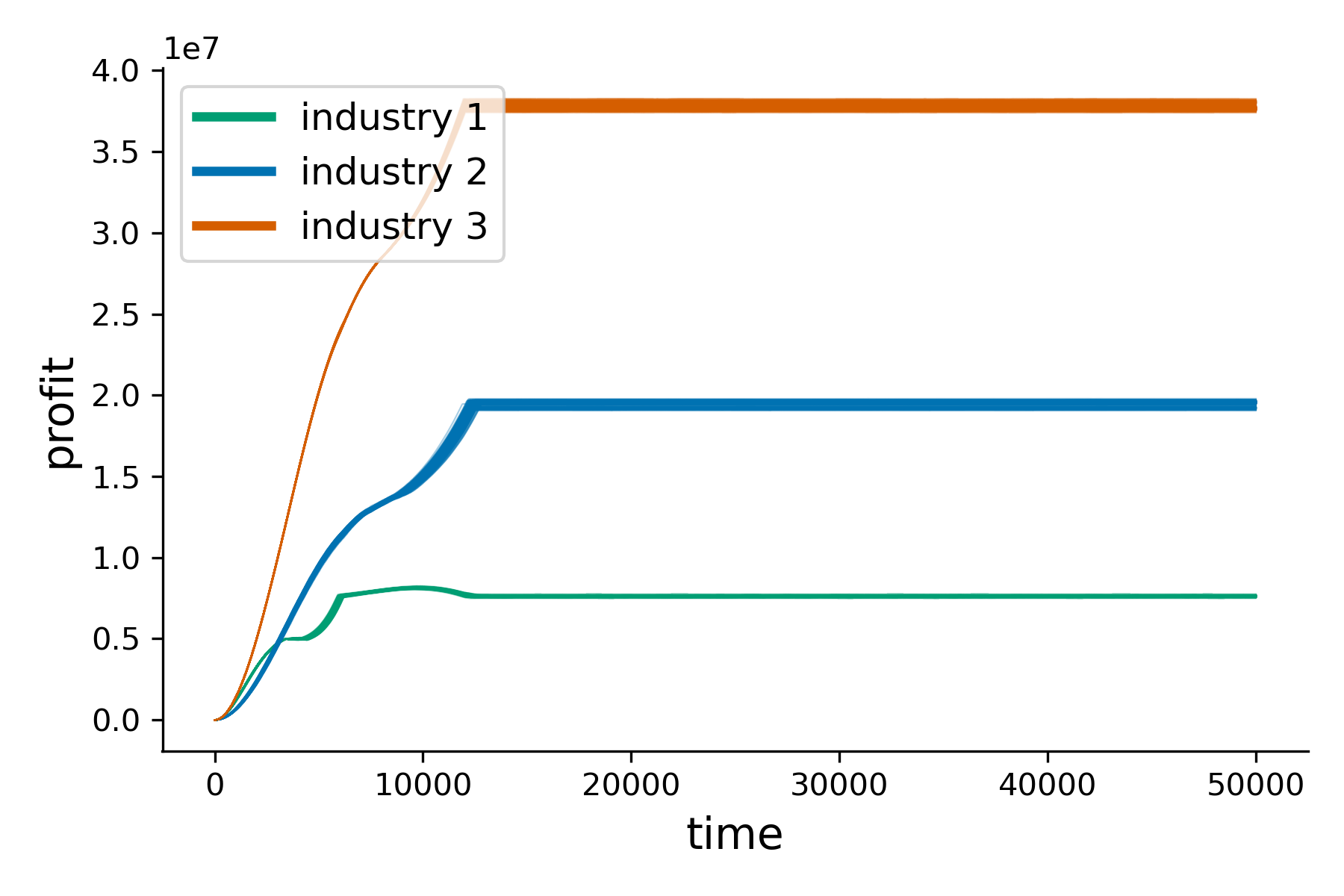}
    \end{minipage}
        \begin{minipage}{0.32\textwidth}
        \subcaption{Price under random ICs}\label{fig:linear_dynamics_sample_random.price}
        \includegraphics[angle=0,width=1.\textwidth]{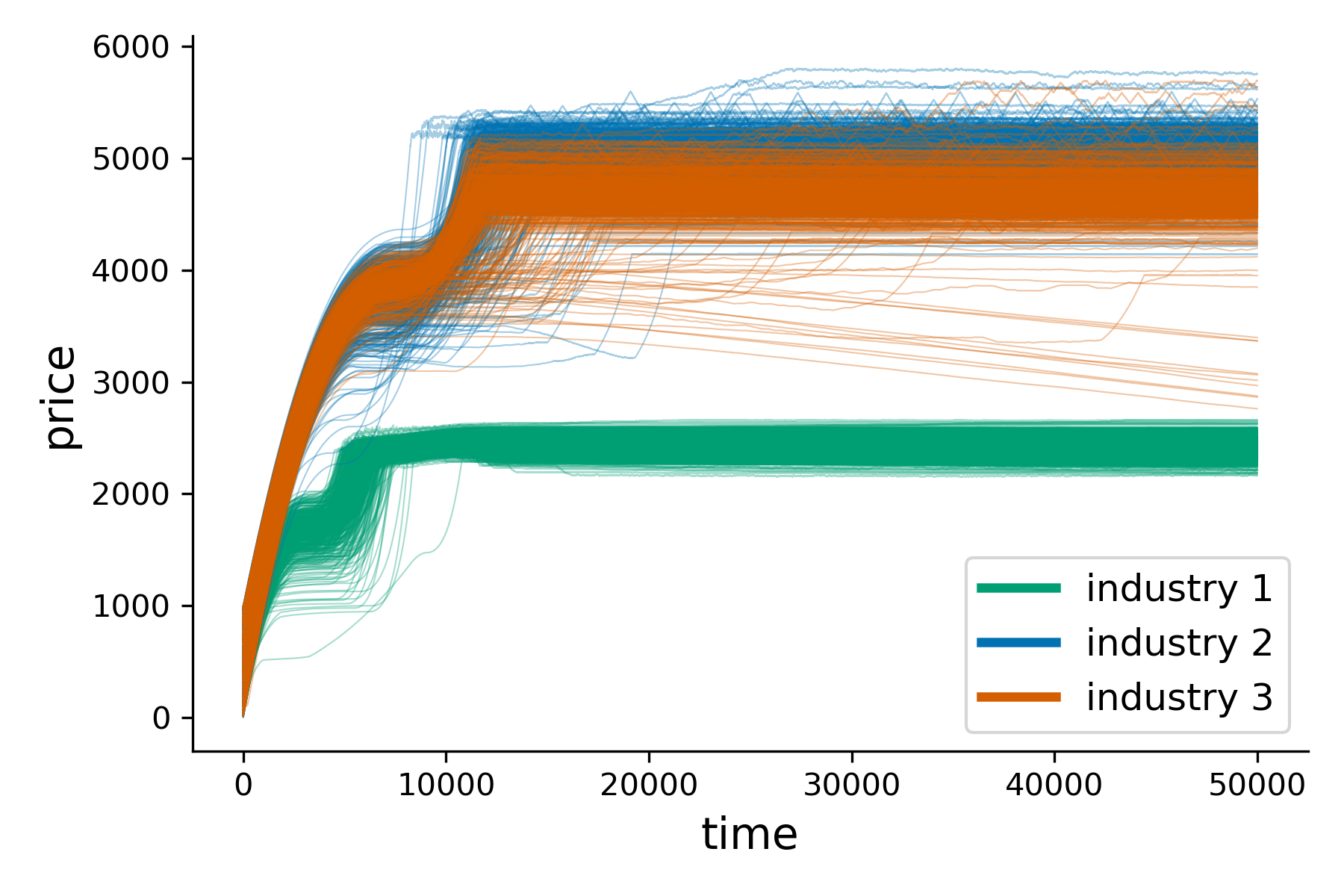}
    \end{minipage}
    \begin{minipage}{0.32\textwidth}
        \subcaption{Output under random ICs}\label{fig:linear_dynamics_sample_random.quantity}
        \includegraphics[angle=0,width=1.\textwidth]{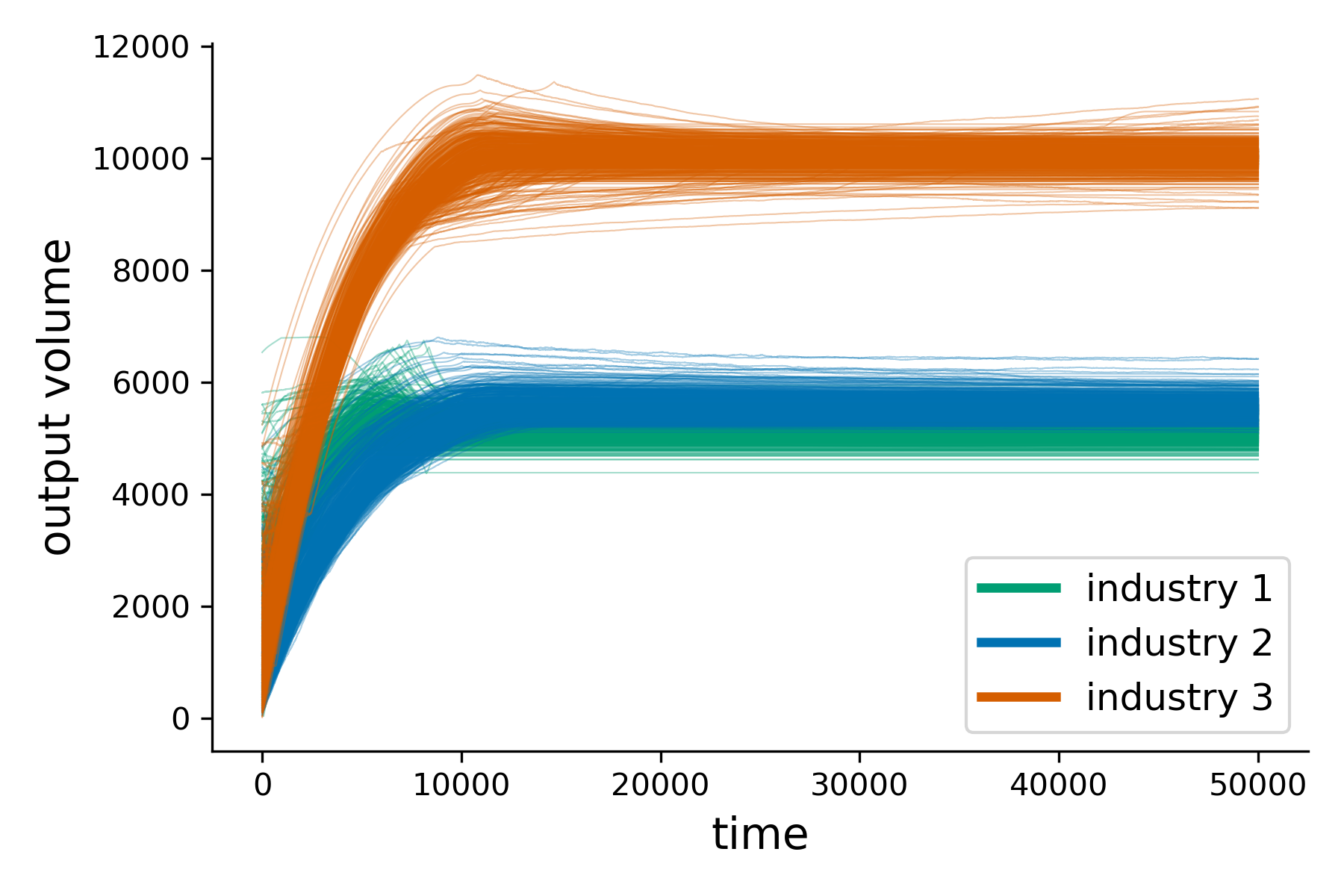}
    \end{minipage}
    \begin{minipage}{0.32\textwidth}
        \subcaption{Profit under random ICs}\label{fig:linear_dynamics_sample_random.profts}
        \includegraphics[angle=0,width=1.\textwidth]{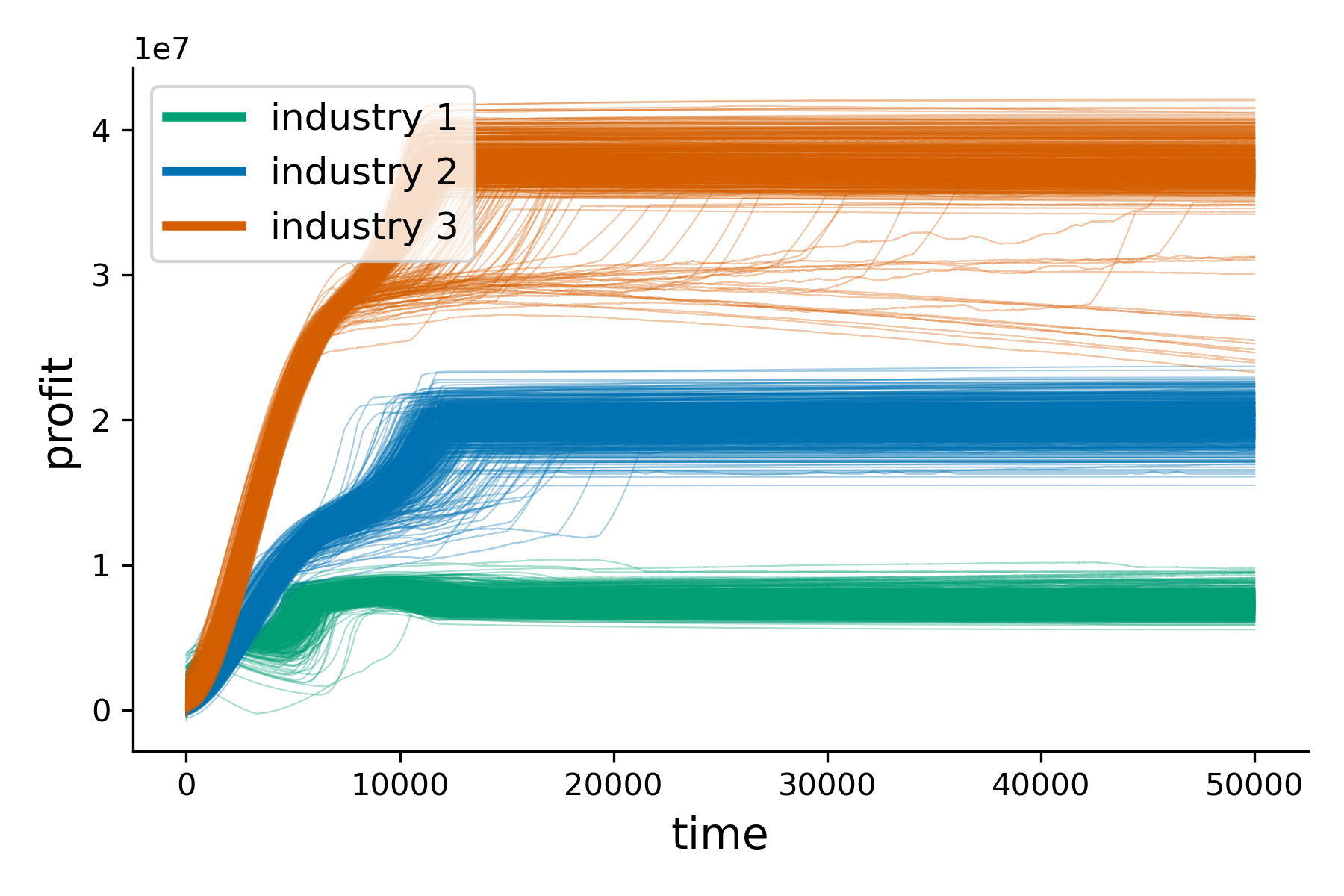}
    \end{minipage}
\caption*{\footnotesize 
\textit{\textbf{Notes}}: The top panels correspond to 1000 independent simulations with the same initial conditions.
The bottom panels show 1000 independent simulations with random initial conditions.
ICs stands for initial conditions.\\}
\end{figure}

The previous example shows the dynamics of the model for a single illustrative model run.
Next, we would like to show that those dynamics are robust across independent simulations with the same parameterization and under randomized initial conditions.
We show a set of 1000 of independent simulations in the top panels of \autoref{fig:linear_dynamics_sample}.
Each of these simulations has the same initial conditions, so the only difference between them is the random seed.
We can see that, overall, the model displays low variation in terms of the realized steady states.
Hence, one could argue that the dynamics of the model are robust under this setup.

The bottom panels of \autoref{fig:linear_dynamics_sample} show a set of independent runs when one randomizes the initial conditions of the model.
In this case, we can see that there is more variation in the realized steady states.
However, numerical verification indicates that the distribution of the outcome variables is unimodal.
Furthermore, a two-sample t-test for difference in means cannot reject the null hypothesis comparing the data of the top panels against that of the bottom panels.
Similar robustness is observed when looking at inter-industry flows.

\subsection{Numerical example with heterogeneous technologies}\label{sec:example_CES}

In economics, the canonical examples of production technologies are of three types: (1) linear, (2) Cobb-Douglas, and (3) Leontief.
Each of these types conveys a different nature on how inputs interact in a production process.
While inputs are fully substitutable in a linear technology, they are impossible to substitute in a Leontief (perfect complements).
The Cobb-Douglas sits between these cases, conveying imperfect substitutability/complementarity.
While there are several other types of production functions \citep{Mishra2010}, the three canonical forms are often integrated into the more general constant-elasticity-substitution (CES) technology, which takes the form

\begin{equation}
    Q_i = A_i \left[ \sum_{j=1}^N a_{i,j} q_{i,j}^{\rho_i} \right]^{\frac{\varphi_i}{\rho_i}},\label{eq:CES_function}
\end{equation}
where $A_i$ is the total factor productivity coefficient, parameters $a_{i,j}$ add up to one, $\varphi_i$ denotes the returns to scale ($\varphi_i=1$ corresponds to constant returns), and $\rho_i$ is the substitution parameter.
It is well known that $\rho_i \rightarrow 1$ leads to a linear function; $\rho_i \rightarrow 0$ to a Cobb-Douglas one; and $\rho_i \rightarrow -\infty$ to a Leontief.
Thus, to specify heterogeneous technologies in the model, we choose values for the substitution parameters that approximate the three canonical cases.
Note, however, that the model could accommodate other non-canonical technology specifications as firms need zero or minimal knowledge about the functional forms to maximize profits.

For the purpose of illustrating the workings of the model, we use the CES specification as the baseline to introduce technological heterogeneity.
More specifically, we parameterize constant-returns-to-scale technologies for the three as follows

\begin{equation}
    \begin{split}
      A_1=10,\quad a_{1,1}=0.12 ,\quad a_{1,2}=0.44,\quad a_{1,3}=0.44,\quad \rho_1=-10,\quad \varphi_1=1.00\\
      A_2=10,\quad a_{2,1}=0.14 ,\quad a_{2,2}=0.72,\quad a_{2,3}=0.14,\quad \rho_2=0.001,\quad \varphi_2=1.00\\
      A_3=10,\quad a_{3,1}=0.50 ,\qquad \qquad \qquad \qquad a_{3,3}=0.50,\quad \rho_3=1.00,\quad \varphi_3=1.00
    \end{split},\label{eq:CES_parameters}
\end{equation}
so firm 1 tends to use inputs as strong complements, firm 2 as imperfect substitutes, and firm 3 as perfect substitutes.
In this specification, we preserve the implied productive structure from the example with linear technologies, i.e., firm 3 does not require inputs from firm 2.

\begin{figure}[ht]
\centering
\caption{Dynamics under heterogeneous technologies}\label{fig:CES_dynamics}
    \begin{minipage}{0.24\textwidth}
        \subcaption{Consistent learning}\label{fig:CES_dynamics.demands}
        \includegraphics[angle=0,width=1.\textwidth]{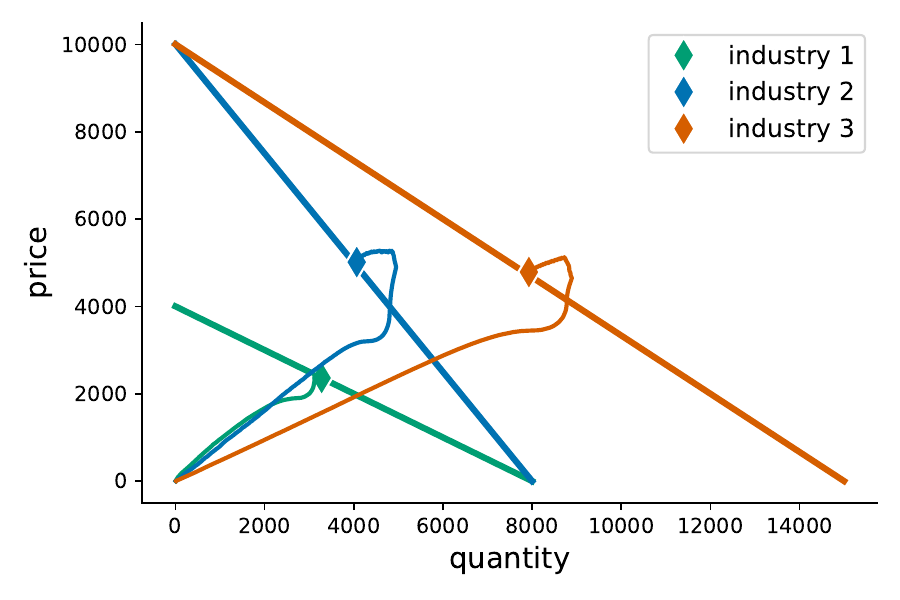}
    \end{minipage}
    \begin{minipage}{0.24\textwidth}
        \subcaption{Prices}\label{fig:CES_dynamics.price}
        \includegraphics[angle=0,width=1.\textwidth]{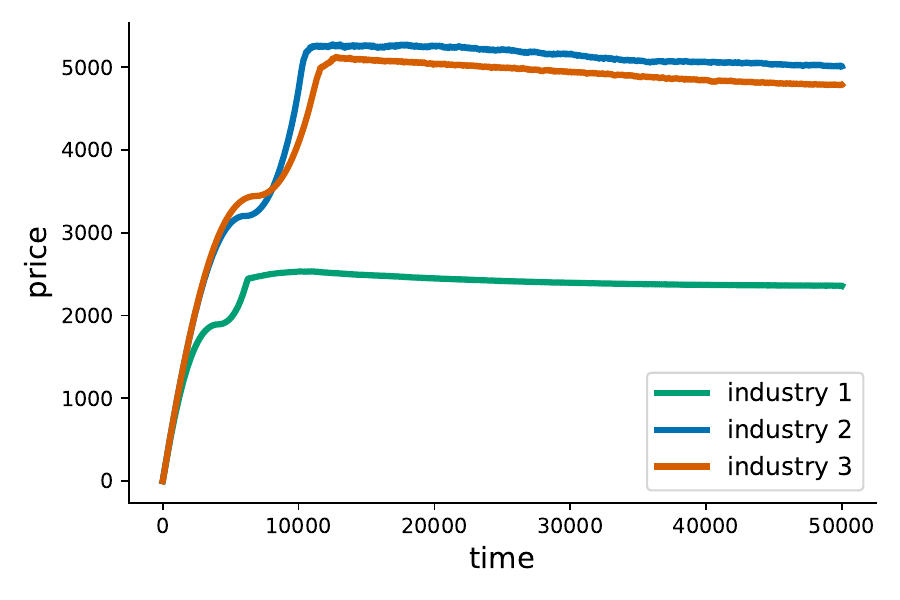}
    \end{minipage}
    \begin{minipage}{0.24\textwidth}
        \subcaption{Output volumes}\label{fig:CES_dynamics.quantity}
        \includegraphics[angle=0,width=1.\textwidth]{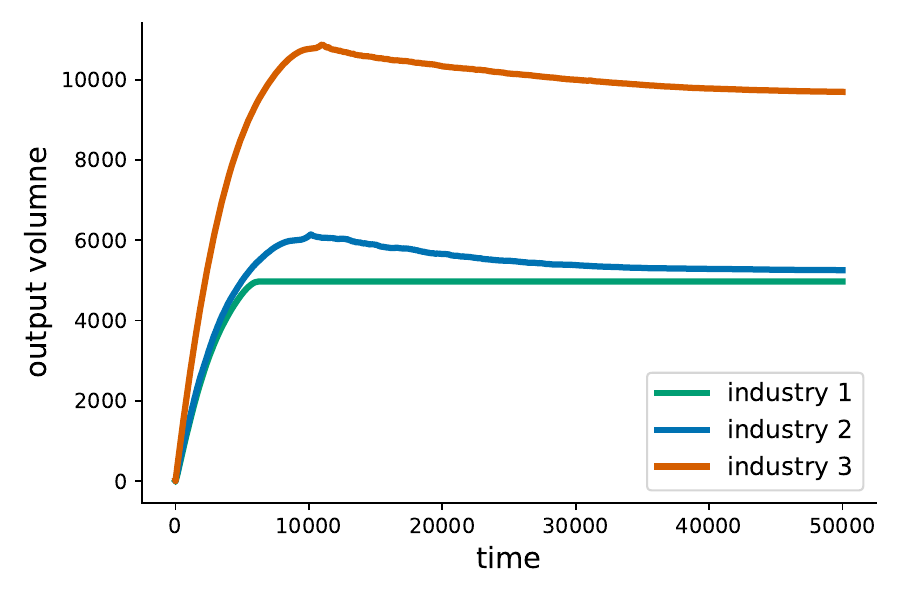}
    \end{minipage}
    \begin{minipage}{0.24\textwidth}
        \subcaption{Profits}\label{fig:CES_dynamics.profts}
        \includegraphics[angle=0,width=1.\textwidth]{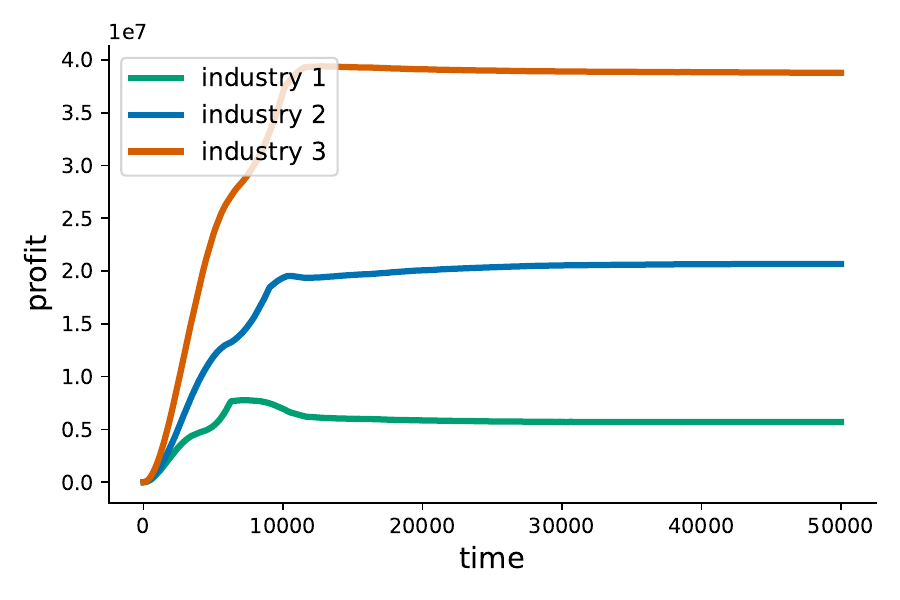}
    \end{minipage}
\end{figure}

\autoref{fig:CES_dynamics} presents the dynamics of the model.
Notice that, with heterogeneous technologies, the three industries achieve consistent learning, as \autoref{fig:CES_dynamics.demands} shows how they set their price and residual quantity on their corresponding demand curves.
Price, volume, and profit variables also stabilize.
It is interesting to see that the ordering of the steady-state outcomes is the same as the case with linear technologies, which owes to preserving the same demand functions.
As we will see ahead, this is not the case when considering other types of firm heterogeneity such as returns to scale.

\begin{figure}[ht]
\centering
\caption{Endogenous inter-industry flows under heterogeneous technologies}\label{fig:ternary_CES}
    \begin{minipage}{0.32\textwidth}
        \subcaption{Purchased volume}\label{fig:ternary_CES.volumes}
        \includegraphics[angle=0,width=1.\textwidth]{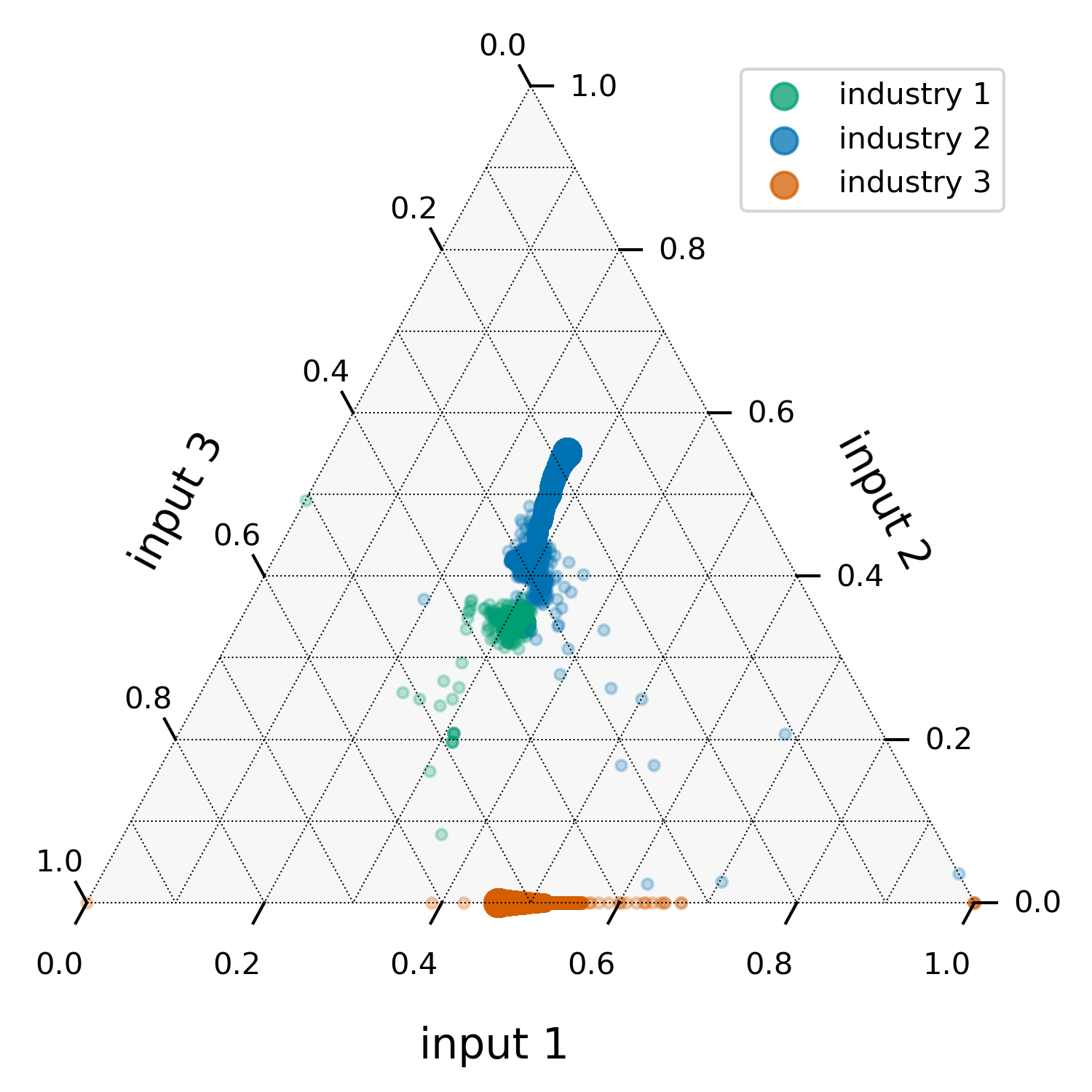}
    \end{minipage}
    \begin{minipage}{0.32\textwidth}
        \subcaption{Costs}\label{fig:ternary_CES.costs}
        \includegraphics[angle=0,width=1.\textwidth]{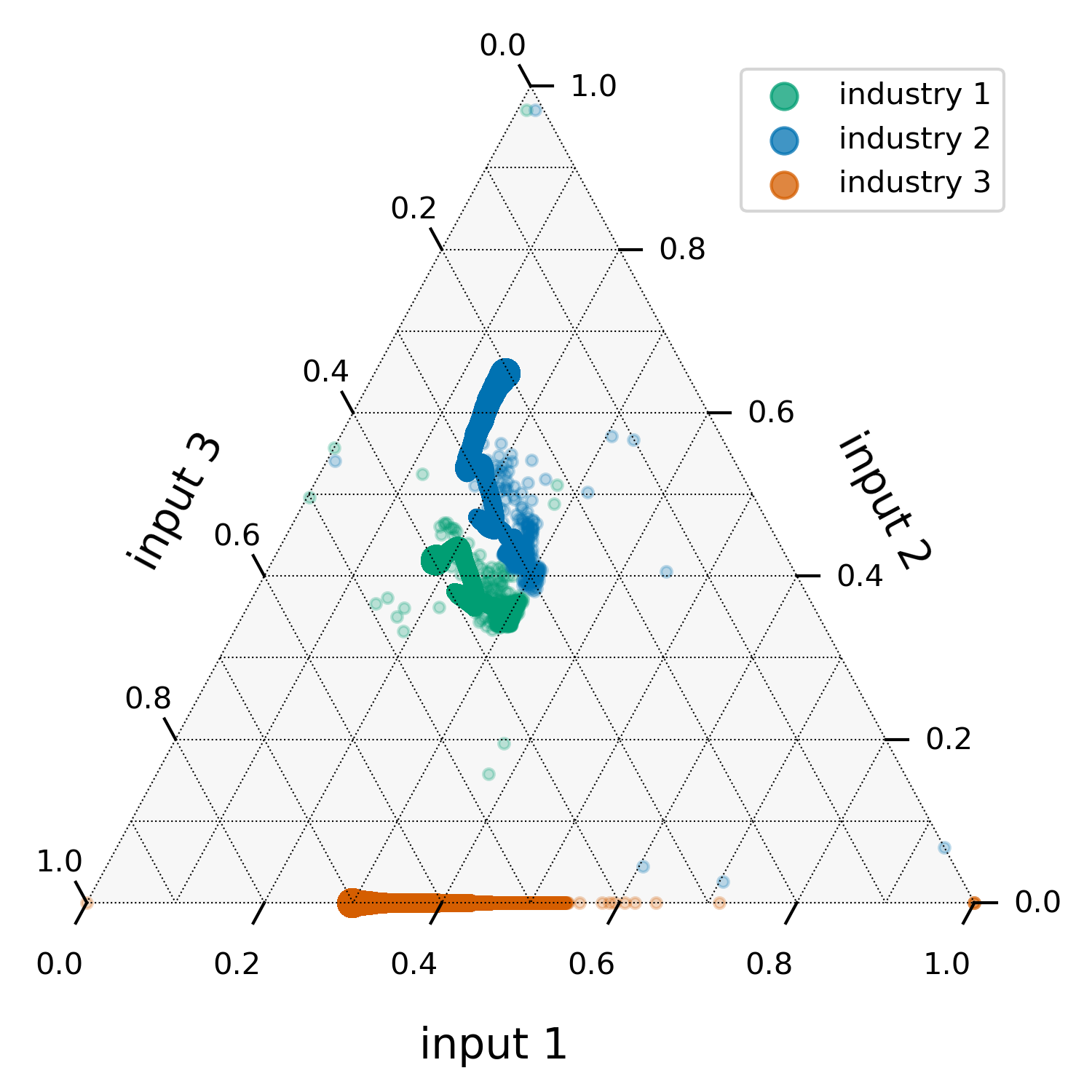}
    \end{minipage}
    \begin{minipage}{0.32\textwidth}
        \subcaption{Sales}\label{fig:ternary_CES.sales}
        \includegraphics[angle=0,width=1.\textwidth]{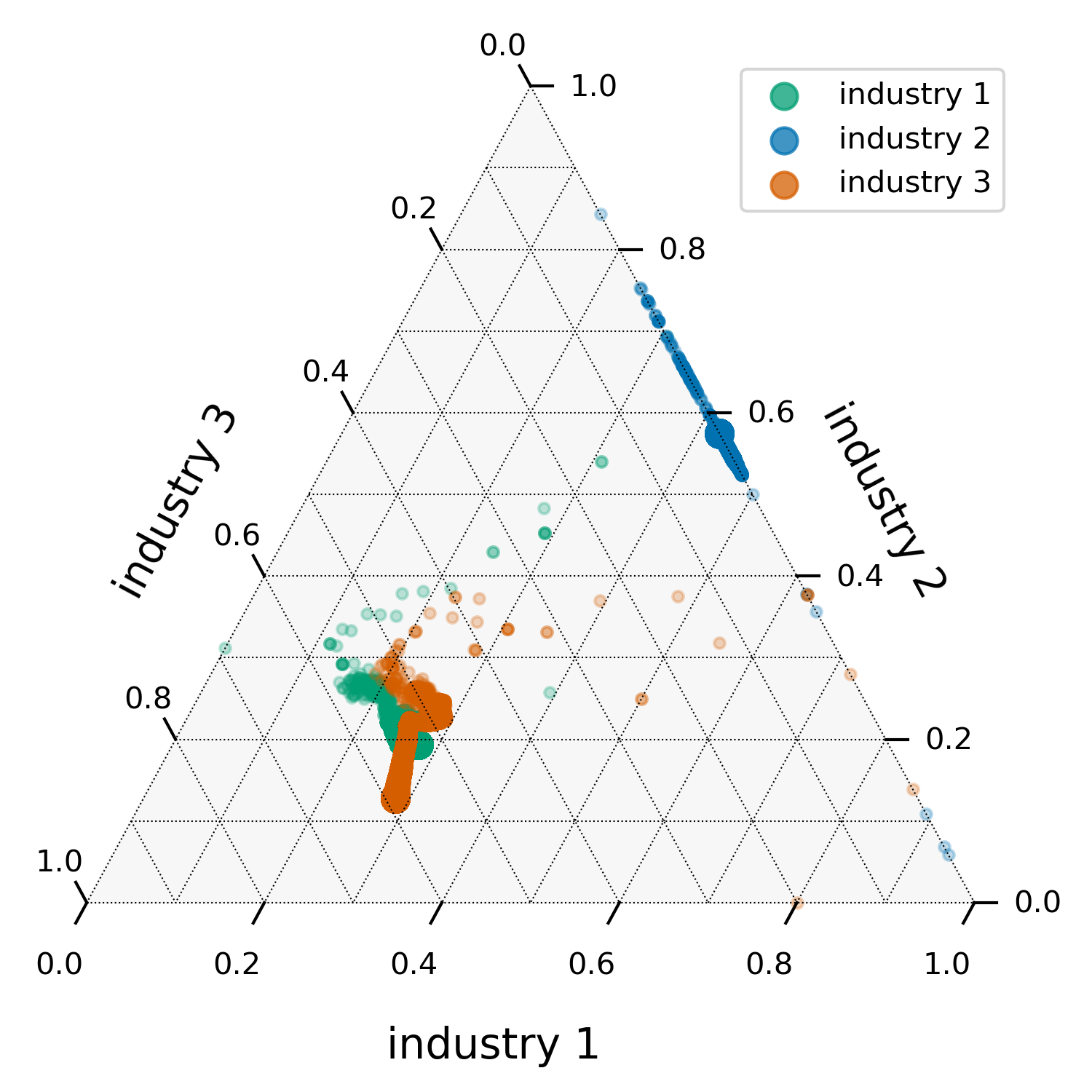}
    \end{minipage}
\caption*{\footnotesize 
\textit{\textbf{Notes}}: Larger and more opaque markers indicate more recent periods.}
\end{figure}

Next, we present the inter-industry flows in \autoref{fig:ternary_CES} under heterogeneous technologies.
In this case, the trajectories differ from the case with linear technologies.
The steady-state inter-industry flows differ as well, suggesting that the endogenous production network that emerges is different from the one with linear technologies.
This difference may have further implications in terms of adaptability to shocks, something that we will cover ahead.

\subsection{Numerical example with heterogeneous returns scale}\label{sec:example_increasing}

Now we look at the model under heterogeneous returns to scale.
Empirical evidence on firm dynamics suggest vast heterogeneity in how firms scale their output as a consequence of using more inputs, i.e., the common assumption of constant returns to scale is unlikely to hold in the real world.
Operating with non-constant returns to scale and, furthermore, with industries that exhibit heterogeneity in the nature of their returns, is an analytical challenge.
Here, we show that our model is able to naturally handle different types of returns to scale.

Recall that parameter $\varphi_i$ from \autoref{eq:CES_function} determines the nature of the returns.
Thus, we continue with the parameter specification presented in \autoref{eq:CES_parameters}, but assigning a different $\varphi_i$ to each industry.
More specifically, we set constant returns for firm 1 ($\varphi_1=1$), decreasing returns to firm 2 ($\varphi_2=0.9$), and increasing returns to firm 3 ($\varphi_3=1.5$).

\begin{figure}[ht]
\centering
\caption{Dynamics under heterogeneous returns to scale}\label{fig:returns_dynamics}
    \begin{minipage}{0.24\textwidth}
        \subcaption{Consistent learning}\label{fig:returns_dynamics.demands}
        \includegraphics[angle=0,width=1.\textwidth]{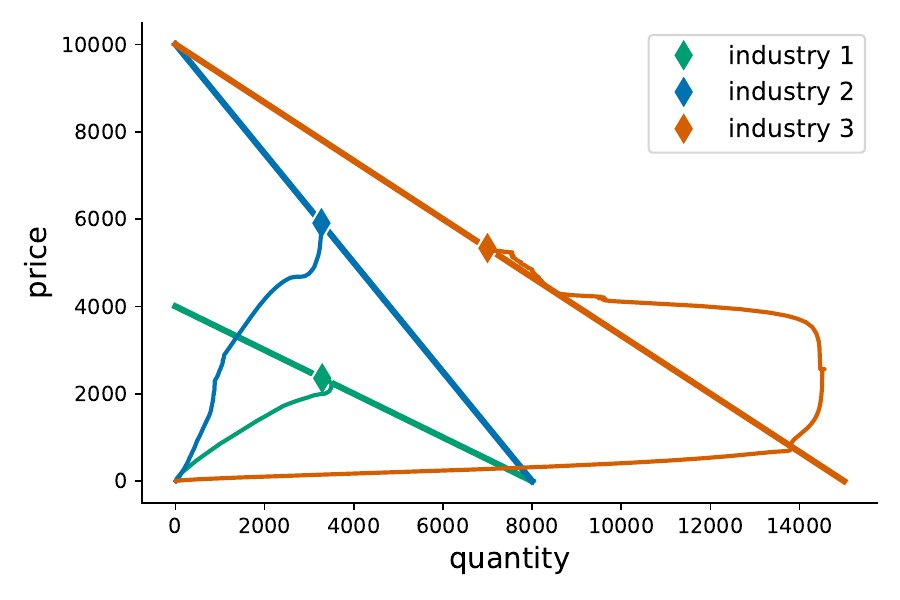}
    \end{minipage}
    \begin{minipage}{0.24\textwidth}
        \subcaption{Prices}\label{fig:returns_dynamics.price}
        \includegraphics[angle=0,width=1.\textwidth]{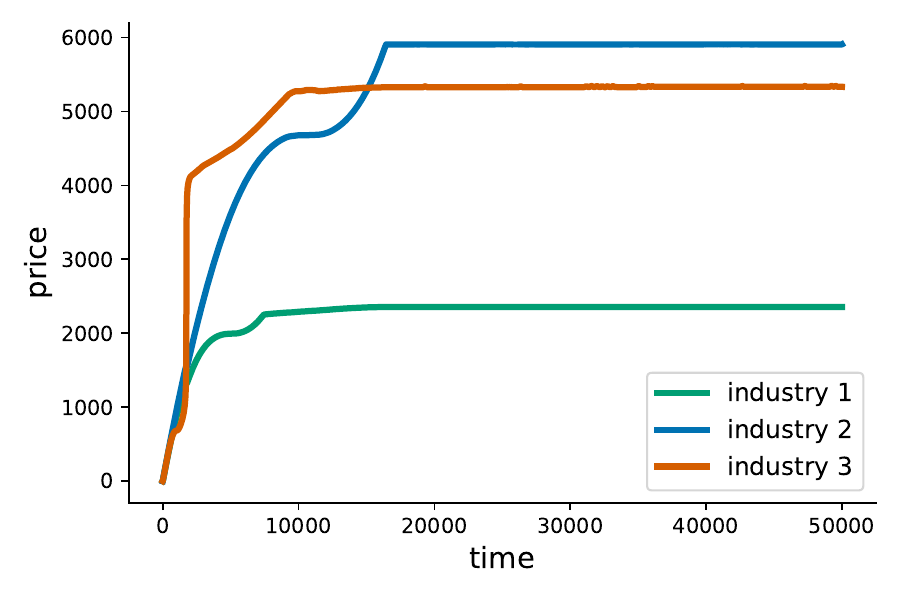}
    \end{minipage}
    \begin{minipage}{0.24\textwidth}
        \subcaption{Output volumes}\label{fig:returns_dynamics.quantity}
        \includegraphics[angle=0,width=1.\textwidth]{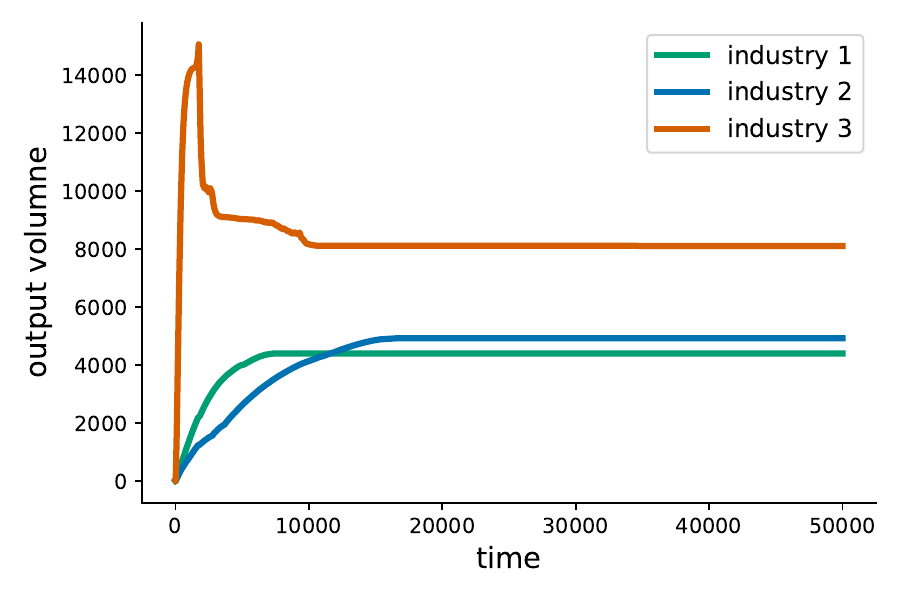}
    \end{minipage}
    \begin{minipage}{0.24\textwidth}
    \subcaption{Profits}\label{fig:returns_dynamics.profts}
        \includegraphics[angle=0,width=1.\textwidth]{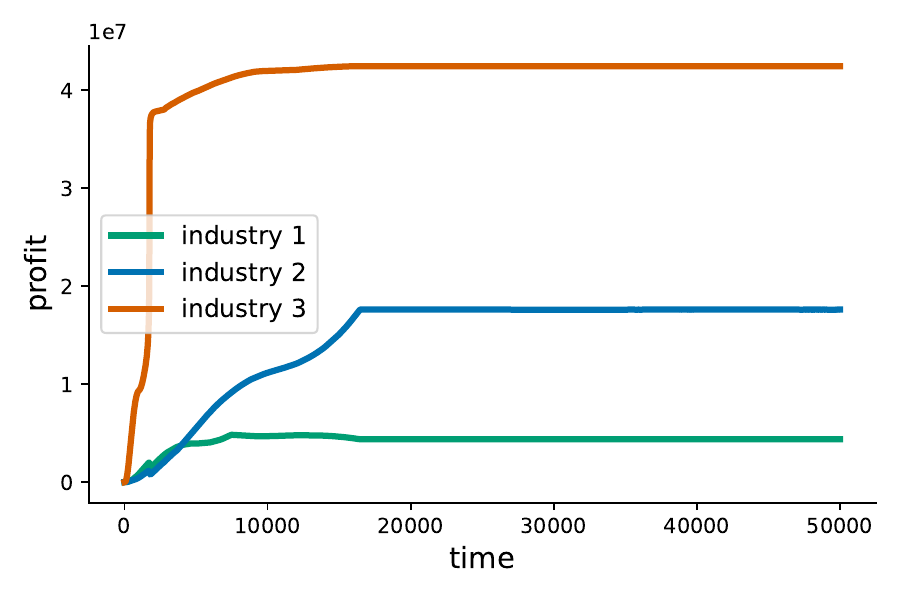}
    \end{minipage}
\end{figure}

\autoref{fig:returns_dynamics} shows the dynamics emerging from a 3-industry economy where each firm has a different type of returns to scale.
First, in \autoref{fig:returns_dynamics.demands} we can see that firms reach consistent learning as well.
However, in the case of firm 3 (increasing returns), learning along the demand curve is achieved through more exploration.
Second, the exploration of firm 3 translates into price volatility in \autoref{fig:returns_dynamics.price}.
Third, the punctuated character of the learning process for firm 3 is accentuated under increasing returns.
For example, \autoref{fig:returns_dynamics.quantity} shows a quick boom and bust in the output volume of firm 3 during the initial learning phase.
Fourth, steady-state profits remain qualitatively consistent with the case of constant returns to scale, with the difference that firm 3 reaches the steady state faster.

\begin{figure}[ht]
\centering
\caption{Endogenous inter-industry flows under heterogeneous returns to scale}\label{fig:ternary_returns}
    \begin{minipage}{0.32\textwidth}
        \subcaption{Purchased volume}\label{fig:ternary_returns.volumes}
        \includegraphics[angle=0,width=1.\textwidth]{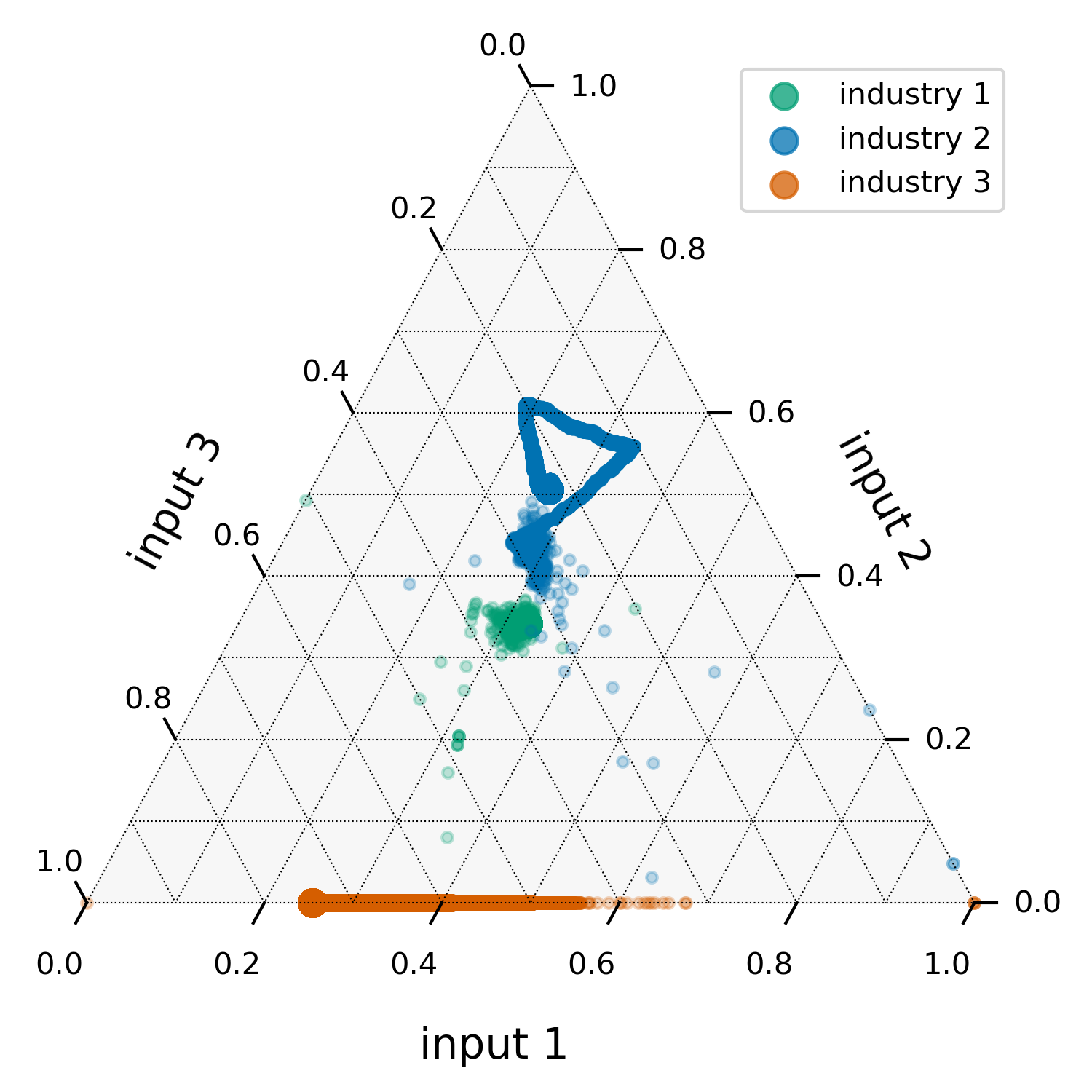}
    \end{minipage}
    \begin{minipage}{0.32\textwidth}
        \subcaption{Costs}\label{fig:ternary_returns.costs}
        \includegraphics[angle=0,width=1.\textwidth]{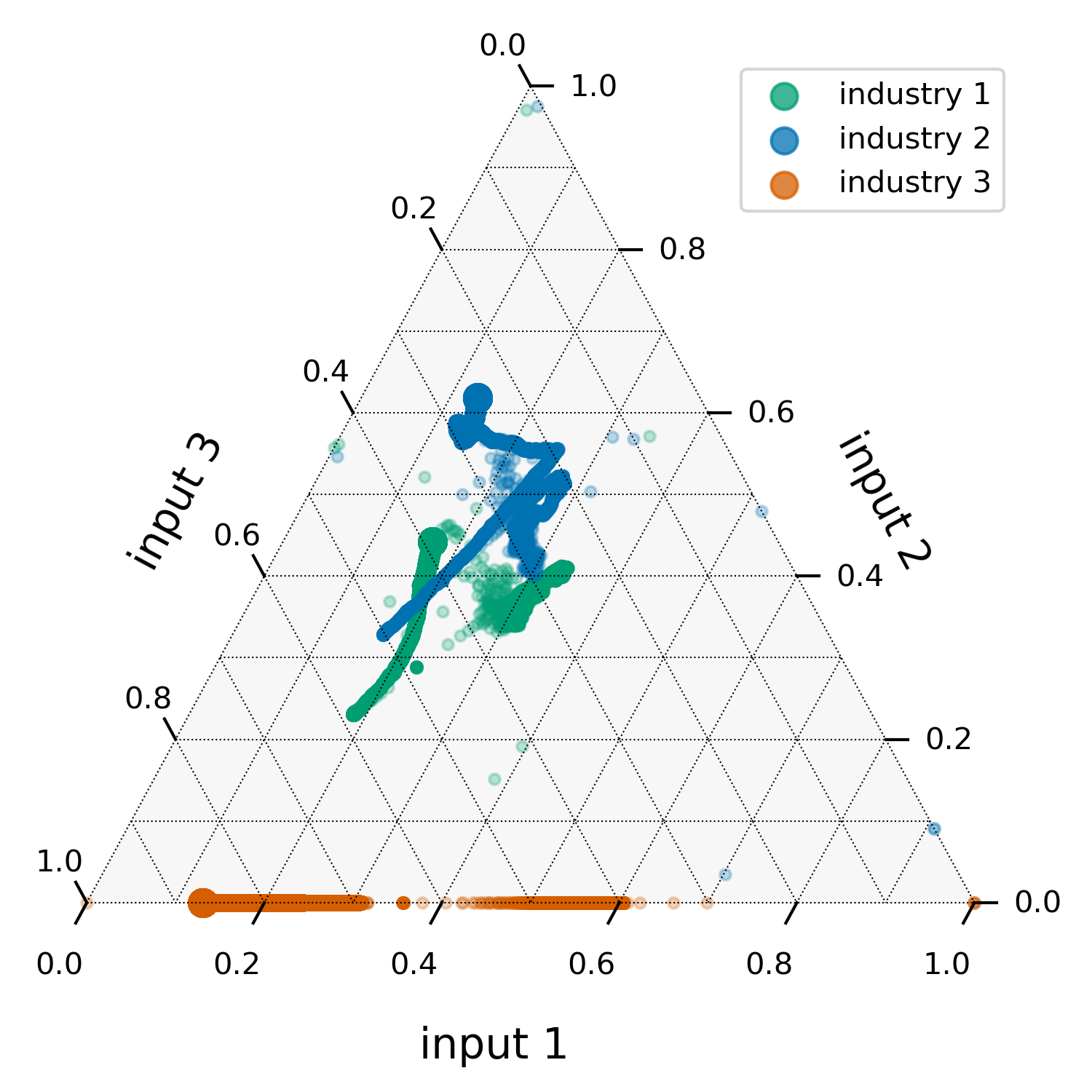}
    \end{minipage}
    \begin{minipage}{0.32\textwidth}
        \subcaption{Sales}\label{fig:ternary_returns.sales}
        \includegraphics[angle=0,width=1.\textwidth]{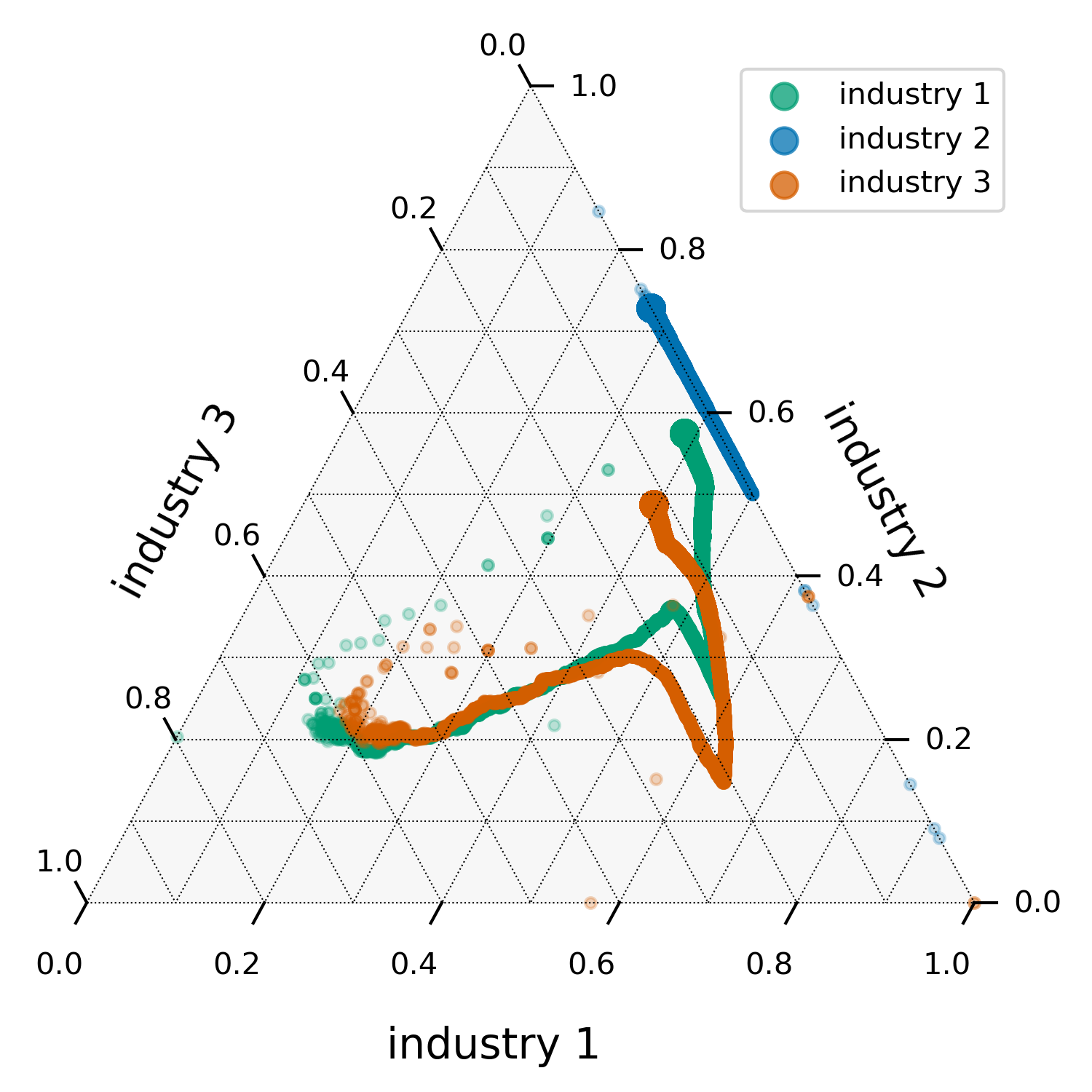}
    \end{minipage}
\caption*{\footnotesize 
\textit{\textbf{Notes}}: Larger and more opaque markers indicate more recent periods.}
\end{figure}

Next, in \autoref{fig:ternary_returns}, we show the endogenous inter-industry flows.
First, the trajectories look qualitatively different from the two previous examples.
For instance, longer distances with opaque markers indicate that exploration and learning in terms of the production network takes longer.
It is interesting to see that this difference is not evident from the more aggregate plot presented in \autoref{fig:returns_dynamics}.
It comes to highlight the importance of focusing on more disaggregate dynamics.
Second, in \autoref{fig:ternary_returns.volumes}, we can see that firm 2 revisits previously explored purchasing proportions, as its trajectory indicates a cycle.
Industry 1 (constant returns to scale), in contrast, exhibits sporadic purchasing configurations and a quick settling in a well-defined neighborhood.
This also suggests that parts of the production network settle in their steady-state value quicker than others, something that may be important to consider when thinking about adaptation to shocks.
Finally, the evolution of profits also exhibit longer trajectories than the previous examples.

\section{Shocks and their propagation}

So far we have demonstrated how endogenous production networks can be explained from simple learning behavior under uncertainty.
From the previous examples, it is clear that various aspects about the dynamics unfolding in the network-formation process are relevant to understand changes and adaptation to exogenous shocks.
We have also argued that the types of shocks that can be operationalized in our model are more varied than those that are permissible in a rational-equilibrium framework; for example, modifying a production function on the run.
In this section, we investigate--with examples--different types of shocks and present their propagation outcomes.
More specifically, we implement shocks to: (1) the aggregate demand, (2) returns to scale, (3) production technologies; and (4) industry shutdown.
To implement these examples, we augment the size of the economy by adding two more industries.
We intentionally design the implied productive structure to place these firms in a semi-isolated position in the economy, as it allows for interesting shock scenarios.
\label{sec:shocks}

\begin{equation}
    \begin{split}
      A_1=10, \quad a_{1,1}=0.13, \quad a_{1,2}=0.31,\quad a_{1,3}=0.31, \quad a_{1,4}=0.25, \qquad \qquad \qquad \quad \rho_1=1,\quad \varphi_1=1\\
      A_2=10,\quad a_{2,1}=0.08, \quad a_{2,2}=0.38,\quad a_{2,3}=0.08,\quad a_{2,4}=0.46, \qquad \qquad \qquad \quad \rho_2=1,\quad \varphi_2=1\\
      A_3=10,\quad a_{3,1}=0.50 ,\qquad \qquad \qquad \quad a_{3,3}=0.50,\quad \qquad \qquad \qquad \qquad \qquad \qquad \rho_3=1,\quad \varphi_3=1\\
      A_4=10, \qquad \qquad \qquad \qquad \qquad \qquad \qquad \qquad \qquad \quad a_{4,4}=0.20, \quad a_{4,5}=0.80,\quad \rho_4=1,\quad \varphi_4=1\\
      A_5=10, \qquad \qquad \qquad \qquad \qquad \qquad \qquad \qquad \qquad \quad a_{5,4}=0.88, \quad a_{5,5}=0.12,\quad \rho_5=1,\quad \varphi_5=1\\
    \end{split},\label{eq:CES_parameters_5}
\end{equation}

\begin{figure}[ht]
\centering
\caption{Implied productive structure in augmented economy}\label{fig:enlarged_net}
        \includegraphics[angle=0,width=.8\textwidth]{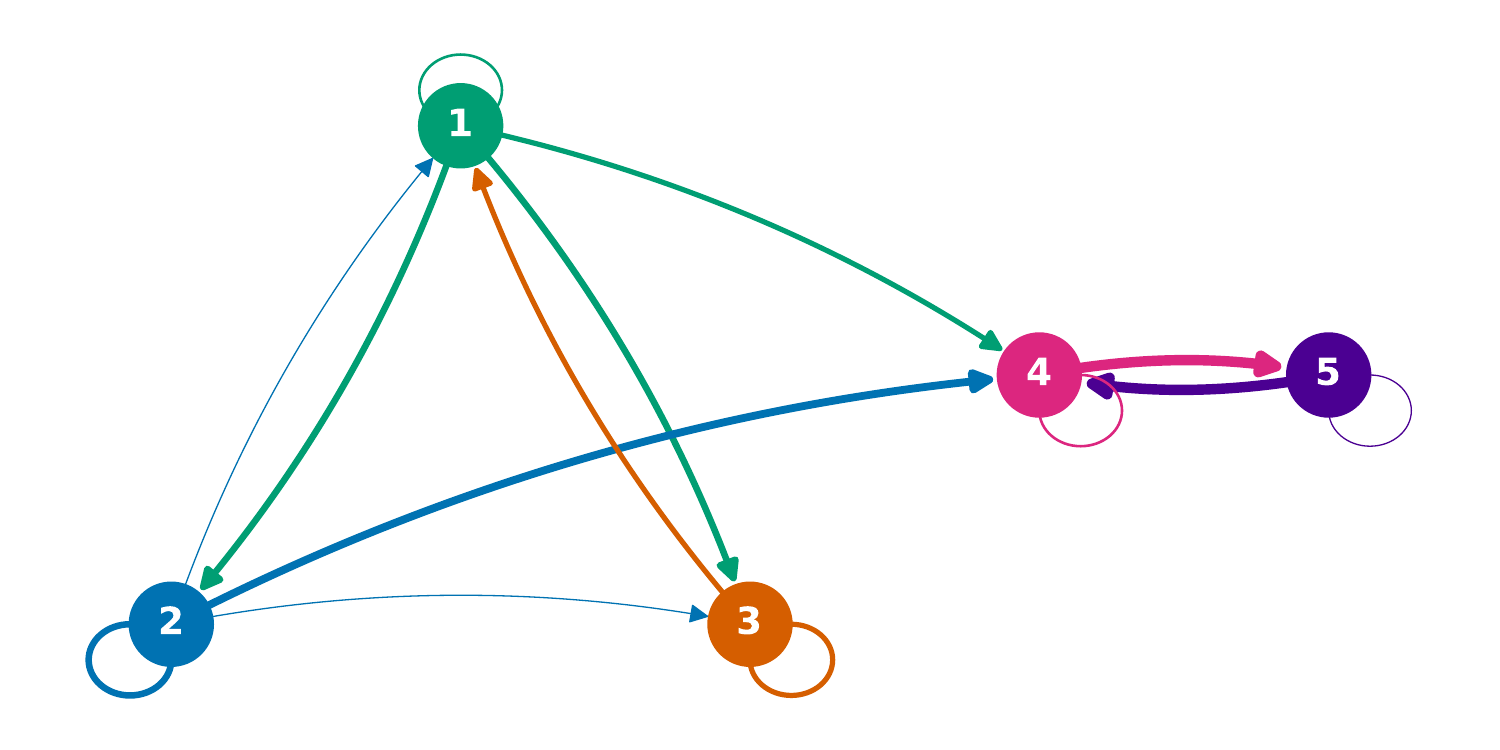}
        \caption*{\footnotesize 
        \textit{\textbf{Notes}}: The arrows indicate purchasing relationships.
        For example $3 \rightarrow 1$ means industry 3 buys inputs from industry 1.
        The circular edges without arrow mean self-loops.
        Nodes 1, 2, and 3 are core industries while 4 and 5 form the periphery.\\}
\end{figure}

The baseline technologies are CES production functions with the parameters specified in \autoref{eq:CES_parameters_5}.
In this setting, inputs are perfect substitutes and yield constant returns to scale.
We illustrate the implied productive structure in \autoref{fig:enlarged_net}.
As it can be seen in this depiction, industry 4 plays an intermediary role between the core of the economy and firm 5. 
However, firm 4 does not need any inputs from the core firms.
Moreover, industry 5 only depends on firm 4 and itself.
Finally, in addition to the demands functions already specified in \autoref{eq:linear_demands}, we add the two new demands $Q^d_4 = 11000 - 2p_3$ and $Q^d_5 = 11000 - p_3$ for the new industries.

\subsection{Baseline dynamics}

\begin{figure}[ht]
\centering
\caption{Baseline dynamics with five industries}\label{fig:five_dynamics}
    \begin{minipage}{0.24\textwidth}
        \subcaption{Consistent learning}\label{fig:five_dynamics.demands}
        \includegraphics[angle=0,width=1.\textwidth]{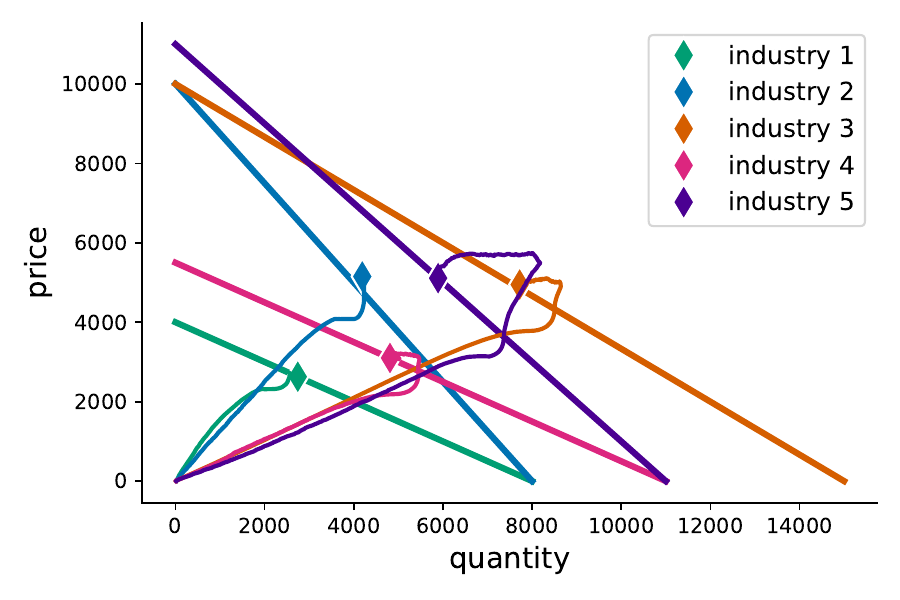}
    \end{minipage}    
    \begin{minipage}{0.24\textwidth}
        \subcaption{Prices}\label{fig:five_dynamics.price}
        \includegraphics[angle=0,width=1.\textwidth]{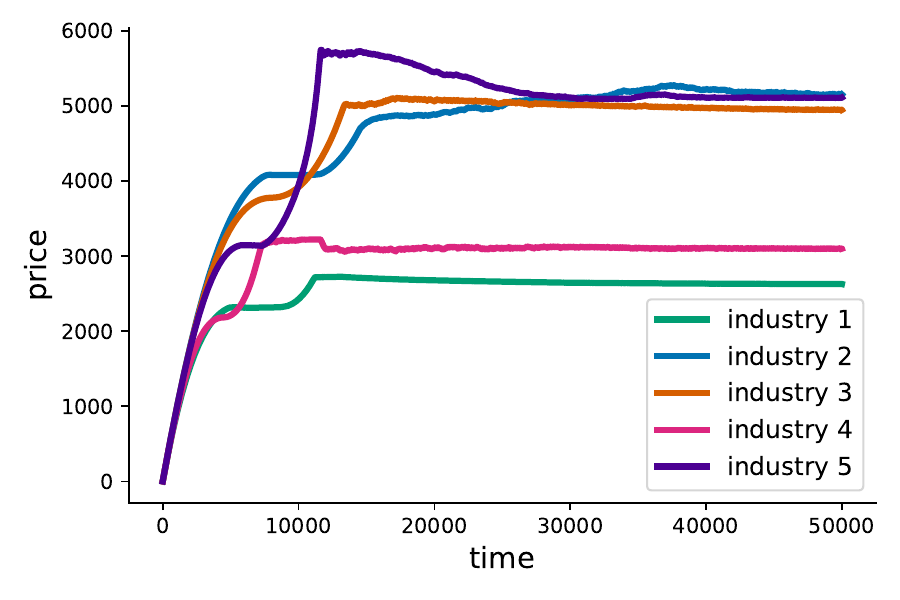}
    \end{minipage}
    \begin{minipage}{0.24\textwidth}
        \subcaption{Output volumes}\label{fig:five_dynamics.quantity}
        \includegraphics[angle=0,width=1.\textwidth]{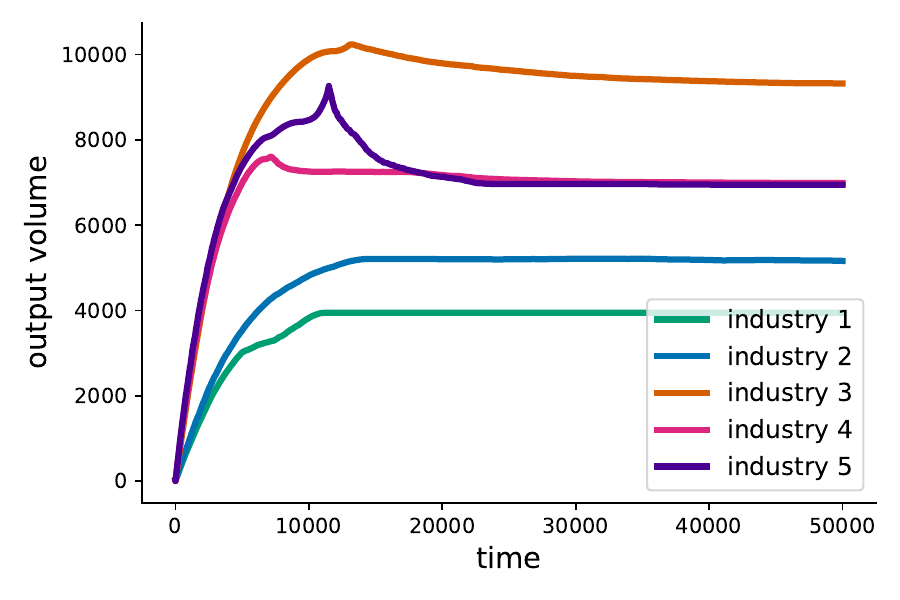}
    \end{minipage}
    \begin{minipage}{0.24\textwidth}
        \subcaption{Profits}\label{fig:five_dynamics.profts}
        \includegraphics[angle=0,width=1.\textwidth]{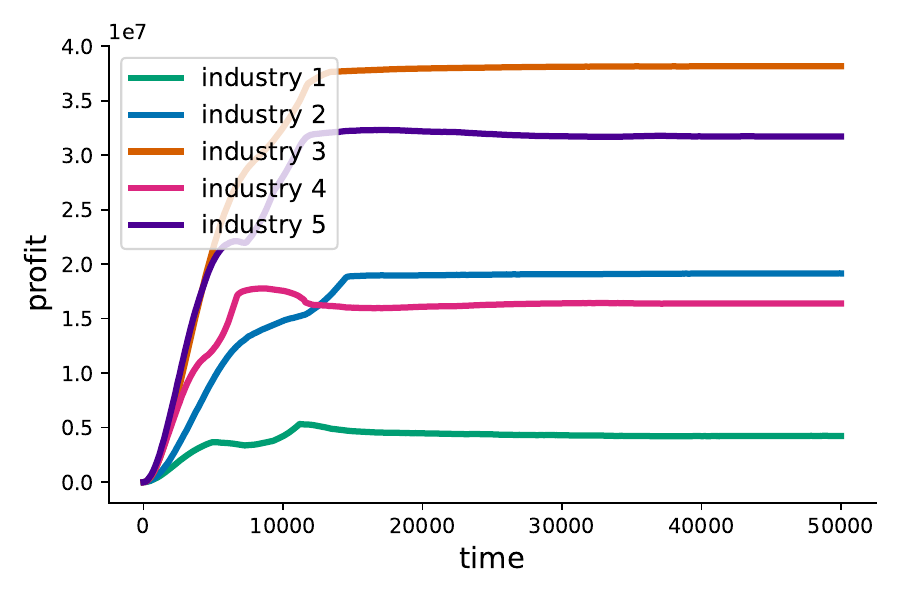}
    \end{minipage}
\end{figure}

First, let us revise the baseline dynamics of the augmented model when no shocks take place.
\autoref{fig:five_dynamics} shows that the enlarged model preserves consistent learning and stability.
Interestingly, industry 5 exhibits some price volatility, mainly due to the potential limit-cycle dynamics induced by engaging only with firm 4 and itself.
Overall, the baseline dynamics are robust and consistent with all the previous examples, providing a benchmark to measure the response of the production network to exogenous shocks.

\subsection{Demand shocks}\label{sec:demand_shocks}

We consider two types of demand shocks: increasing and decreasing its elasticity by 50\%.
On the one hand, an example of an increase in elasticity is when more substitutes are made available to the consumers, for instance, due to lower trade tariffs or a patent's expiration that enables more firms to produce the same or similar product.
On the other, an decrease in elasticity can be expected when a good becomes more essential to consumers, for example, the ubiquity of smart phones as support tools in everyday life.
The shock is implemented by, first, letting the model reach its baseline dynamics and, then, inducing the change.
Since firms need to adapt their prices and quantities to the new demands, it is important to verify that learning remains consistent.
In other words, industries should be able to `find' the location of the new demand and establish prices and quantities in a new neighborhood on the affected demand curves.
Once the new steady state is reached, we measure the difference with respect to the baseline case.

\begin{figure}[ht]
\centering
\caption{Example of two shocks to firm 4 under different technologies}\label{fig:shock_one_prod}
    \begin{minipage}{0.49\textwidth}
        \subcaption{Perfect substitutes}\label{fig:shock_one_prod.linear}
        \includegraphics[angle=0,width=1.\textwidth]{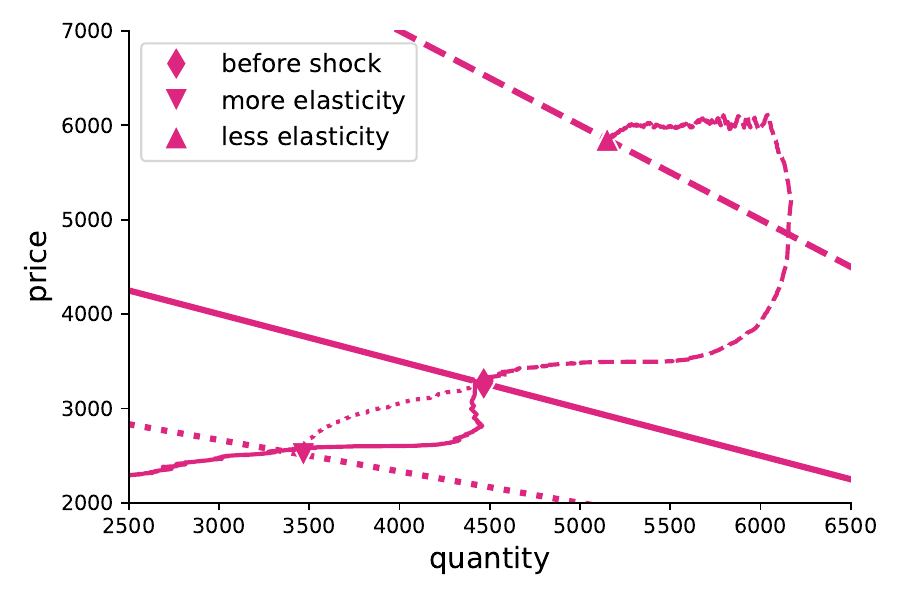}
    \end{minipage}
    \begin{minipage}{0.49\textwidth}
        \subcaption{Imperfect substitutes}\label{fig:shock_one_prod.leontief}
        \includegraphics[angle=0,width=1.\textwidth]{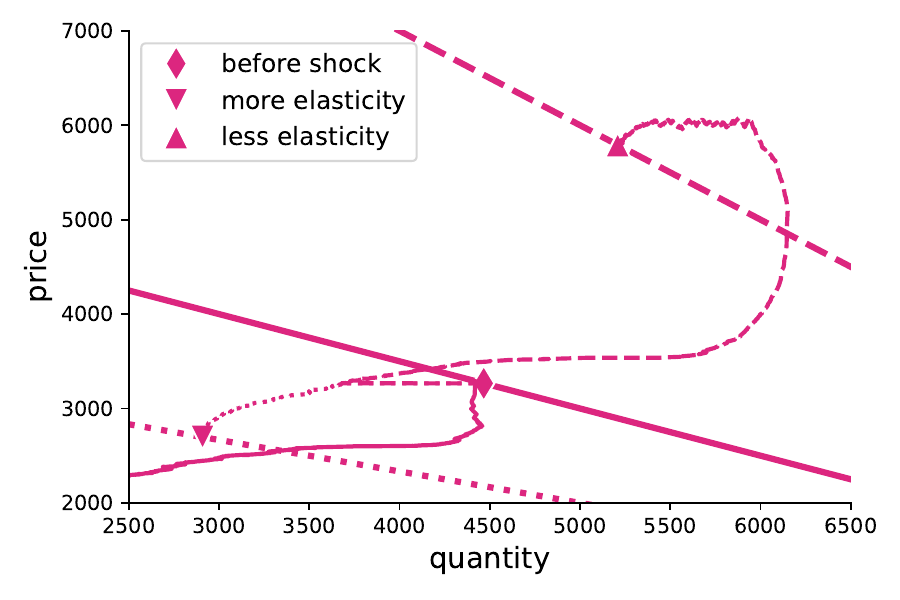}
    \end{minipage}
\end{figure}

As a first example, let us demonstrate how firm 4 adapts to demand shocks.
\autoref{fig:shock_one_prod.linear} shows the baseline and the two shocked demand curves.
Industry 4 is able to adapt and learn consistently under both types of shocks.
When the demand becomes less elastic, the firm takes longer to learn and generates some price volatility.

In a second example, we change the technology of firm 4 by setting $\rho_4 = 0.001$, so its inputs become imperfect substitutes.
Notice how, while the firm learns consistently, the trajectory through which it adapts to a less elastic demand is different from its linear-technology counterpart.
\autoref{fig:shock_one_prod.leontief} shows that, when the inputs are imperfect substitutes, the industry adjusts output volume much less aggressively than when they are perfect substitutes.
For the shock that increases elasticity, the new steady state exhibits a higher price and lower volume in industry 4 than in the first scenario.

\begin{figure}[ht]
\centering
\caption{Example of demand shocks to firm 4}\label{fig:shock_one}
    \begin{minipage}{0.32\textwidth}
        \subcaption{Prices}\label{fig:shock_one_ela.price}
        \includegraphics[angle=0,width=1.\textwidth]{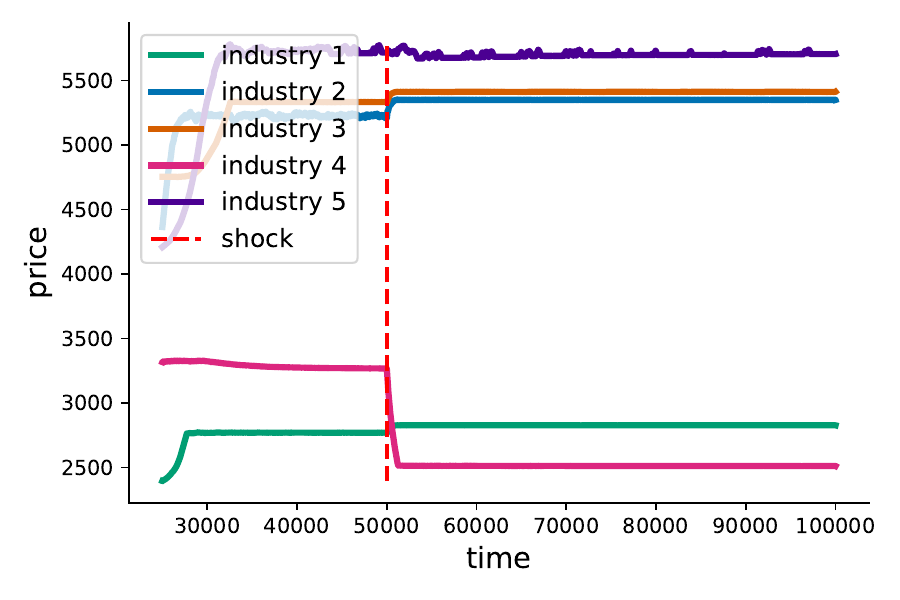}
    \end{minipage}
    \begin{minipage}{0.32\textwidth}
        \subcaption{Output volumes}\label{fig:shock_one_ela.quantity}
        \includegraphics[angle=0,width=1.\textwidth]{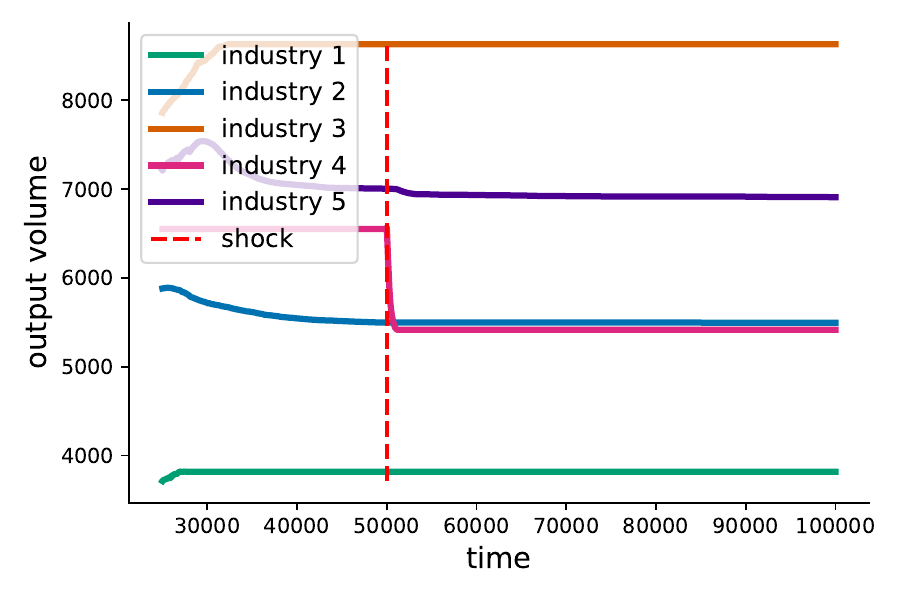}
    \end{minipage}
    \begin{minipage}{0.32\textwidth}
        \subcaption{Profits}\label{fig:shock_one_ela.profts}
        \includegraphics[angle=0,width=1.\textwidth]{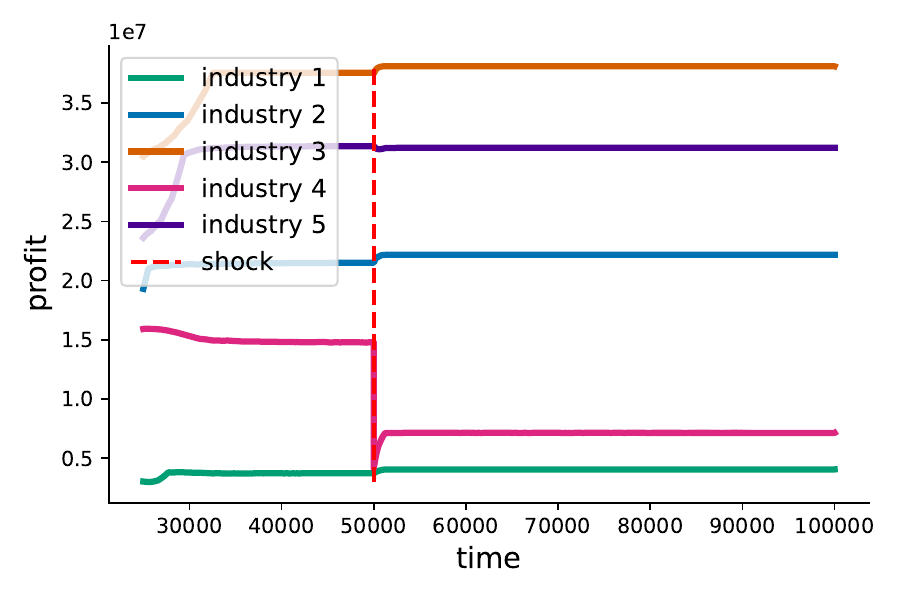}
    \end{minipage}
    \begin{minipage}{0.32\textwidth}
        \subcaption{Prices}\label{fig:shock_one_inela.price}
        \includegraphics[angle=0,width=1.\textwidth]{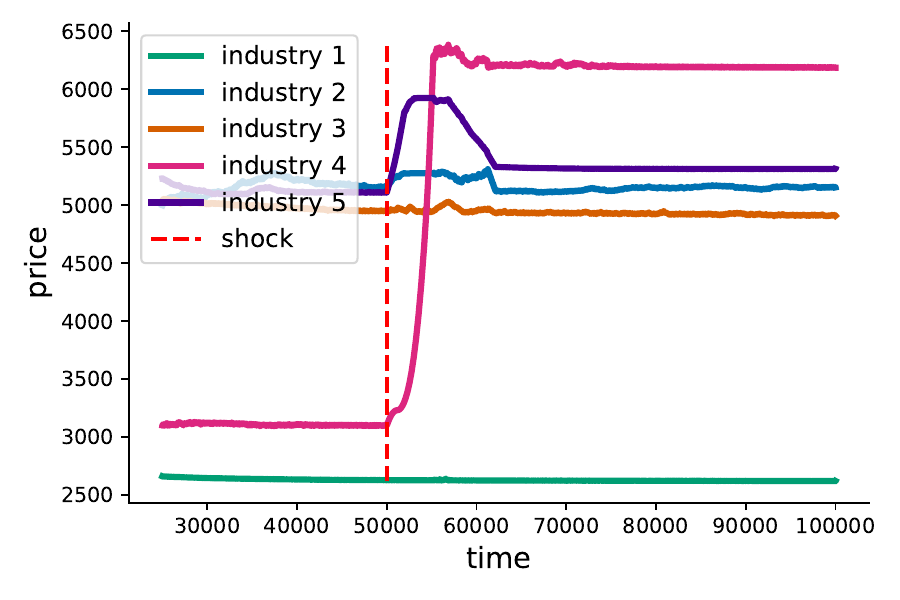}
    \end{minipage}
    \begin{minipage}{0.32\textwidth}
        \subcaption{Output volumes}\label{fig:shock_one_inela.quantity}
        \includegraphics[angle=0,width=1.\textwidth]{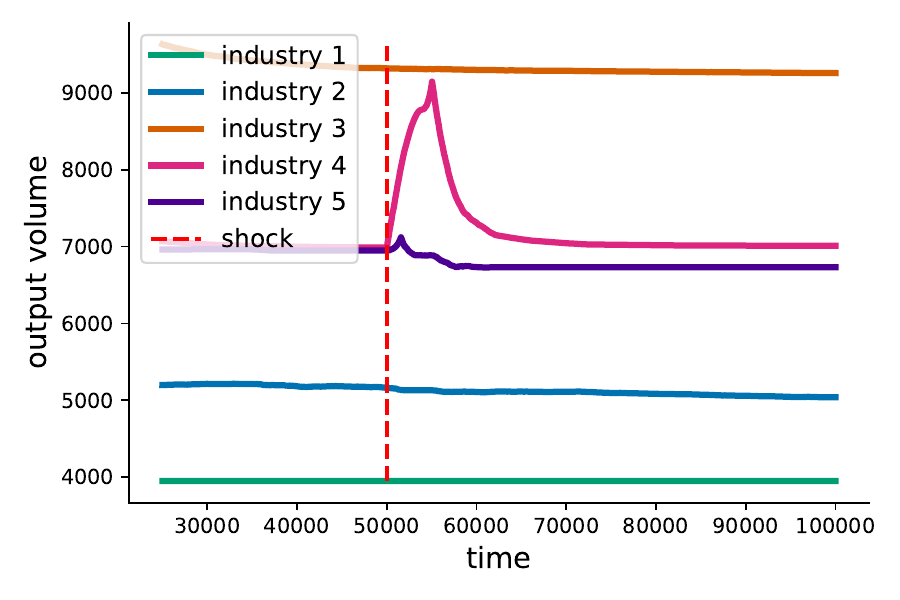}
    \end{minipage}
    \begin{minipage}{0.32\textwidth}
        \subcaption{Profits}\label{fig:shock_one_inela.profts}
        \includegraphics[angle=0,width=1.\textwidth]{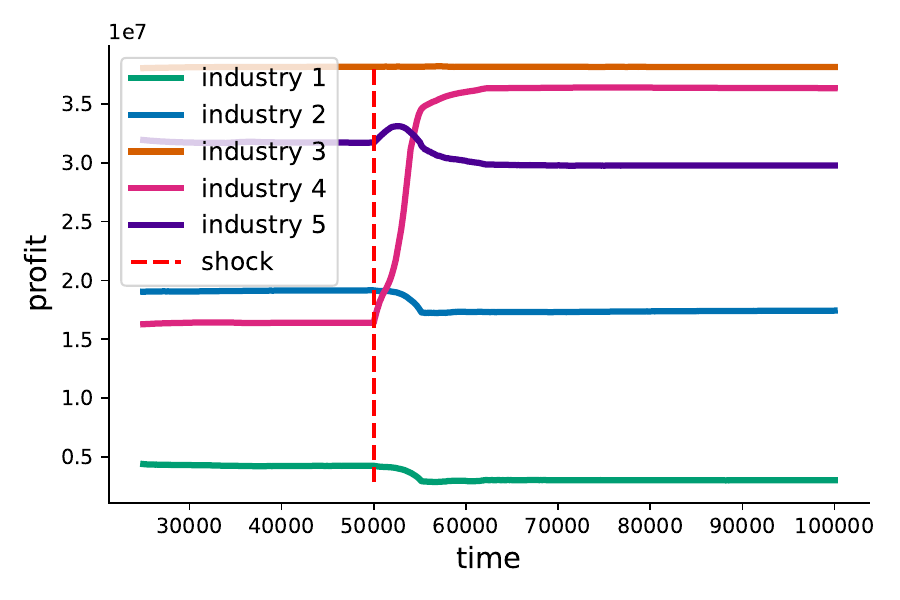}
    \end{minipage}
\caption*{\footnotesize 
\textit{\textbf{Notes}}: Top panels correspond to an increase in demand elasticity through a 50\% change in the slope coefficient.
Bottom panels correspond to a reduction in demand elasticity through a 50\% change in the slope coefficient.\\}
\end{figure}

The previous examples show that, while learning remains consistent in the presence of demand shocks, factors such as the type of production technology play an important role in shaping the adaptation path and, hence, how a production network endogenously reorients itself.
Next, let us examine the impact that the shock to the demand of industry 4 has on the other firms.
In \autoref{fig:shock_one}, we present the time series of the main output variables of every industry.
The vertical dotted line indicates the period in which the demand shock is introduced.

The top panels display the responses to an increase in demand elasticity.
As expected, firm 4 is the first to adjust price and quantity.
The shock has a noticeable effect in the profits of industry 4's clients: firms 2 and 5.
Firm 1 also exhibits some response but it is hardly noticeable.
Interestingly, industry 3, which is not a client of 4, displays a noticeable increment in profits.
Overall, these indirect impacts seem modest in the scales of the top panels, something quite different from the outcomes of a reduced demand elasticity.

The bottom panels in \autoref{fig:shock_one} show the change in the fundamental variables after a shock that reduces demand elasticity in the final market of industry 4.
Fist, notice how the adaptation of firm 4 to a reduced demand elasticity is qualitatively distinct from that displayed in the top panels.
Under this shock, industry 4 increases its price substantially and adjusts its output volume in a `boom and bust' fashion, resulting in a modest net adjustment.
Hence, most of the profit gains from firm 4 come from price increases.
In contrast with the top panels, industry 3 experiences no change in profits, while all the other firms exhibit some reduction.
This illustrates the nuanced adaptation that can emerge in an endogenous production network setting with consistent learning and minimal knowledge.
Appendix~A provides detailed results on all inter-industry impacts from both types of demand shocks.

\subsection{Changes in returns to scale}

In this example, we examine the inter-industry impacts of deviating from the constant-returns baseline: first towards increasing returns and, then, to decreasing returns to scale.
A technological shock consists of modifying parameter $\varphi_i=1$ from \autoref{eq:CES_function}, which prompts constant returns to scale in the baseline.
More specifically, we induce a change to increasing returns through $\varphi_i=1.05$ and to decreasing returns through$\varphi_i=0.95$.

\begin{figure}[ht]
\centering
\caption{Inter-industry impacts from changes in returns to scale}\label{fig:shock_returns_matrix}
    \begin{minipage}{0.32\textwidth}
        \subcaption{Prices}\label{fig:shock_returns_matrix_inc.price}
        \includegraphics[angle=0,width=1.\textwidth]{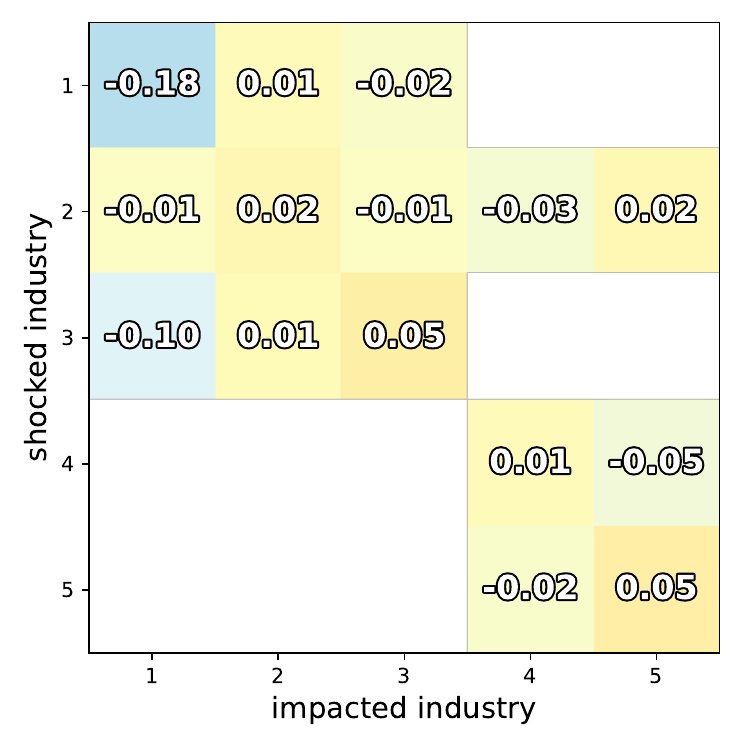}
    \end{minipage}
    \begin{minipage}{0.32\textwidth}
        \subcaption{Output volumes}\label{fig:shock_returns_matrix_inc.quantity}
        \includegraphics[angle=0,width=1.\textwidth]{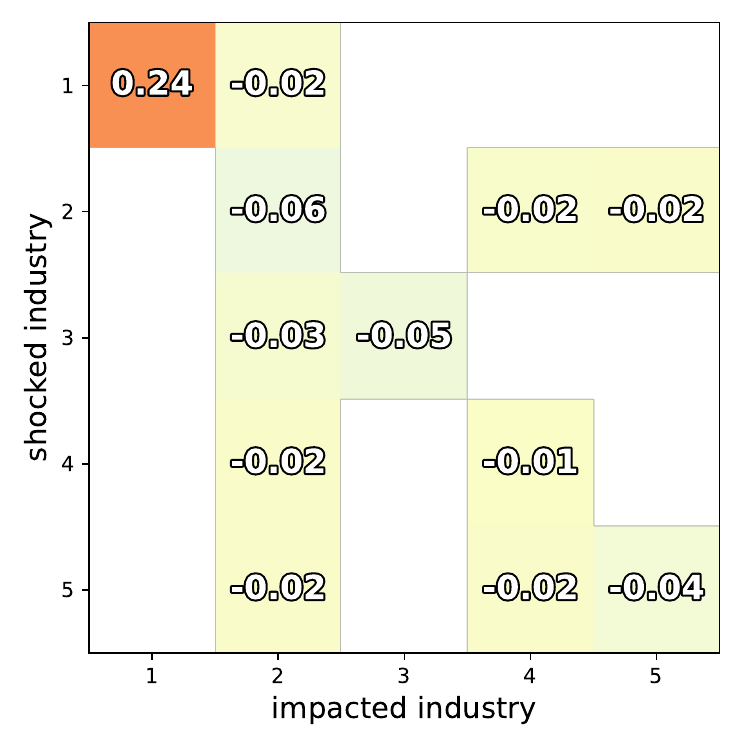}
    \end{minipage}
    \begin{minipage}{0.32\textwidth}
        \subcaption{Profits}\label{fig:shock_returns_matrix_inc.profts}
        \includegraphics[angle=0,width=1.\textwidth]{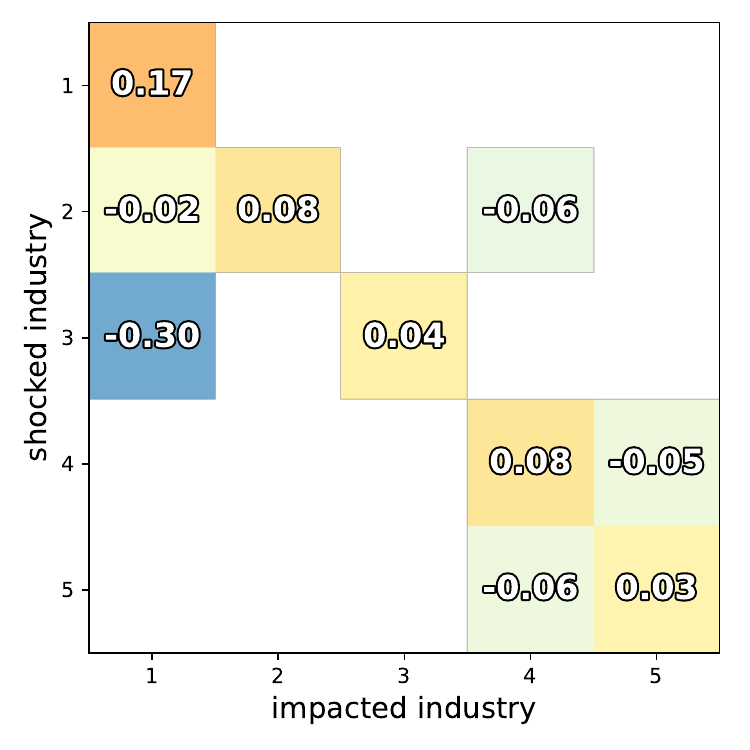}
    \end{minipage}
     \begin{minipage}{0.32\textwidth}
        \subcaption{Prices}\label{fig:shock_returns_matrix_dec.price}
        \includegraphics[angle=0,width=1.\textwidth]{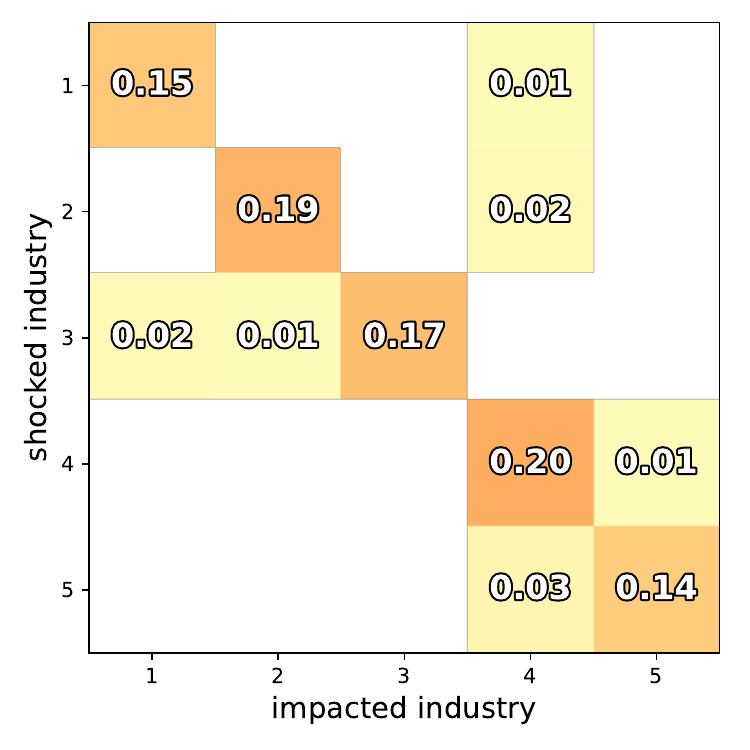}
    \end{minipage}
    \begin{minipage}{0.32\textwidth}
        \subcaption{Output volumes}\label{fig:shock_returns_matrix_dec.quantity}
        \includegraphics[angle=0,width=1.\textwidth]{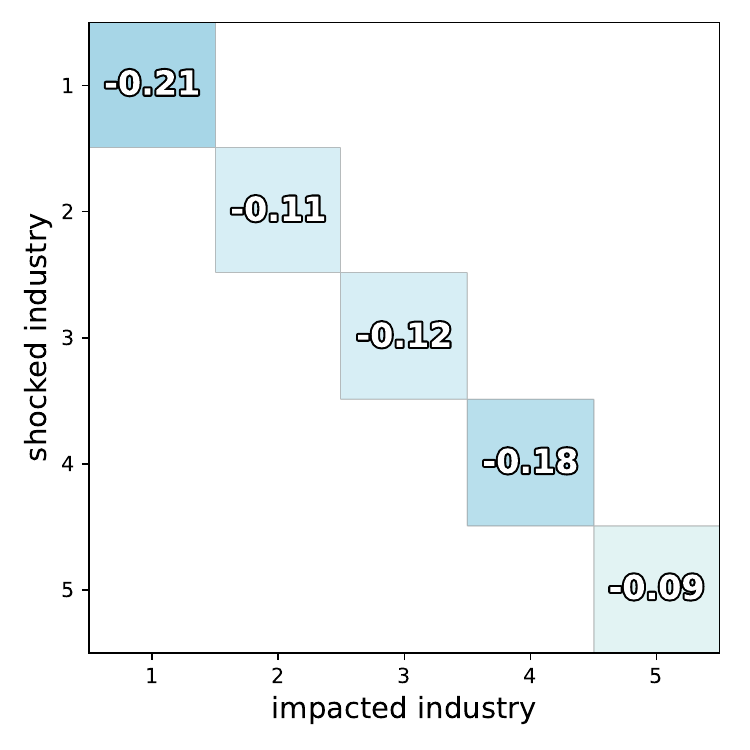}
    \end{minipage}
    \begin{minipage}{0.32\textwidth}
        \subcaption{Profits}\label{fig:shock_returns_matrix_dec.profts}
        \includegraphics[angle=0,width=1.\textwidth]{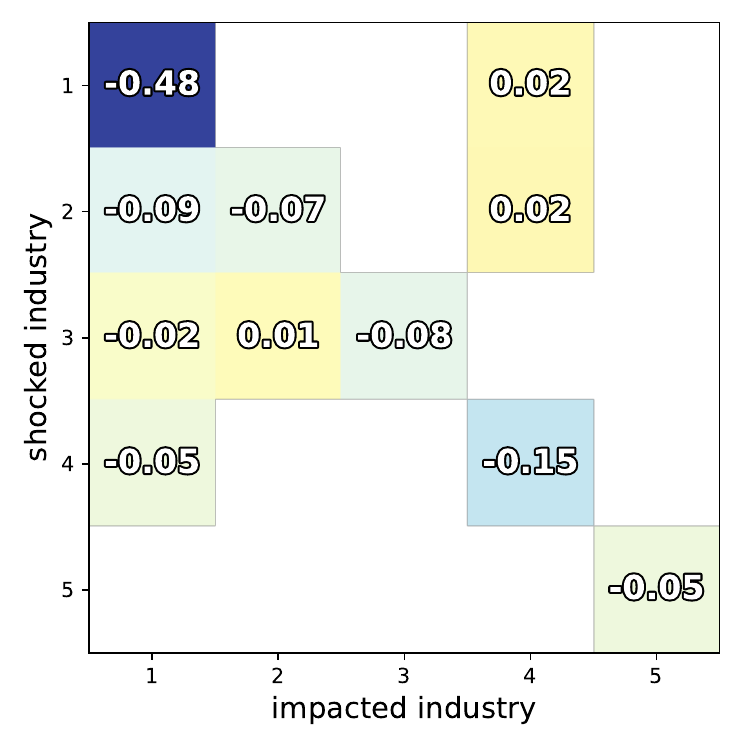}
    \end{minipage}
\caption*{\footnotesize 
\textit{\textbf{Notes}}: The entries in the matrices denote proportional changes in the steady-state outcomes after the shock.
They show changes that are greater than  0.01 (1\%) in absolute value.
Top panels correspond to increasing returns.
Bottom panels denote the impact from switching to decreasing returns to scale.\\}
\end{figure}

\autoref{fig:shock_returns_matrix} shows the inter-industry impacts of shocks to the returns to scale.
The top panels correspond to a switch to increasing returns while the bottom panels to a shift to decreasing returns.
There are various interesting outcomes that are worth highlighting in this example.
First, focusing on increasing returns, we can see that direct and indirect impacts in terms of the sign of price changes are heterogeneous (see \autoref{fig:shock_returns_matrix_inc.price}).
For instance, industry 1 reduces its price by 18\%, while all other firms increase it.
Indirect price impacts also depend on which industry is being shocked; and firm 2 stands out as the one whose technology impacts most industries.
In \autoref{fig:shock_returns_matrix_inc.quantity} we can see that the direction in which industries adjust quantities is the opposite to prices.
This highlights how specific firms may choose a different adaptation strategy to the same type of shock.
In the case of output volumes, it is clear that firm 2 is the most impacted, as it systematically adjusts its quantity in more than 1\% when any of the other industries experiences a switch to increasing returns to scale.
In terms of profits, \autoref{fig:shock_returns_matrix_inc.profts} shows the winners and losers in each experiment, with industry 1 being the most negatively affected (a loss of 30\% in profit) when firm 3 (its direct but not only client) starts operating under increasing returns.

When an industry switches to decreasing returns, one naturally expects a drop in profits.
This is the case for all industries (see \autoref{fig:shock_returns_matrix_dec.profts}), although the size of the proportional loss varies significantly (between 5 and 30\%).
Nevertheless, we can also see that the indirect impacts lead to winners and losers.
For instance, firm 1 loses profits when either industries 2, 3, or 4 operate under decreasing returns.
In contrast, industry 4 makes more profits when firms 1 or 2 switch to decreasing returns.
Arguably, this is because, under decreasing returns, firms 1 and 2 need more inputs from industry 4, increasing the demand for good 4 and, hence, its price (see price change of industry 4 in \autoref{fig:shock_returns_matrix_dec.price}).
Another interesting outcome is that, when industries switch to decreasing returns, all indirect impacts of more than 1\% happen through the price channel, not through output volumes.
These results highlight heterogeneity in outcomes, responses, and strategies adopted by firms.
While they are derived from a particular hypothetical case, they are insightful with regards to the complexity of the formation of production networks and their adaptation to shocks.

\subsection{Technological innovation}

A way to conceptualize technological change is by modifying the nature of the interaction between the inputs of a productive process, i.e., changing the production function of an industry.
As we have previously argued, specific functional forms of production technologies are typically assumed in the literature, in part, to make models solvable.
Furthermore, given the high reliance of such modeling choices, rational-equilibrium models become fragile to modifications to the production function, and unable to analyze, for example, the impact of shifts in the production technology during the dynamics of the model.

In this section, we demonstrate that, through our learning approach, the model remains robust to `swaps' of production technologies as firms consistently learn new prices and quantities as they reach new steady states.
For the sake of this example, we only show the case of a technological disruption in industry 4, and provide all the other cases in Appendix~A.
The shock consists if swapping the production function from one where inputs are substitutes to another where they are closer to complements.
We implement this change by modifying parameter $\rho_4$ from 1 to 4.
Of course, this example of a sudden change in the nature of technology is not realistic in the sense that production technologies evolve gradually, often through an imitation/emulation process; although there could be examples in which a technology is suddenly replaced for strategic reasons of the firm.

\begin{figure}[ht]
\centering
\caption{Example of firm 4 changing its technology from substitutes to complements}\label{fig:shock_one_tech}
    \begin{minipage}{0.32\textwidth}
        \subcaption{Price with increasing returns}\label{fig:shock_one_tech.price}
        \includegraphics[angle=0,width=1.\textwidth]{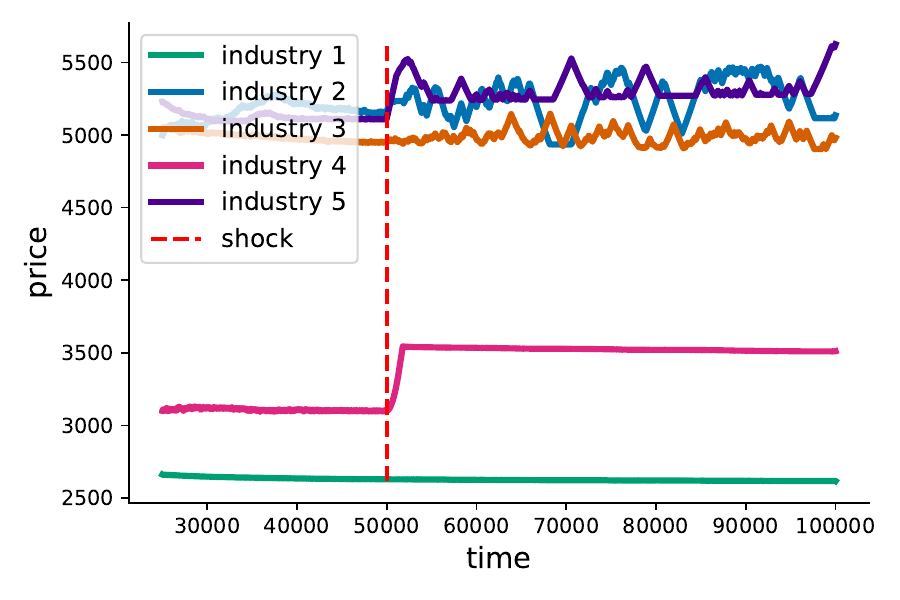}
    \end{minipage}
    \begin{minipage}{0.32\textwidth}
        \subcaption{Output with increasing returns}\label{fig:shock_one_tech.quantity}
        \includegraphics[angle=0,width=1.\textwidth]{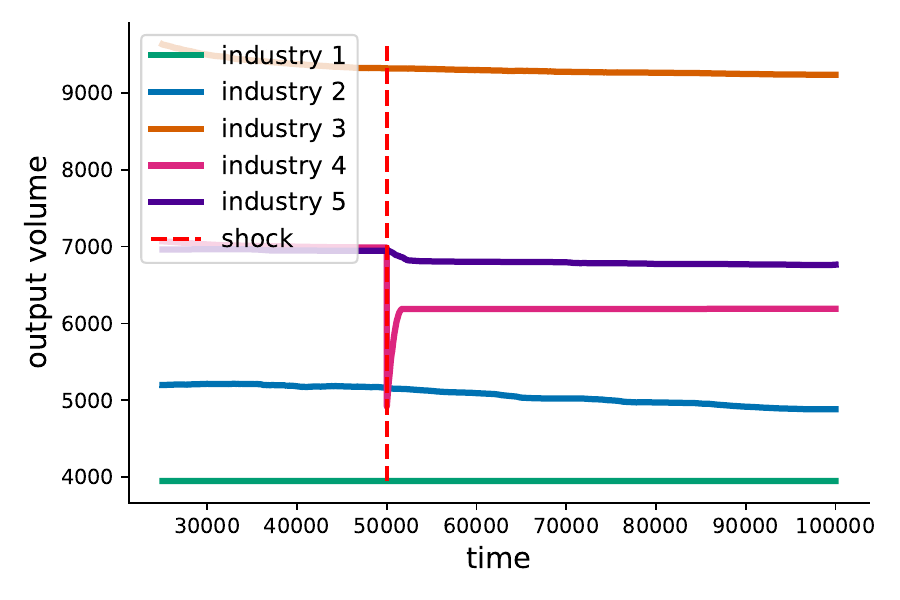}
    \end{minipage}
    \begin{minipage}{0.32\textwidth}
        \subcaption{Profit with increasing returns}\label{fig:shock_one_tech.profts}
        \includegraphics[angle=0,width=1.\textwidth]{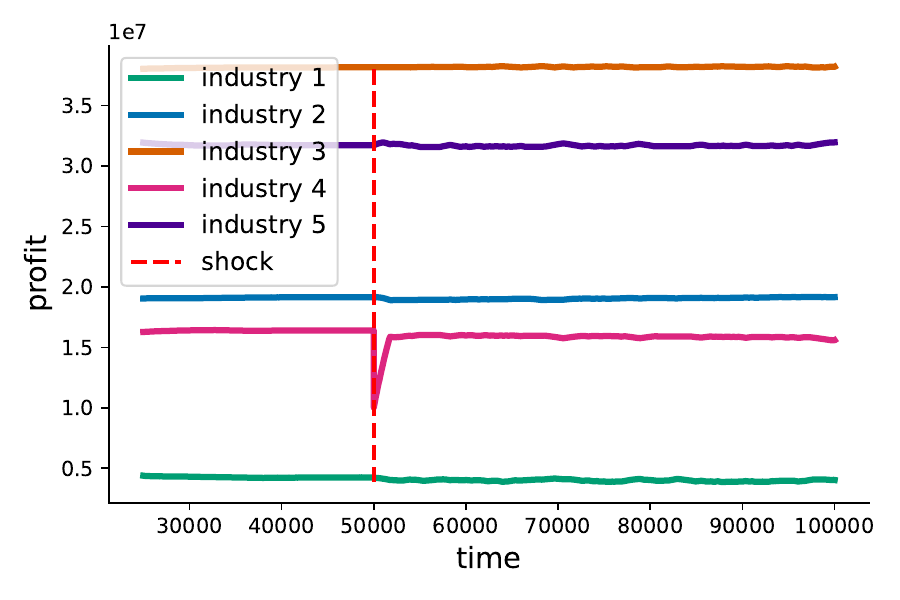}
    \end{minipage}
\end{figure}

\autoref{fig:shock_one_tech} shows the dynamics induced by the technological change in industry 4.
Notice that, in terms of profit, industry 4 experiences a sudden drop but quickly recovers to a slightly lower lever.
This drop is due to the fact that the technological change is exogenous, for example, as it would take place when a new regulation forces companies to adopt specific machines and practices to comply with new environmental standards.
In addition, notice that the shock introduces high price volatility in industries 2, 3, and 5.
When firm 4's inputs become more complementary, it leads to a higher price and output volume; and to a slight drop in the output of firms 2 and 5.
Overall, despite the indirect impacts through an output reduction and higher price volatility , all firms adjust well in terms of profits.

\subsection{Industry shutdown}

Now we look at the case in which an industry closes down and the endogenous production networks adapts.
First, we demonstrate this shock with the example of closing down industry 4.
Then, we provide the inter-industry impacts of shutting down each of the firms.
As in our previous example, one could argue that an industry closure would be a gradual process.
Nevertheless, and while a gradual closure would be easy to implement, it is still useful to learn from a stylized example when firm 4 closes overnight.

\begin{figure}[ht]
\centering
\caption{Example of firm 4 suddenly shutting down}\label{fig:shock_one_exit}
    \begin{minipage}{0.32\textwidth}
        \subcaption{Prices}\label{fig:shock_one_exit.price}
        \includegraphics[angle=0,width=1.\textwidth]{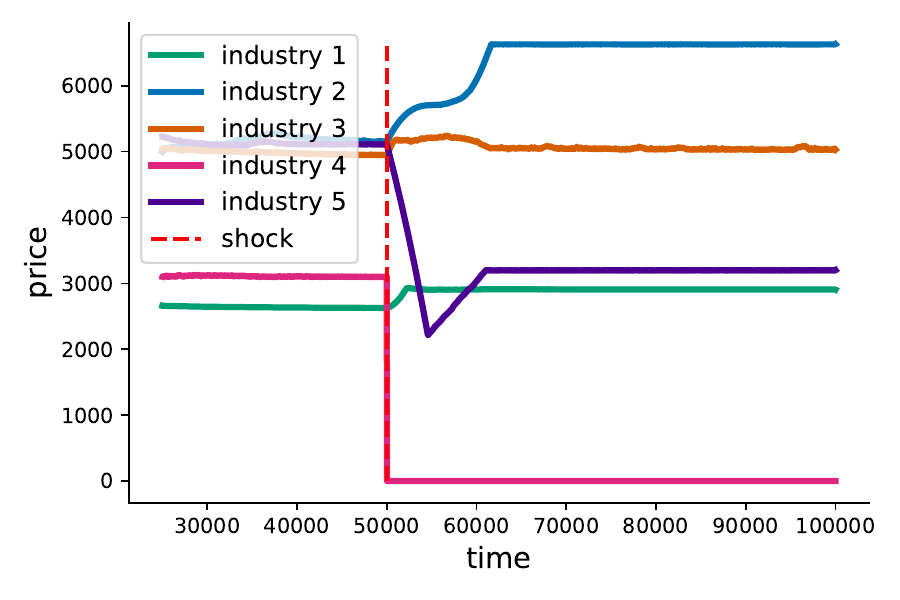}
    \end{minipage}
    \begin{minipage}{0.32\textwidth}
        \subcaption{Output volumes}\label{fig:shock_one_exit.quantity}
        \includegraphics[angle=0,width=1.\textwidth]{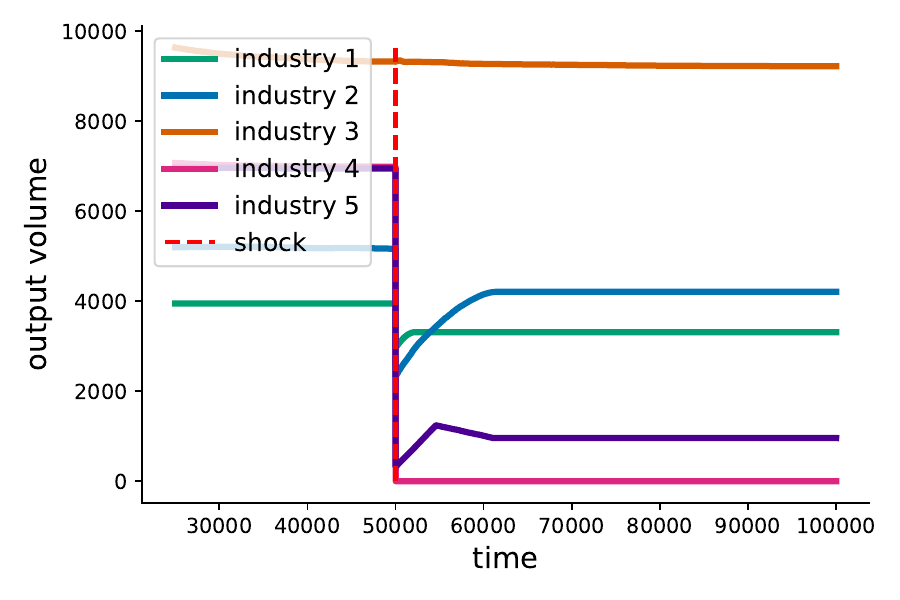}
    \end{minipage}
    \begin{minipage}{0.32\textwidth}
        \subcaption{Profits}\label{fig:shock_one_exit.profts}
        \includegraphics[angle=0,width=1.\textwidth]{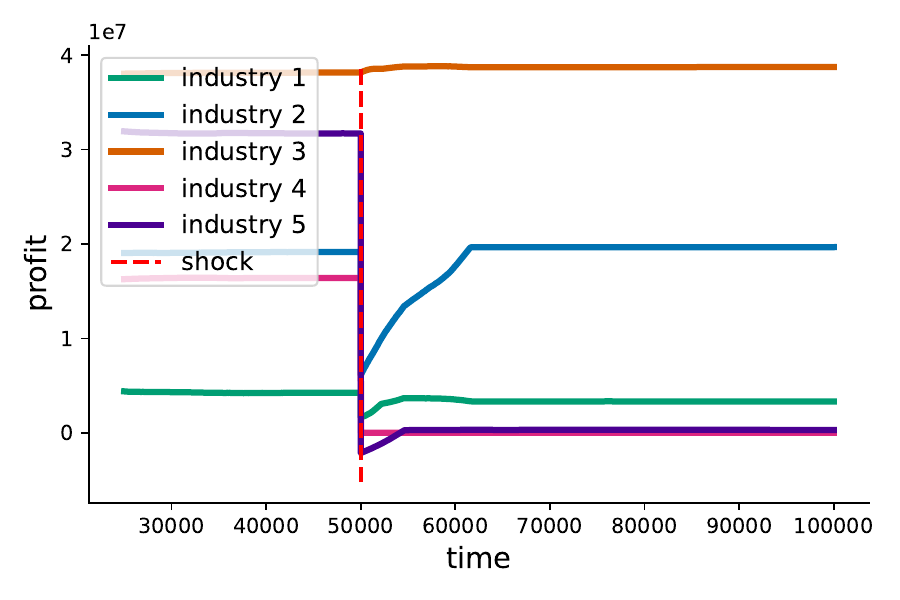}
    \end{minipage}

    \begin{minipage}{0.32\textwidth}
        \subcaption{Prices}\label{fig:shock_exit_matrix.price}
        \includegraphics[angle=0,width=1.\textwidth]{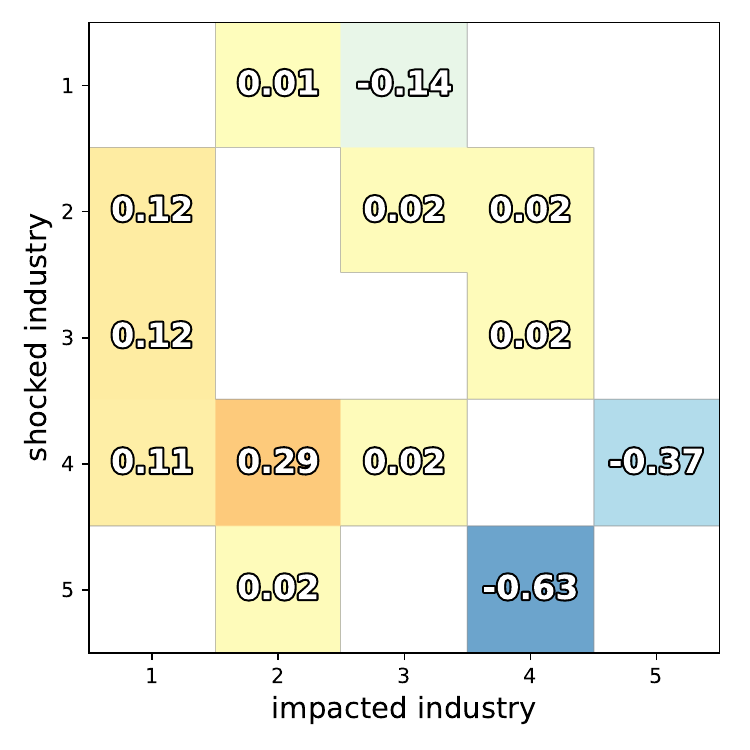}
    \end{minipage}
    \begin{minipage}{0.32\textwidth}
        \subcaption{Output volumes}\label{fig:shock_exit_matrix.quantity}
        \includegraphics[angle=0,width=1.\textwidth]{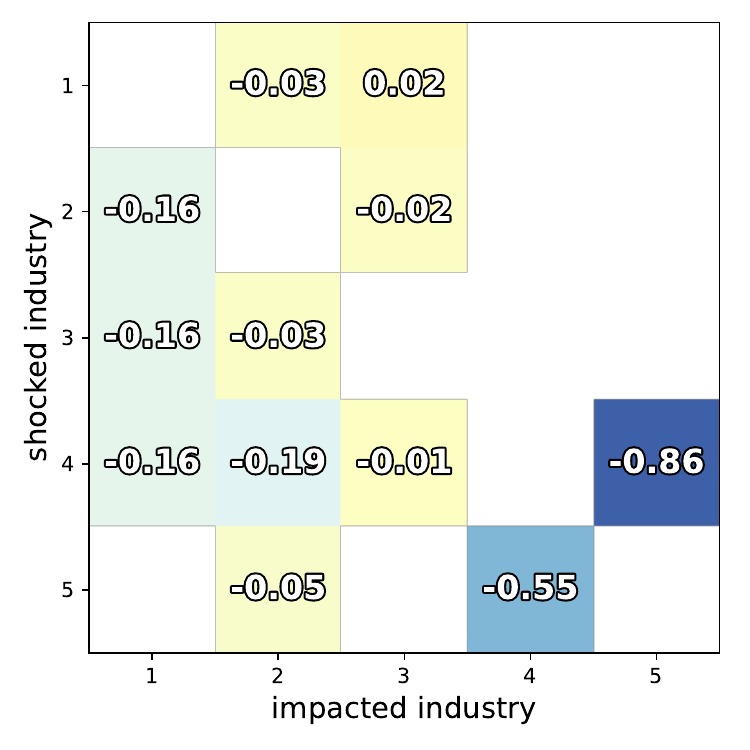}
    \end{minipage}
    \begin{minipage}{0.32\textwidth}
        \subcaption{Profits}\label{fig:shock_exit_matrix.profts}
        \includegraphics[angle=0,width=1.\textwidth]{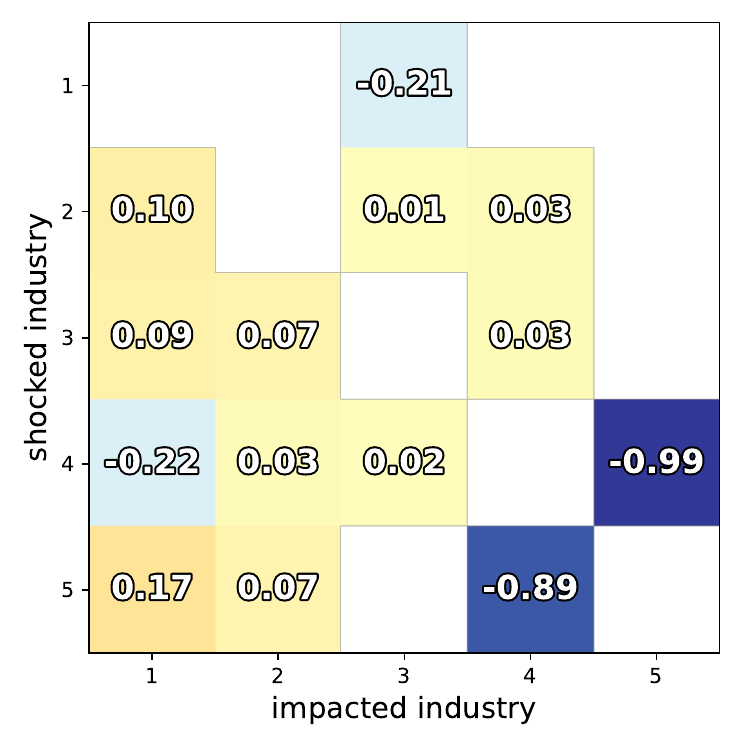}
    \end{minipage}
\caption*{\footnotesize 
\textit{\textbf{Notes}}: Top panels correspond to industry 4 shutdown.
Bottom panels display inter-industry impacts.
The entries in the matrices denote proportional changes in the steady-state outcomes after the shock.
They show changes that are greater than 0.01 (1\%) in absolute value.\\}
\end{figure}

The top panels in \autoref{fig:shock_one_exit} show the dynamics of the fundamentals when industry 4 shuts down.
In terms of prices, we can see that all industries adjust, some with a positive and others with a negative direction, and in varying magnitudes.
The responses of industries 2 and 5 stand out as they seem to follow a two-phase process; apparently monotonic for industry 2 and punctuated for industry 5.
In terms of output volumes, as expected, industry 5 is the one that experiences the largest change.
Industry 2 also exhibits a substantial drop in quantity, while firms 1 and 3 display a moderate change.
Finally, in terms of profits, we can see that firm 5 drops almost to zero, while firm 2 experiences a severe transient drop, with an eventual recovery to nearly its pre-shock profits.

The bottom panels in \autoref{fig:shock_one_exit} show the inter-industry impacts of shutting down each one at a time.
The diagonal elements in these matrices are empty because they correspond to the closing industry.
Overall, we can see that impacts are heterogeneous across industries and transmission channels.
For example, prices can increase or decrease depending on which industry shuts down and who are the impacted ones.
With exception of industry 3 during the closure of firm 1, industries usually reduce their output volumes.
In terms of profits, an industry shutdown can translate into gains or losses, again, depending on who closes and who is impacted.

\subsection{Transient shocks}\label{sec:recovery}

So far, shocks have been implemented as structural changes that remain after the fact.
However, another important phenomenon that tests the resilience of production networks is a transient shock.
This is inherently a dynamical problem, and one for which procedurally explicit models are well suited.
Through the following examples, we would like to show how industries adapt to a transient demand shock and how these adaptations propagate to the other industries.

\begin{figure}[ht]
\centering
\caption{Example of responses to transient demand shocks}\label{fig:shock_recovery}

    \begin{minipage}{0.32\textwidth}
        \subcaption{Prices (shocking 1)}\label{fig:shock_recovery.price.1}
        \includegraphics[angle=0,width=1.\textwidth]{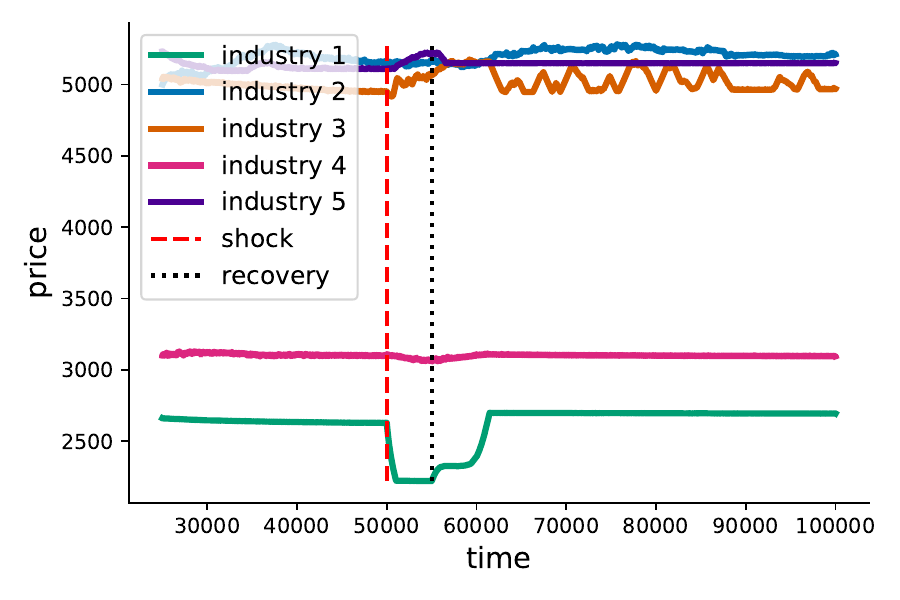}
    \end{minipage}
    \begin{minipage}{0.32\textwidth}
        \subcaption{Output volumes (shocking 1)}\label{fig:shock_recovery.quantity.1}
        \includegraphics[angle=0,width=1.\textwidth]{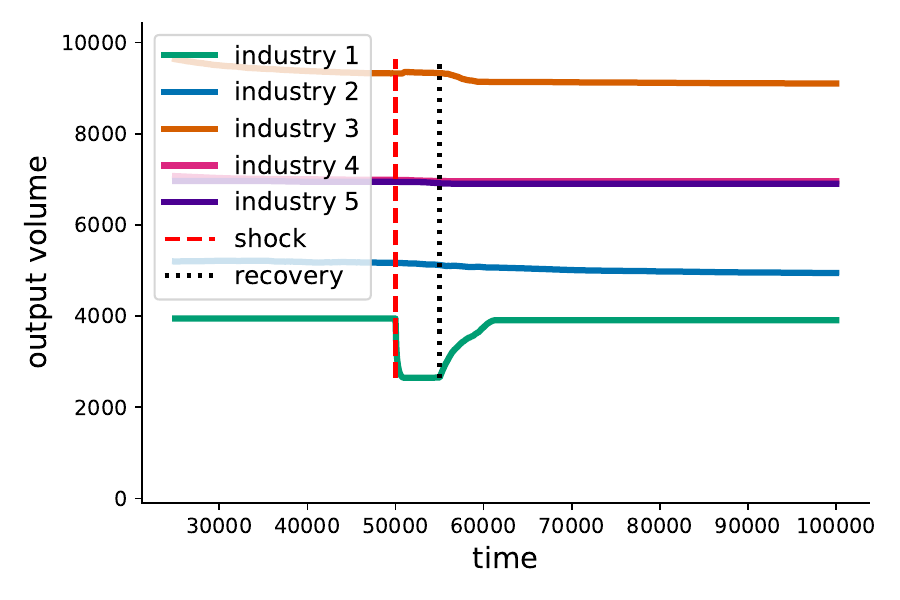}
    \end{minipage}
    \begin{minipage}{0.32\textwidth}
        \subcaption{Profits (shocking 1)}\label{fig:shock_recovery.profts.1}
        \includegraphics[angle=0,width=1.\textwidth]{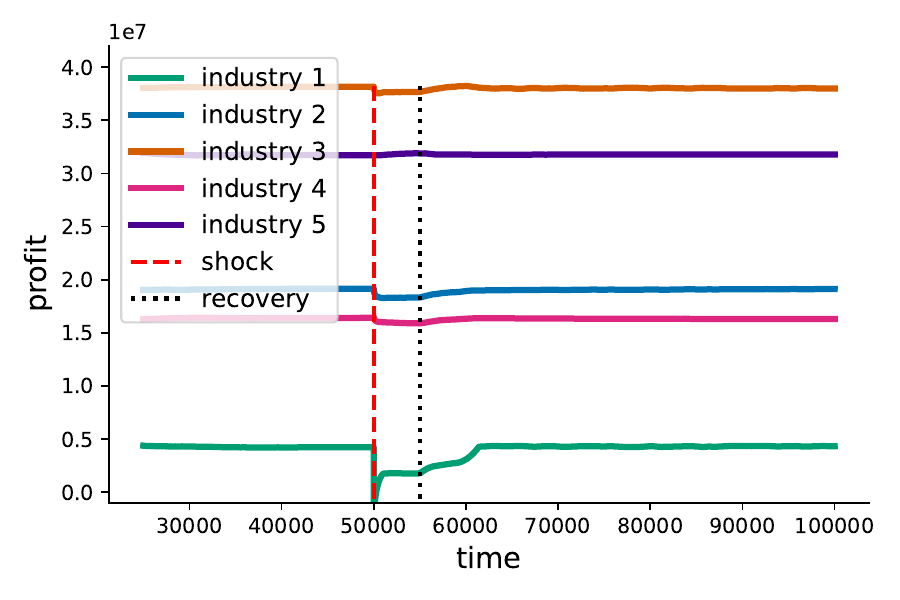}
    \end{minipage}

    \begin{minipage}{0.32\textwidth}
        \subcaption{Prices (shocking 2)}\label{fig:shock_recovery.price.2}
        \includegraphics[angle=0,width=1.\textwidth]{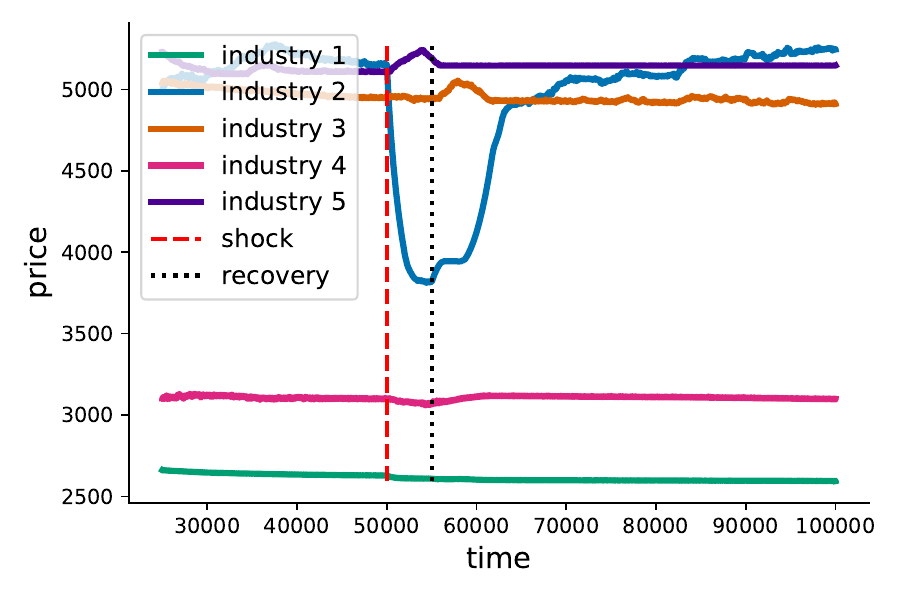}
    \end{minipage}
    \begin{minipage}{0.32\textwidth}
        \subcaption{Output volumes (shocking 2)}\label{fig:shock_recovery.quantity.2}
        \includegraphics[angle=0,width=1.\textwidth]{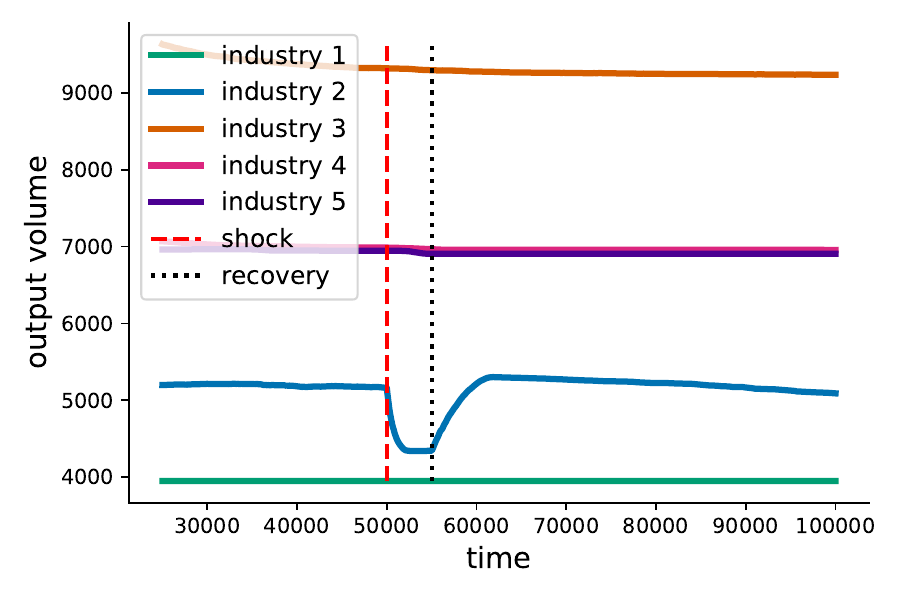}
    \end{minipage}
    \begin{minipage}{0.32\textwidth}
        \subcaption{Profits (shocking 2)}\label{fig:shock_recovery.profts.2}
        \includegraphics[angle=0,width=1.\textwidth]{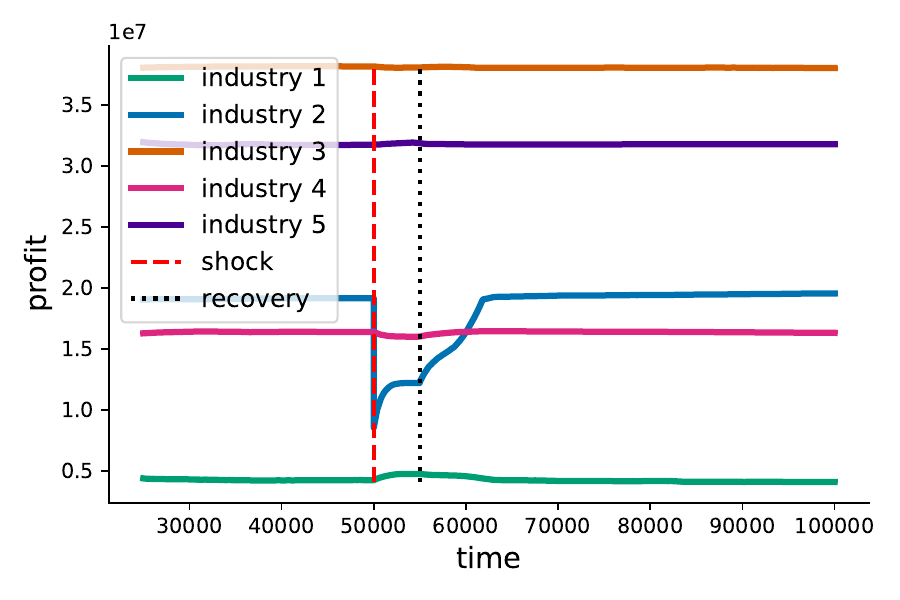}
    \end{minipage}

    \begin{minipage}{0.32\textwidth}
        \subcaption{Prices (shocking 3)}\label{fig:shock_recovery.price.3}
        \includegraphics[angle=0,width=1.\textwidth]{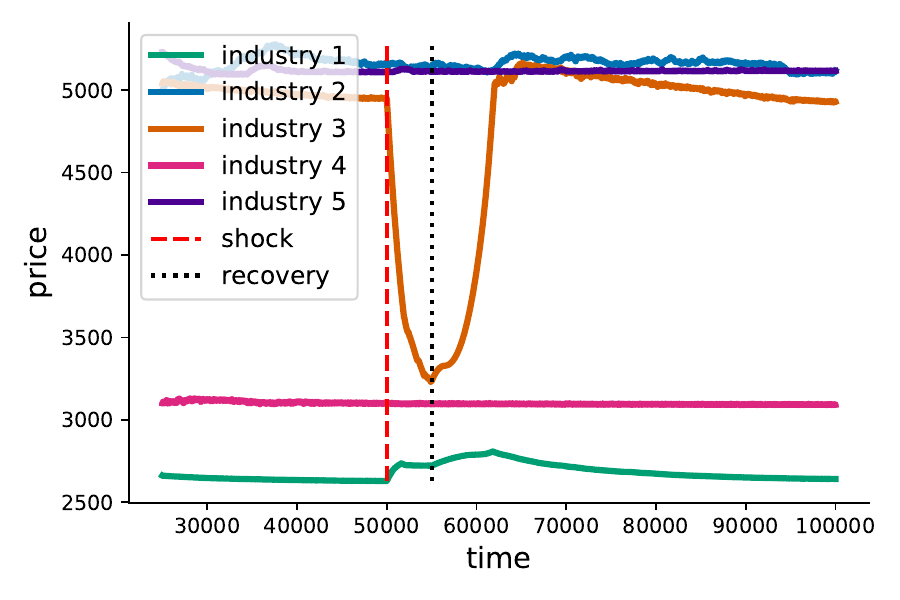}
    \end{minipage}
    \begin{minipage}{0.32\textwidth}
        \subcaption{Output volumes (shocking 3)}\label{fig:shock_recovery.quantity.3}
        \includegraphics[angle=0,width=1.\textwidth]{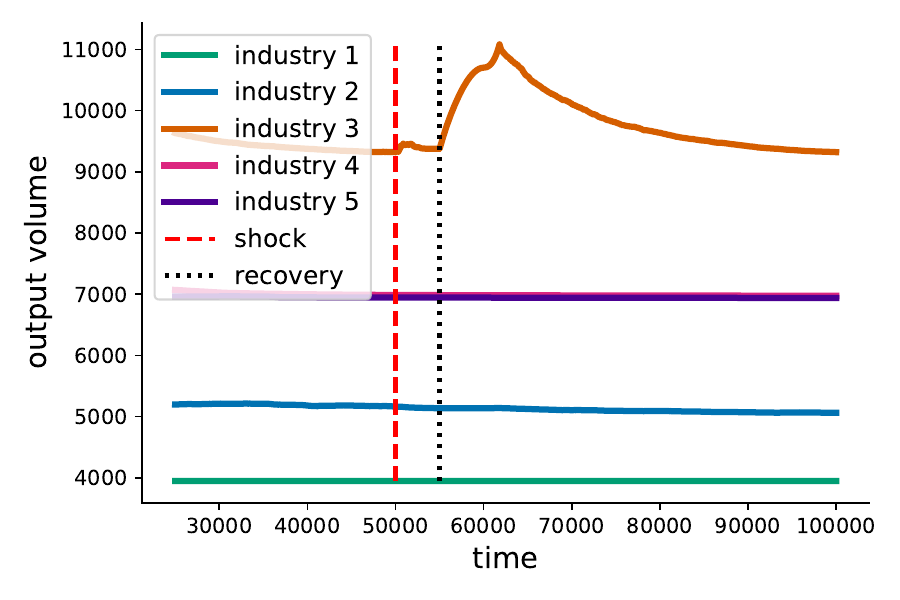}
    \end{minipage}
    \begin{minipage}{0.32\textwidth}
        \subcaption{Profits (shocking 3)}\label{fig:shock_recovery.profts.3}
        \includegraphics[angle=0,width=1.\textwidth]{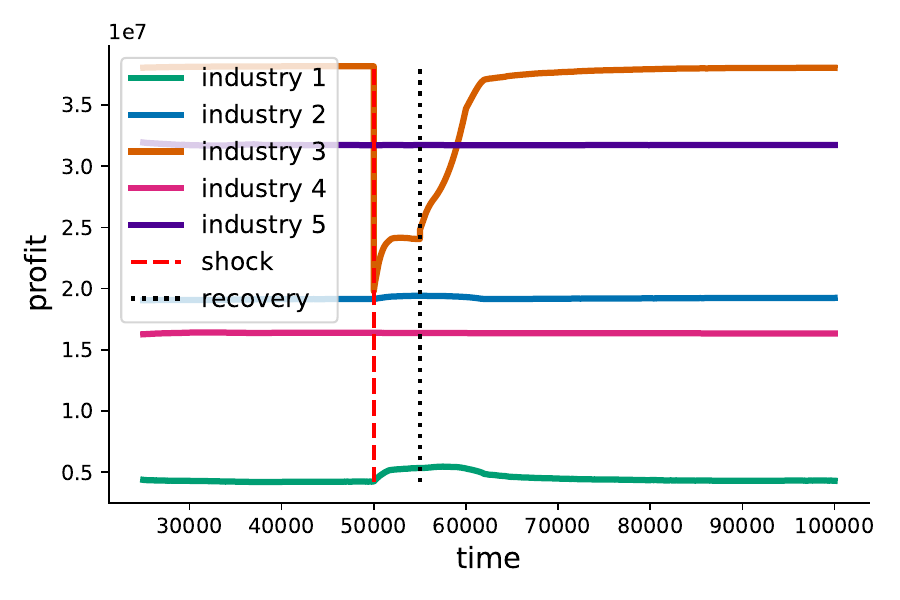}
    \end{minipage}

    \begin{minipage}{0.32\textwidth}
        \subcaption{Prices (shocking 4)}\label{fig:shock_recovery.price.4}
        \includegraphics[angle=0,width=1.\textwidth]{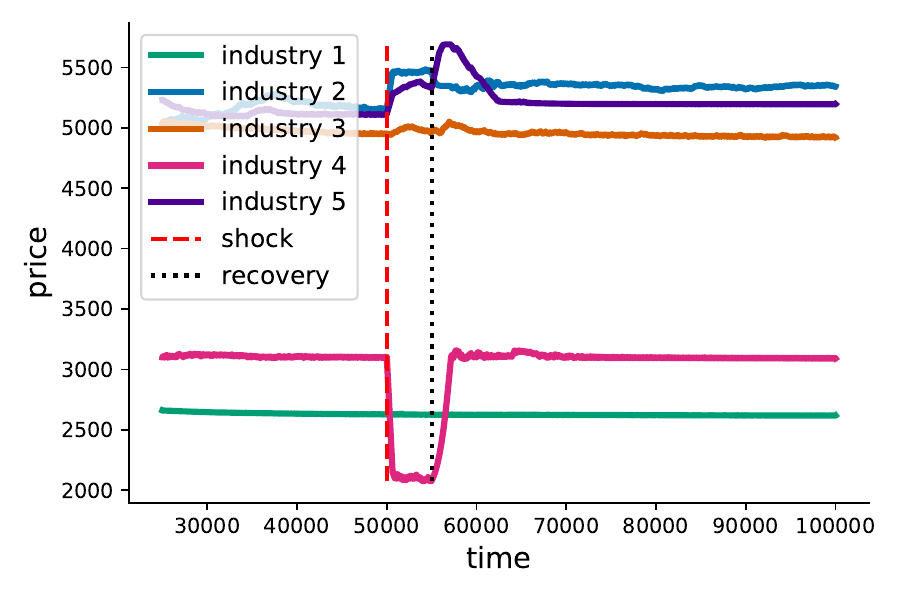}
    \end{minipage}
    \begin{minipage}{0.32\textwidth}
        \subcaption{Output volumes (shocking 4)}\label{fig:shock_recovery.quantity.4}
        \includegraphics[angle=0,width=1.\textwidth]{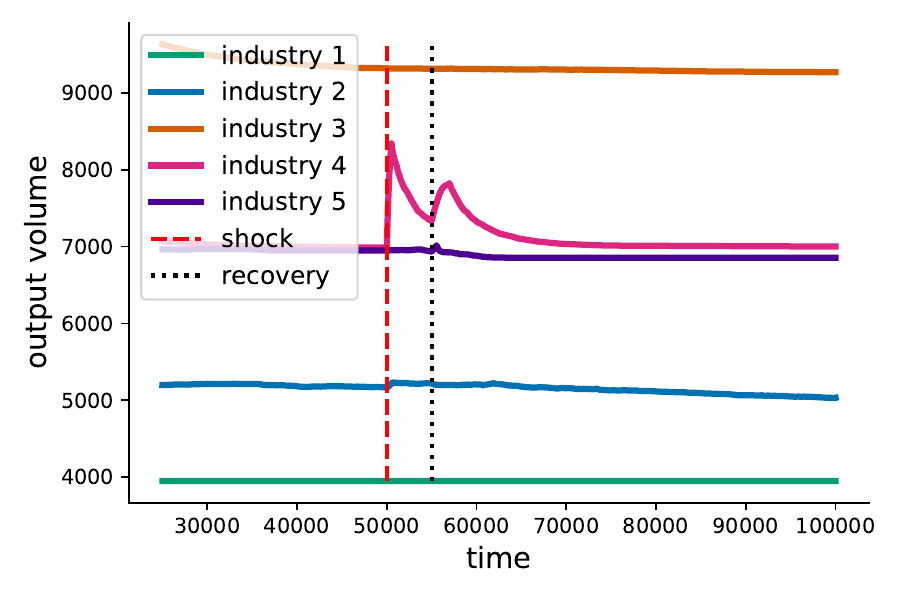}
    \end{minipage}
    \begin{minipage}{0.32\textwidth}
        \subcaption{Profits (shocking 4)}\label{fig:shock_recovery.profts.4}
        \includegraphics[angle=0,width=1.\textwidth]{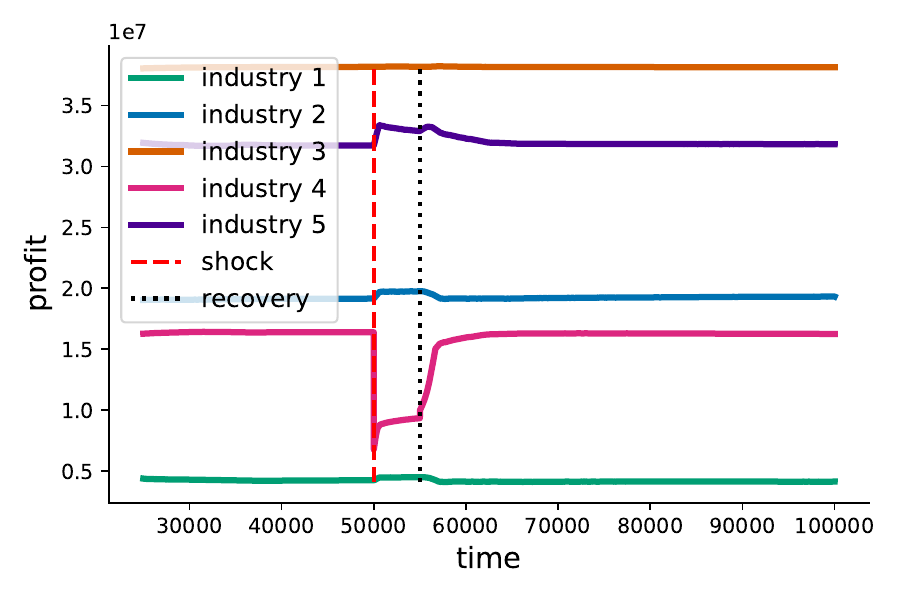}
    \end{minipage}

    \begin{minipage}{0.32\textwidth}
        \subcaption{Prices (shocking 5)}\label{fig:shock_recovery.price.5}
        \includegraphics[angle=0,width=1.\textwidth]{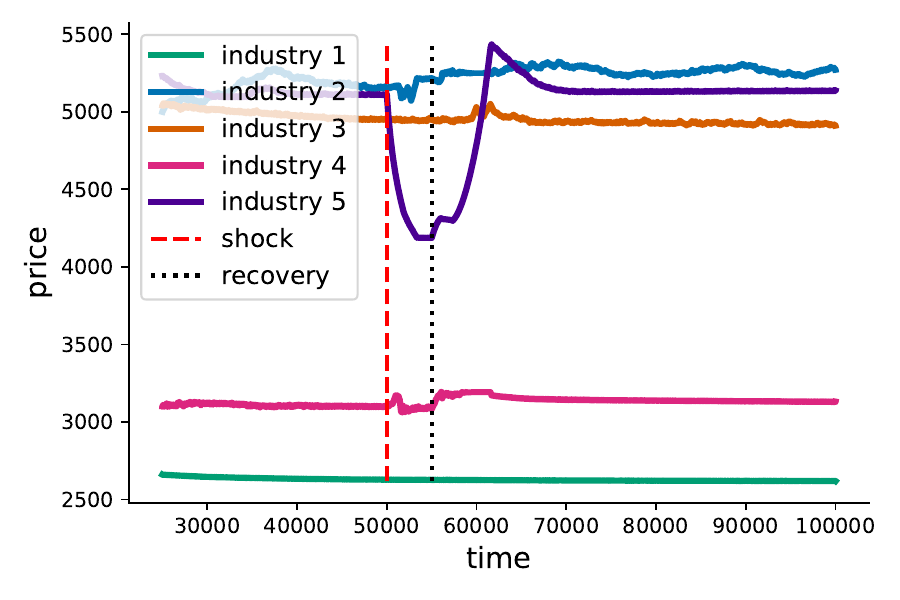}
    \end{minipage}
    \begin{minipage}{0.32\textwidth}
        \subcaption{Output volumes (shocking 5)}\label{fig:shock_recovery.quantity.5}
        \includegraphics[angle=0,width=1.\textwidth]{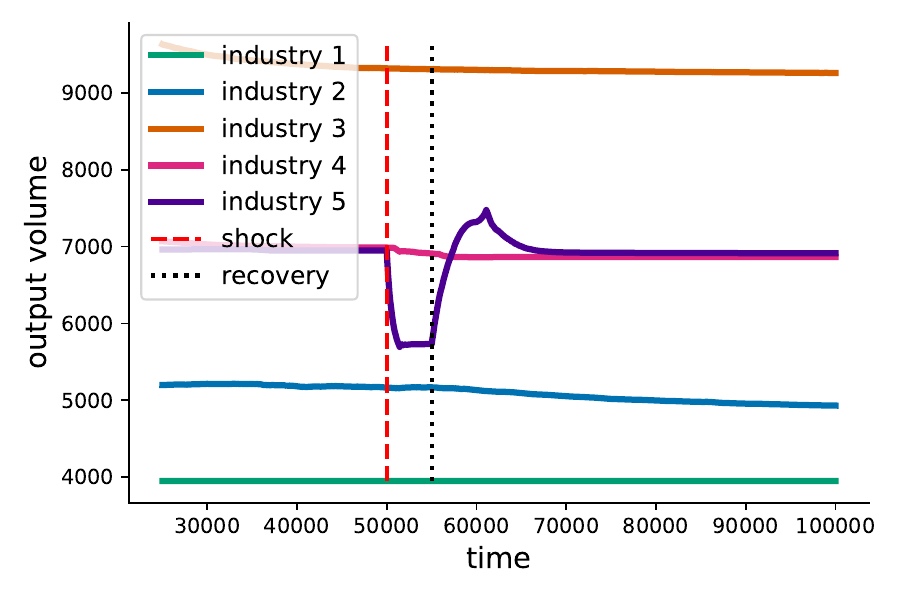}
    \end{minipage}
    \begin{minipage}{0.32\textwidth}
        \subcaption{Profits (shocking 5)}\label{fig:shock_recovery.profts.5}
        \includegraphics[angle=0,width=1.\textwidth]{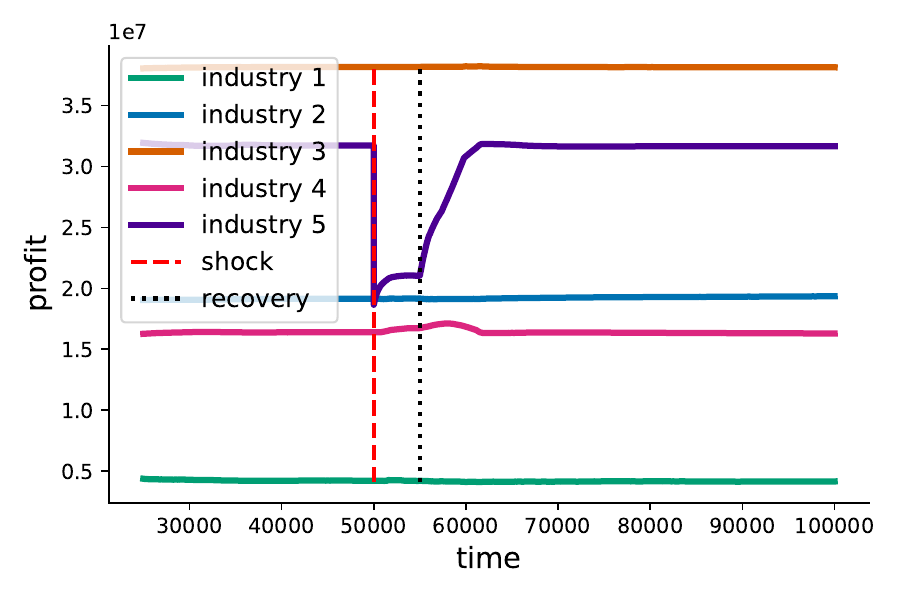}
    \end{minipage}
\end{figure}

For this example, we increase demand elasticity temporarily.
After 500 periods, the shocked demand reverts to its original elasticity.
\autoref{fig:shock_recovery} shows, in each row, the time series of the fundamental variables as a single industry's demand is being temporarily shocked.
The transient shock is active during the periods that fall between the two dashed vertical lines.
Here, we are interested in seeing differences in responses between variables, direct impacts, indirect impacts, and ability to re-establish the pre-shock steady state.

Let us begin with the direct impacts on prices.
The first column of \autoref{fig:shock_recovery} shows that all industries whose demand has been shocked drop their prices and recover once the demand goes back to its original elasticity.
The price responses, however, are heterogeneous in both quantitative and qualitative terms.
Quantitatively, it is clear that industry 3 is the most reactive, while industry 1 is the least sensitive one.
Qualitatively speaking, some industries quickly reach a new steady state within the shock period, and then learn another one once the shock has been removed.
Other industries do not learn so fast.
This yields dip-and-recovery trajectories that denote the differences in learning rates.
In terms of indirect price effects, a transient shock to industry 1 introduces price volatility in industry 3.
Interestingly, this volatility does not dissipate after the shock has been removed.
Similar dynamics can be seen in the indirect impact that demand for industry 5 has on firm 2.

Output volumes exhibit very interesting dynamics as, in contrast to prices, industries may adapt upwards or downwards.
For instance, industries 1, 2, and 5 drop their quantities during the transient and increase them afterwards, with the distinction that industry 5 overshoots production after its demand recovers its original elasticity.
Industry 3, in contrast, increases its production slightly during the transient, then overshoots production after the shock, to finally reach a new steady state with higher volume than the pre-shock level.
Industry 4, on the other hand, drastically increases production immediately after the shock, then adjusts downwards, and then repeats the pattern after the demand elasticity is reestablished.
Finally, in terms of indirect impacts, we can see that the responses of non-shocked industries are modest compared to those generated through prices.
However, in cases like industry 2, it is possible to observe a slow declining response when shocking the demands of firms 4 and 5.

Overall, industries seem to be resilient in terms of profits in this example.
In most cases, the post-shock profits reach levels similar to the pre-shock ones.
One case that stands out is industry 4 in the bottom right panel of \autoref{fig:shock_recovery}, where the demand of industry 5 is shocked.
In this case, we can see that the indirect impact of this shock on industry 4 has a permanent effect in terms of lower profits after recovery.
Appendix~A provides detailed quantitative information on the inter-industry impacts generated in this example.
What is important to take from this exercise is that, when production networks emerge in a context with minimal knowledge and learning, transient shocks can lead to new steady states with long-term implications.
Furthermore, the transmission channels and the nature of the response of a firm or industry seems highly dependent on factors such as the implied productive structure, the nature of technology, the current steady-state, and the relevant demands.

\section{Investigating upstream and downstream shock propagation}

The analysis and assessment of the downstream versus upstream propagation of shocks has a long tradition. Downstream propagation refers to how shocks to a supplier affect its customers, while upstream propagation involves how shocks to a customer impact its supplier.
In economists have approached this problem through rational-equilibrium models and Leontief-inverse models (e.g., \citet{Acemoglu2015, ferrari2023inventories} for a comprehensive survey) as well as procedural ones (e.g., \citet{inoue2019firm, inoue2019propagation, Pichler2021, pichler2022simultaneous, bugert2023analyzing}).
The typical questions in this topic concern whether shocks are amplified or dissipated in either direction.
According to \citep{Acemoglu2015}, in special case of Cobb-Douglas production function and customer preference, upstream propagation is stronger than downstream propagation for demand-side shocks, while downstream propagation is stronger for the supply side. Empirically, \citet{kisat2020consumer} concluded that the demand shock of India's 2016 demonetization affects the downstream customers substantially, while has no meaningful effect on upstream suppliers. Other studies did not compare upstream versus downstream but focus on a single type of propagation, and investigate the different firm size or order of disruptions. \citet{yoshiyuki2022demand} studies the upstream propagation caused by demand shocks in different firm size. \citet{boehm2019input} investigates downstream negative shock caused by the Tohoku earthquake in Japan in 2011 and show how this shock propagates to customers in unaffected regions.
In summary, there seems to be much more nuance to this problem and no one-fit-all solution can be applied to all contexts.
\label{sec:propagation}

To contribute to this discussion, we apply our model to a larger network where the notion of distances between industries is explicit.
The experiments consists of implementing demand and supply shocks to study their propagation across firms that are at different distances from the source of the shock.
Our intention is to exemplify the level of nuance that can emerge in the context of the endogenous formation of production networks under bounded rationality and limited information.

\subsection{Implied productive structure}

\begin{figure}[ht]
\centering
\caption{Synthetic network structure}\label{fig:syn_network_structure}
    \begin{minipage}{0.45\textwidth}
        \subcaption{Edges weighted by input shares of production function}\label{fig:syn_network_structure.alphas}
        \includegraphics[angle=0,width=1.\textwidth]{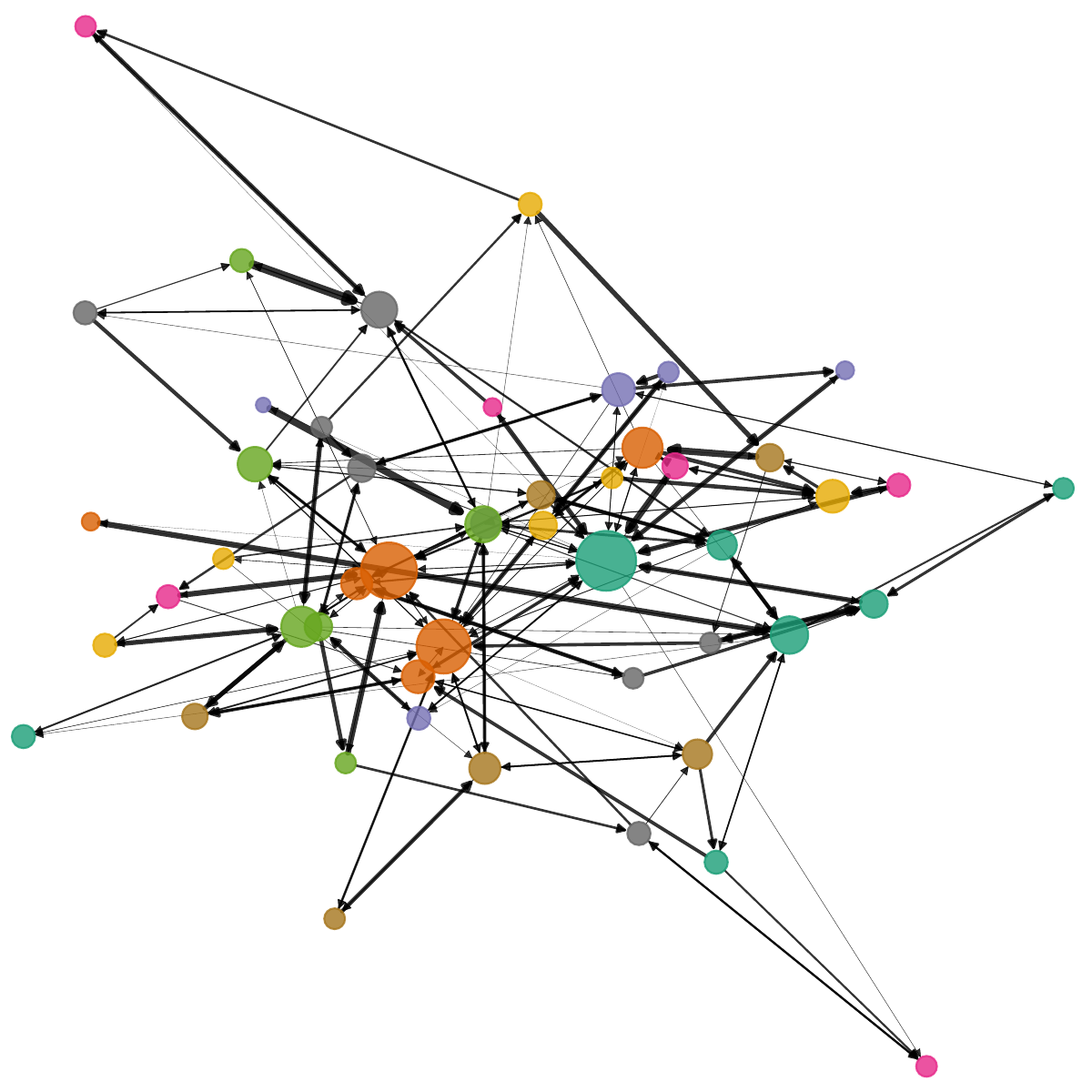}
    \end{minipage}
    \begin{minipage}{0.45\textwidth}
        \subcaption{Edges weighted by normalized volume inflows of the emerged production network}\label{fig:syn_network_structure.flows}
        \includegraphics[angle=0,width=1.\textwidth]{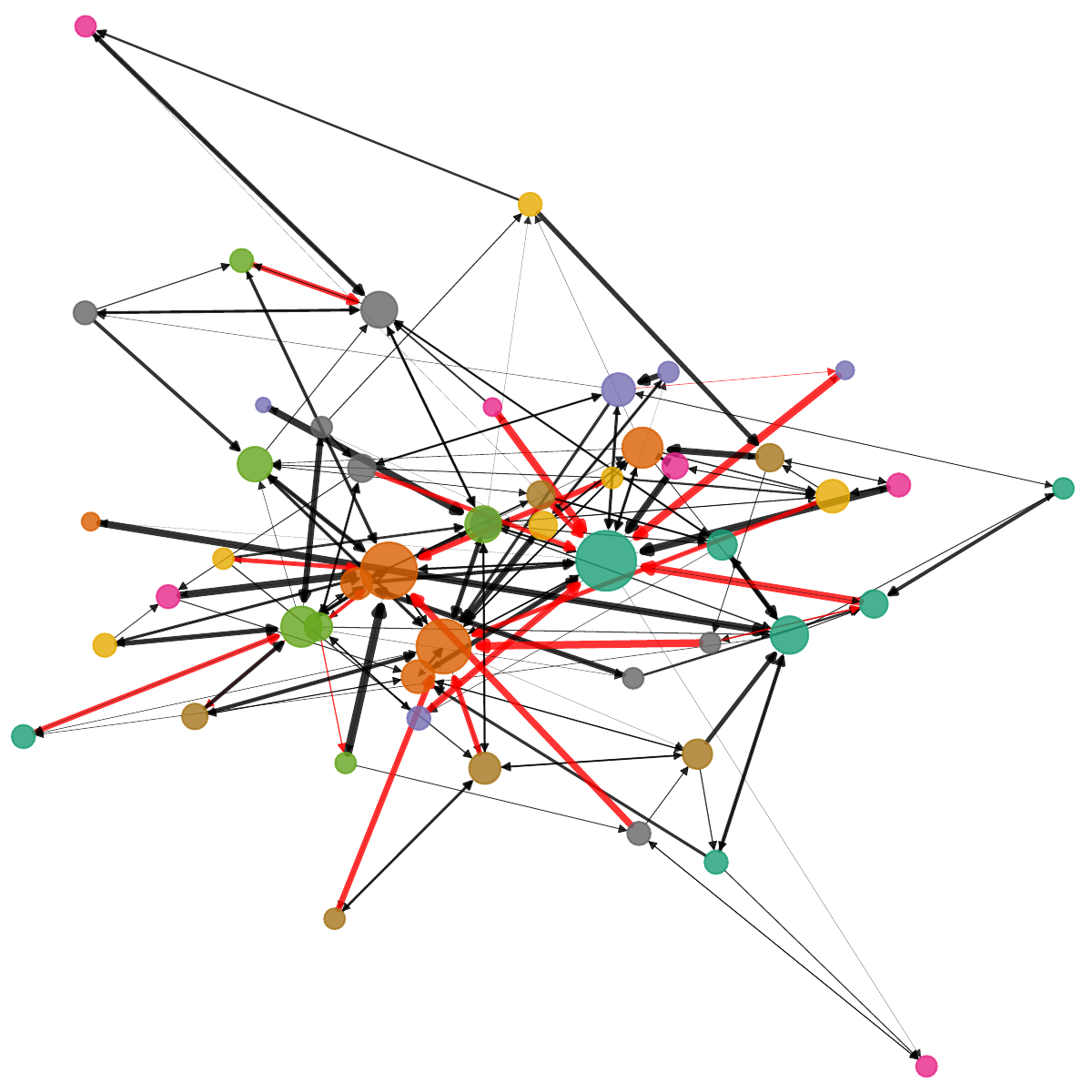}
    \end{minipage}
    \caption*{\footnotesize 
\textit{\textbf{Notes}}: The left panel shows the implied productive structure, with each arrow representing the need for a specific input.
The arrow width is proportional to the corresponding parameter in the production function.
The right panel shows the emerging production network.
The edge width is proportional to the normalized value inflow of the corresponding input.
The red edges highlight the largest discrepancies (top 10\%) between the endogenous production network and the implied productive structure.\\}
\end{figure}

We specify an economy with 50 industries and a sparse productive structure.
By sparse productive structure we mean that each firm uses a reduced subset inputs.
Hence, the distance between two firms can be measured through the smallest number of intermediate inputs flowing from one firm to the other (the distance may vary depending if the flow is upwards of downwards).
We specify the implied productive structure so that it preserves real-world features observed in supply chains, for example, sparsity and power-law input distribution \citep{Atalay2011}.\footnote{More complex or sophisticated industries require more types of inputs, which turn them into hubs.}
\autoref{fig:syn_network_structure.alphas} illustrates the implied productive structure.
The width of each arrow represents the magnitude of input share coefficients in the production function.
\autoref{fig:syn_network_structure.alphas} shows the emerging production network, which we discuss ahead.

Each industry can use its own inputs as well. 
The diameter of the implied productive structure is 6 (i.e., the largest minimum amount of inputs between two industries).
In total, there are 249 supplier-client potential relationships.\footnote{They are potential relationships because firms may decide not to purchase all their available input types as they may be substitutable.}
The average number of inputs used by an industry is 3.98.
The firm with most diversity of potential inputs could be supplied by 15 different firms.

\subsection{Production technologies}

\begin{figure}[ht]
\centering
\caption{Dynamics in the synthetic economy}\label{fig:synthetic_dynamics}
    \begin{minipage}{0.32\textwidth}
        \subcaption{Prices}\label{fig:synthetic_dynamics.price}
        \includegraphics[angle=0,width=1.\textwidth]{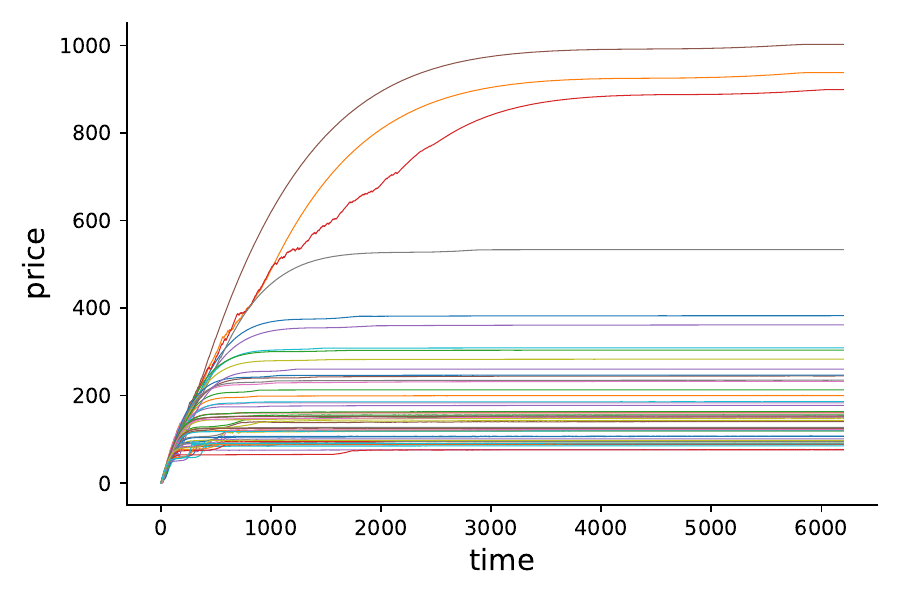}
    \end{minipage}
    \begin{minipage}{0.32\textwidth}
        \subcaption{Output volumes}\label{fig:synthetic_dynamics.quantity}
        \includegraphics[angle=0,width=1.\textwidth]{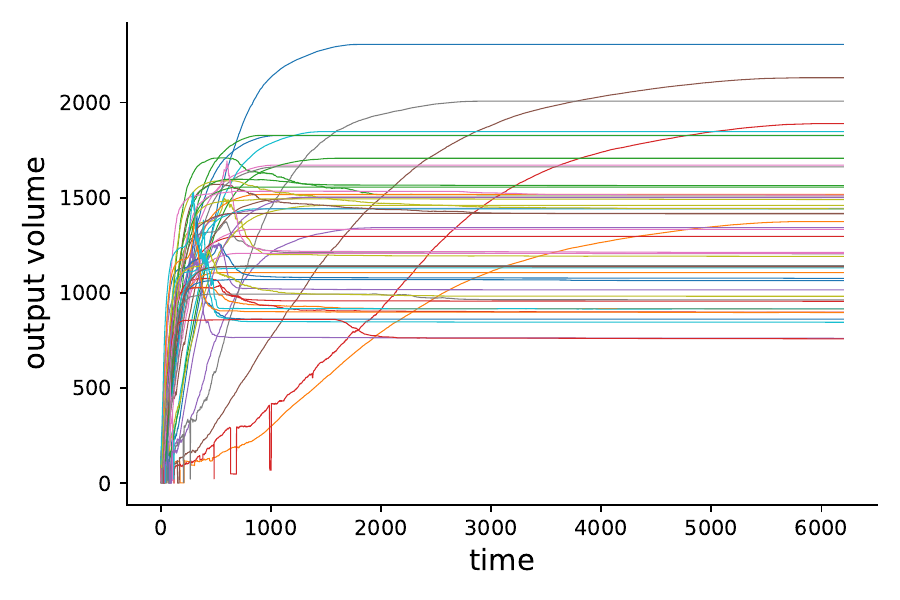}
    \end{minipage}
    \begin{minipage}{0.32\textwidth}
        \subcaption{Profits}\label{fig:synthetic_dynamics.profit}
        \includegraphics[angle=0,width=1.\textwidth]{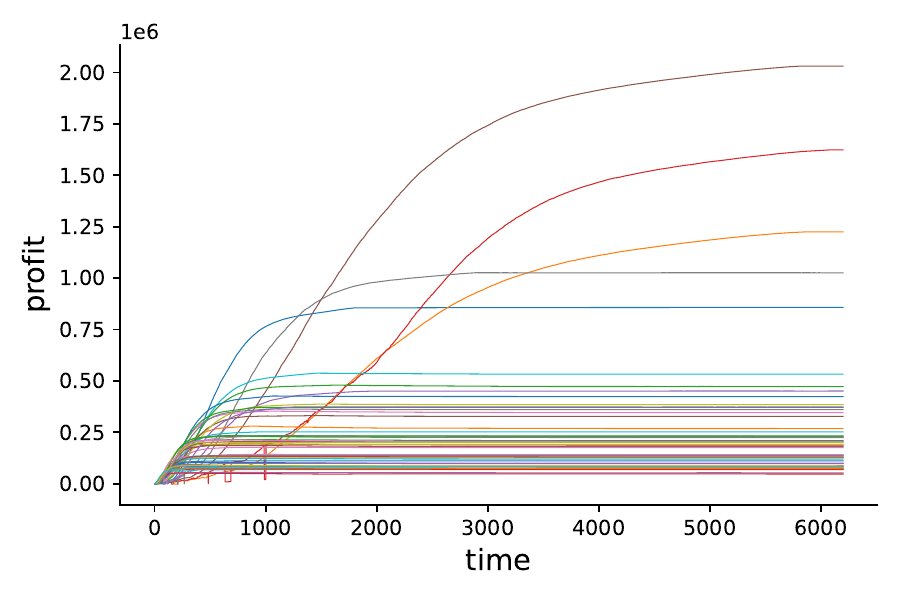}
    \end{minipage}
\end{figure}

We use CES production functions for all the firms.
Once the set of inputs are determined for a firm, we draw random parameters $a_{i,j}$ from a log-normal distribution and normalize them so they add up to one. 
We set $A_i=100$ for all firms, $\varphi_i=0.01$ to induce imperfect substitutability/complementarity, and $\rho_i=1$ for constant returns to scale.
Finally, each industry has a linear demand function given by $Q^d_i(P_{i}) = a_i - b_i P_{i} $, where $b_i=10$ and $a_i$ is uniformly sampled from [1000, 15000] and normalized by the number of inputs that the firm uses (the normalization accelerates consistent learning). 
Under this parameterization, baseline simulations converge to a steady state. 
\autoref{fig:synthetic_dynamics} illustrates the evolution of prices, quantities, and profits.

\autoref{fig:syn_network_structure.flows} shows the emerging production network in terms of steady-state value flows between industries.
The red arrows highlight the top 10\% largest discrepancies between the production network and the implied productive structure, suggesting that imputing production-function parameters directly from observed value flows (a common practice) could be misleading.
These differences emphasize the importance of modeling endogenous production networks to capture real-world responses and reorientation behavior.

\subsection{Experimental design}

We design two sets of experiments for supply-side and demand-side shocks. 
For supply shocks, we increase the productivity of a single industry via total factor productivity and analyze its downstream propagation (to customer industries). 
For demand shocks, we increase the aggregate demand through the intercept parameter and study upstream impacts (to supplier industries).
The magnitude of the parameter change in every experiment is 20\%.
We perform one supply and one demand shock for each industry in the dataset, with each experiment consisting of 30 independent realizations.
We present our results in terms of the average values across simulations.

The outcome variables of interest are the same as in our previous examples: price, output volumes, and profit.
We report proportional differences between the baseline (calibrated) and the shocked scenarios.\footnote{Proportional differences are necessary due to large differences in flow values between pairs of industries.
However, focusing on a particular pair could benefit from looking at raw differences.}
To study the propagation of shocks, we stratify the results according to the distance from the source by using the implied productive structure (i.e., the smallest number of intermediate goods from one industry to another).\footnote{The stratification works in the following way.
For a given distance, we identify those industry pairs with a one-way shortest path (in the implied productive structure) to the source industry (the intervened firm) and measure the outcome of the experiment.
To measure the distances, we count the minimum number of intermediate industries required to connect two industries in a specific direction.
Given a directed arrow from $i$ to $j$, where industry $i$ is the customer and industry $j$ is the supplier, distance~1 indicates a direct customer-supplier link with no intermediates.
Distance 2 implies firm $i$ buys from one intermediate industry which, in turn, purchases from industry $j$, and so forth.
The identified shortest paths in the implied productive structure are analyzed separately for upstream and downstream direction.}

\subsection{Results}

\begin{figure}[ht]
    \centering
    \caption{Proportional-change responses across intervened industries} \label{fig:synthetic_network_own_effect}
    \includegraphics[width=0.9\textwidth]{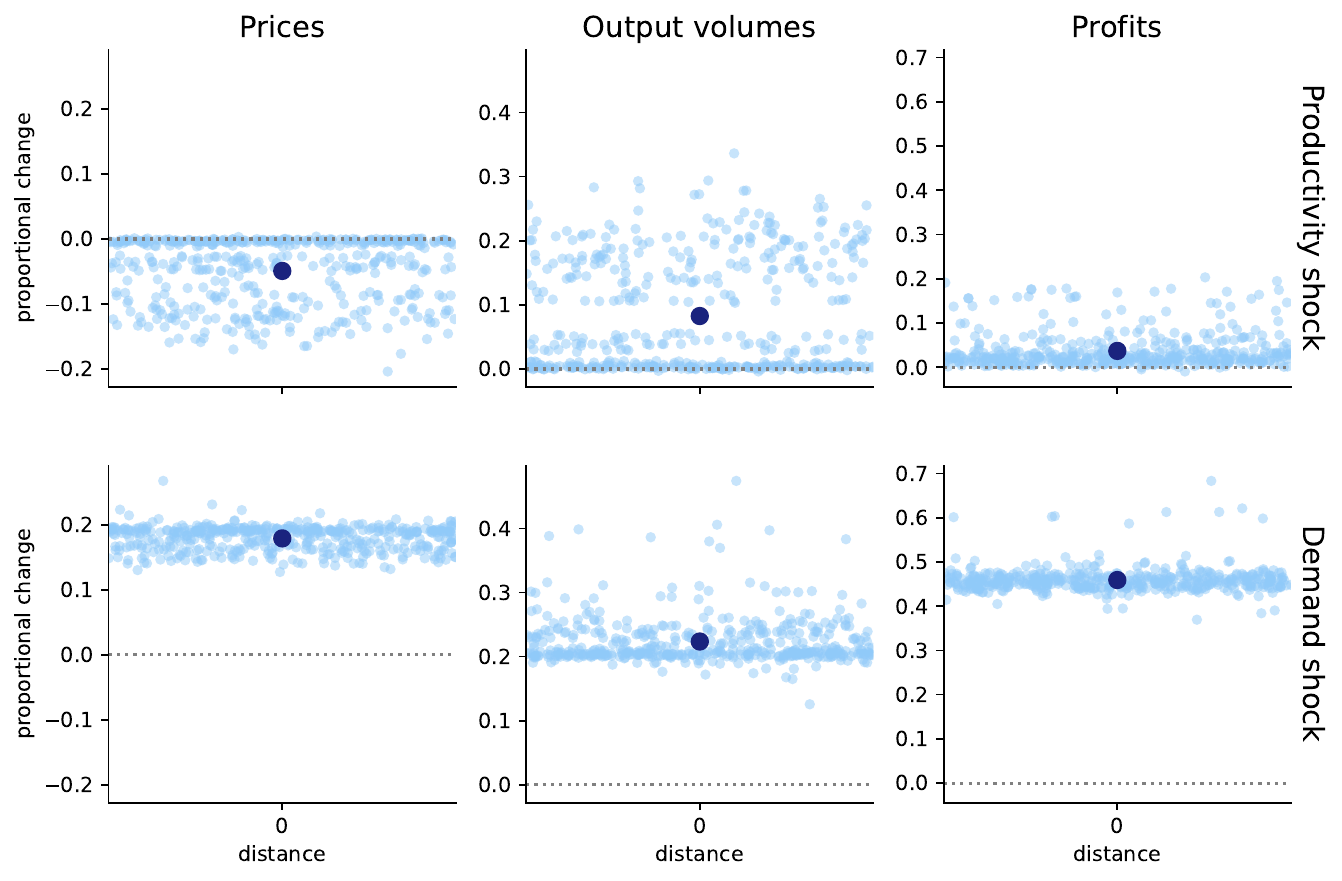}
\end{figure}

First, we look at the impact of the shocks on the intervened firms.
These results show differentiated responses across firms and outcome variables.
\autoref{fig:synthetic_network_own_effect} compares the mean proportional change on prices, output volumes, and profits of the intervened firms.
The top panels show responses to productivity shocks and the bottom ones display responses to demand shocks.
The mean is represented by a solid point.
The translucid dots correspond to individual firms, and these points are randomly placed along the horizontal axis to exhibit the structure of the data.

Following a positive supply shock, the intervened firms increase productivity, leading to lower marginal costs.
This cost reduction enables firms to decrease prices while simultaneously expand output volumes, resulting in increased profits and higher sales.
In contrast, when firms experience positive demand shocks, they generate profit increases through a different channel: the simultaneous increase of price and output volume.

\begin{figure}[ht]
    \centering
    \caption{Proportional-change responses across non-intervened industries\protect\footnotemark} \label{fig:synthetic_network_network_effect}
    \includegraphics[width=0.9\textwidth]{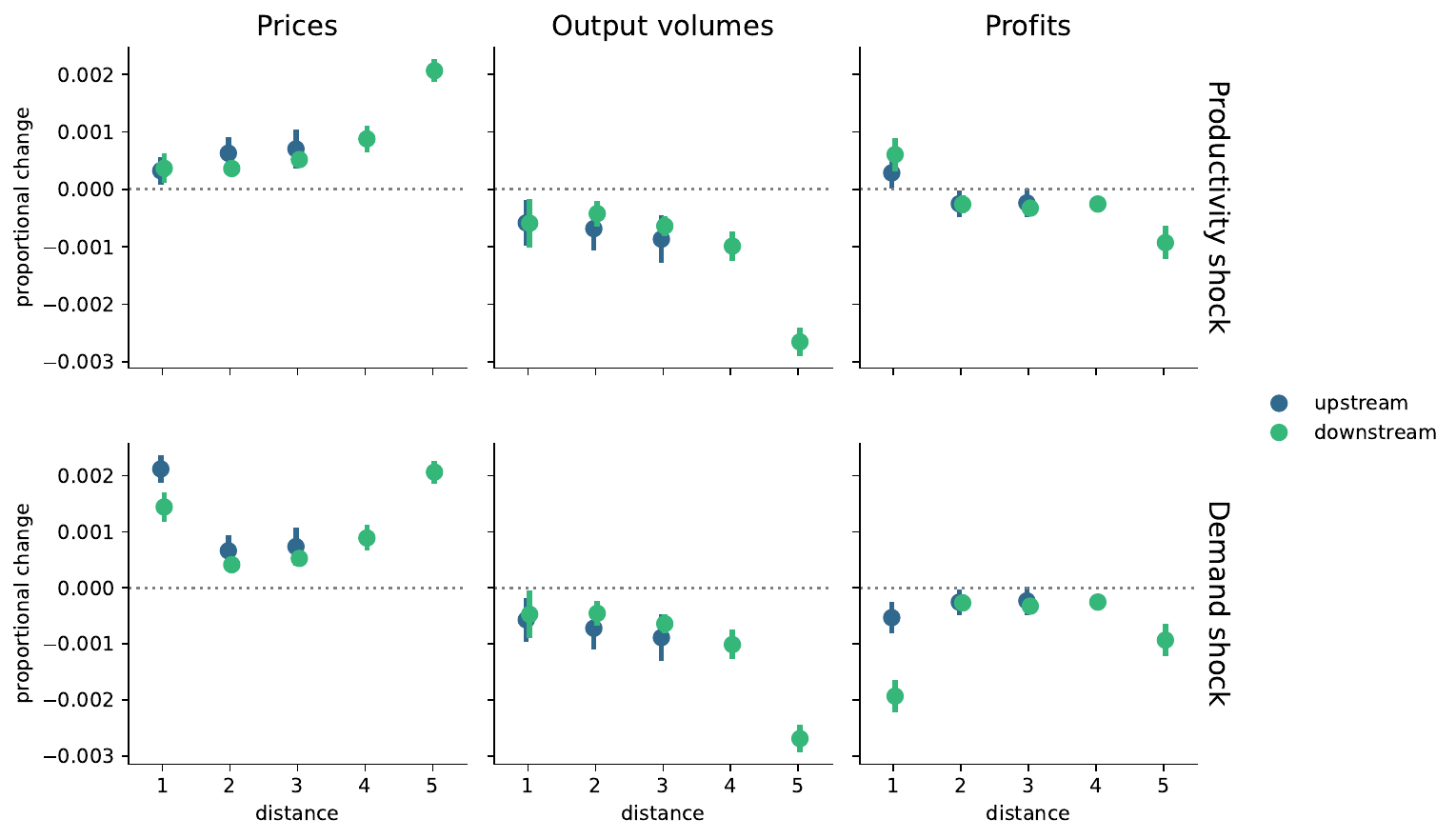}
\end{figure}

Next, let us shift our attention to indirect effects through propagation dynamics.
\autoref{fig:synthetic_network_network_effect} shows the average proportional change across non-intervened firms, pooled in groups defined by the distance from the source (means are dots and standard errors are denoted with vertical lines).
\footnotetext{Due to limited observations, we excluded data points at distance~4 for upstream shocks.
This filtering ensures reliable statistical estimates across distance groups.
}

The magnitude of the indirect effects is smaller compared to direct effects.
The results presented in \autoref{fig:synthetic_network_network_effect} reveal distinct propagation patterns across productivity and demand shocks, and upstream and downstream propagation.
For instance, in terms of prices, the proportional impact of productivity shocks increases monotonically with the distance from the source, supporting the idea of the amplification of indirect effects.
However, demand shocks show a different--U-shaped--pattern, suggesting that the relationship between indirect impact and distance may be non-linear.
Both productivity and demand shocks exhibit a negative association between indirect impact and distance, indicating a consistent amplification in terms of magnitude.
Furthermore, for downstream propagation, the absolute size of the impact among distance-5 industries if considerably larger than among firms in shorter distances.
Finally, another interesting result is that, in this example, a productivity increase generates positive average impacts in terms of profits among the direct clients and suppliers of the intervened firms.
This is not the case among industries with distances of more than one from the source. 

\begin{figure}[ht]
    \centering
    \caption{Input diversity versus absolute difference between demand and productivity impacts}
    \label{fig:corr_degree_abs_demand_minus_productivity}
    \includegraphics[width=0.7\textwidth]{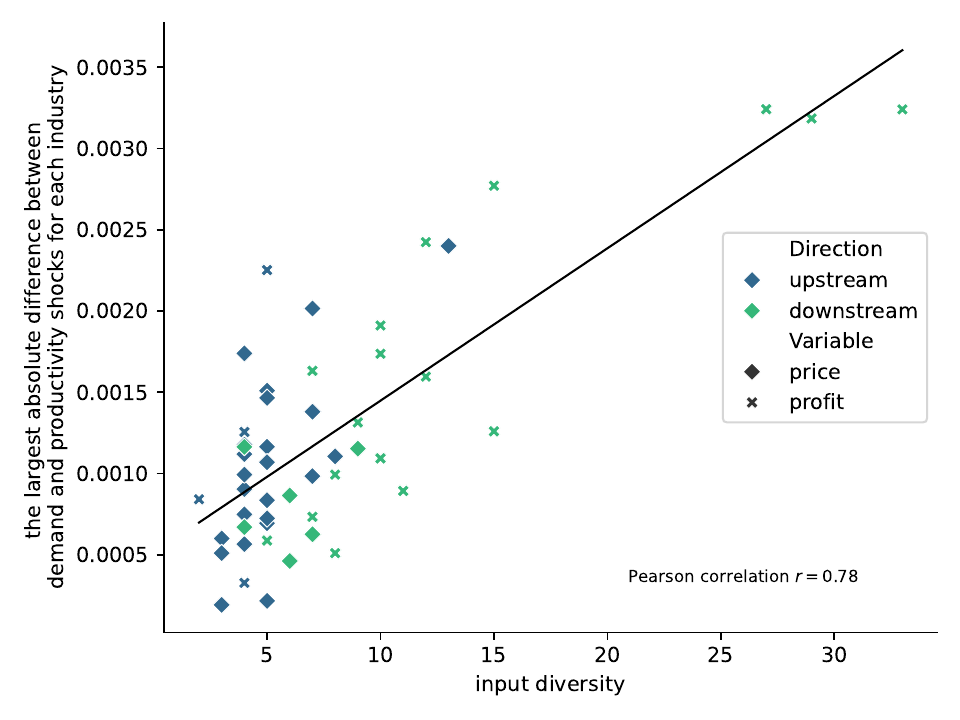}
    \caption*{\footnotesize 
    \textit{\textbf{Notes}}: Each data point represents an individual industry and is calculated by taking the absolute difference between proportional changes by demand shocks and productivity shocks to, finally, obtain the maximum value across all distances and variables (price, volume, and profit).
    For each data point, the color distinguishes between upstream and downstream, while the marker differentiates between price, volume, or profit change being the largest effect caused by the intervened industry.\\}
\end{figure}

Comparing the two types of shocks, \autoref{fig:synthetic_network_network_effect} shows that demand shocks generate proportional effects with larger magnitude, especially in prices and profits at distance~1. 
Further analysis reveals a strong correlation ($r=0.78$) between intervened firms' input diversity and their differential sensitivity to demand versus productivity shocks. 
\autoref{fig:corr_degree_abs_demand_minus_productivity} illustrates this relationship through a scatter plot, where the x-axis represents input diversity and y-axis shows the largest absolute difference between proportional effects caused by demand and productivity shocks.

Industries with higher input diversity (particularly above 20) show the largest disparities in their responses through profit changes rather than price changes, particularly in downstream industries. 
However, in less connected industries (particular under 10), these disparities are heterogeneous and can appeared in prices or profits, and in upstream or downstream.
There is an important lesson to be taken from \autoref{fig:corr_degree_abs_demand_minus_productivity}.
It is commonly argued that economic complexity--understood as economic processes that use more diverse inputs--is a key driver of growth and, hence, industrialization policies should aspire to increase it.
While such policies, indeed, may augment economic sophistication, they may also expose production networks and the economy to more vulnerabilities, given that industries that use more diverse inputs tend to be more susceptible to indirect shocks.
Therefore, it is important to think in terms of the resilience that policies may destroy when solely focusing on promoting economic complexity.
In doing so, endogenous production network models should play a key role in assessing such risks and other unintended consequences that may arise as a result of industrial policy.

\section{Discussion, limitations, and conclusion}

This study introduces a novel model for understanding the formation of production networks.
By incorporating reinforcement learning and allowing firms to operate with minimal knowledge about production functions, our model overcomes major limitations of existing approaches such as relying on equilibrium assumptions, fixing exogenous networks, assuming rationality and perfect knowledge, and setting fixed and mostly homogeneous production functions.
This approach offers significant insights into the dynamics of network formation and the impact of different types shocks, their transmission channels (upstream or downstream), and their adjustment mechanisms (prices and volumes).
The model is robust across different technology specifications and allows the construction and testing of nuanced counterfactual scenarios.

We introduce the concept of consistent learning, which entails that firms find a price-quantity point on their end-consumer demand curve under uncertainty and operate in its neighborhood.
Consistent learning allows steady-state production networks to emerge endogenously.
After investigating these network with numerical examples of three and five firms, and across different types of production functions and shocks, we bring our model closer to important issues about supply chains such as the upstream and downstream transmission of shocks. 
Our experiments with a larger theoretical economy show that, on average, the potential amplification of indirect impacts depends on the adjustment mechanism, the nature of the shock (productivity or demand), and may not be linear.

The differential impacts of demand and productivity shocks reveal important characteristics of shock propagation in production networks.
Demand shocks generate larger network effects in magnitude compared to productivity shocks, particularly evident in price and profit responses.
This asymmetry could be attributed to several mechanisms. 
First, demand shocks directly affect revenue streams, triggering immediate adjustments in firms' pricing and production decisions.
Second, a stronger response to demand shocks among highly connected nodes suggests that these shocks are amplified through network linkages.
This amplification might occur because demand changes cascade both upstream through input requirements and downstream through price adjustments.
These findings highlight how the network structure not only transmits but also transforms different types of economic shocks, with implications for understanding aggregate fluctuations and designing targeted economic interventions.

While this is the first model that is able to explain the endogenous formation of production network with minimal knowledge and learning, it also has limitations that need to be addressed in future work, especially with regard to its empirical application.
Fitting the model to real-world data would assume some knowledge about the parameters of the production functions of the industries.
Potentially, these parameters could be estimated by fitting the model to an empirical network such as an IO table.
However, this exercise can prove challenging as the interactions between inputs can make the emerging network topology highly sensitive to the choice of parameters.
Furthermore, one needs to consider constraints to the parameter space in order to avoid extreme values/volatility in the outputs that may hinder consistent learning and, hence, reaching a steady state.
The specification of these constraints needs to be done careful as it may prevent finding the set of parameters that yields good fitting.
Hence, an adequate fitting procedure needs to be carefully considered if one does not know the firm's production functions parameters; something that we leave for follow-up work.

Overall, this study contributes to the understanding of production network dynamics by offering a model that brings us closer to real-world conditions and behavioral realities. 
It provides a robust framework for exploring how firms adapt to various shocks and technological changes, paving the way for future research and policy implications aimed at enhancing the resilience and efficiency of production networks.

\newpage

\setcounter{page}{1}
\counterwithin{figure}{section}
\counterwithin{table}{section}

\section{Appendix A: Inter-industry impacts}
\label{app:inter-impacts}

\begin{figure}[ht]
\centering
\caption{Inter-industry impacts from demand shocks}\label{fig:shock_demand_matrix}
    \begin{minipage}{0.32\textwidth}
        \subcaption{Prices}\label{fig:shock_demand_matrix_ela.price}
        \includegraphics[angle=0,width=1.\textwidth]{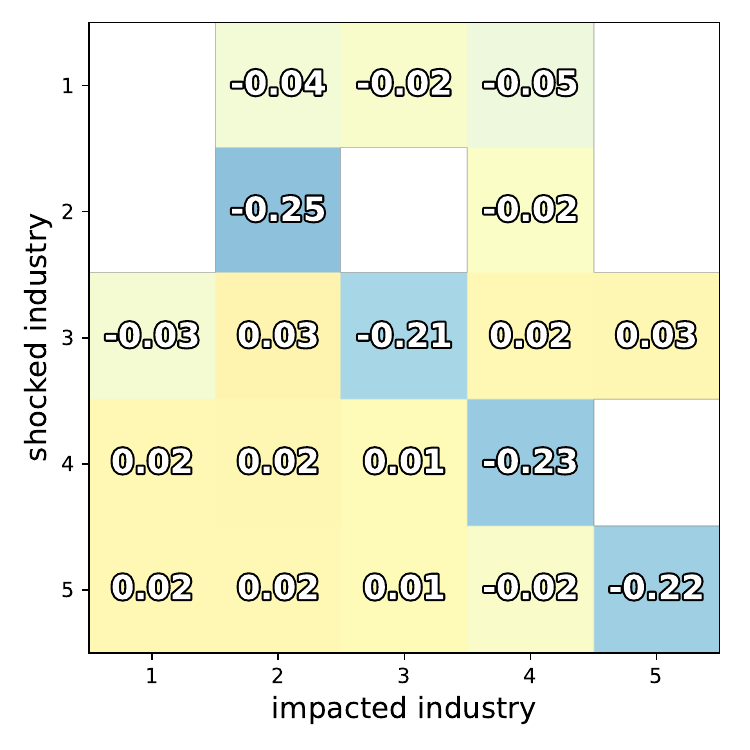}
    \end{minipage}
    \begin{minipage}{0.32\textwidth}
        \subcaption{Output volumes}\label{fig:shock_demand_matrix_ela.quantity}
        \includegraphics[angle=0,width=1.\textwidth]{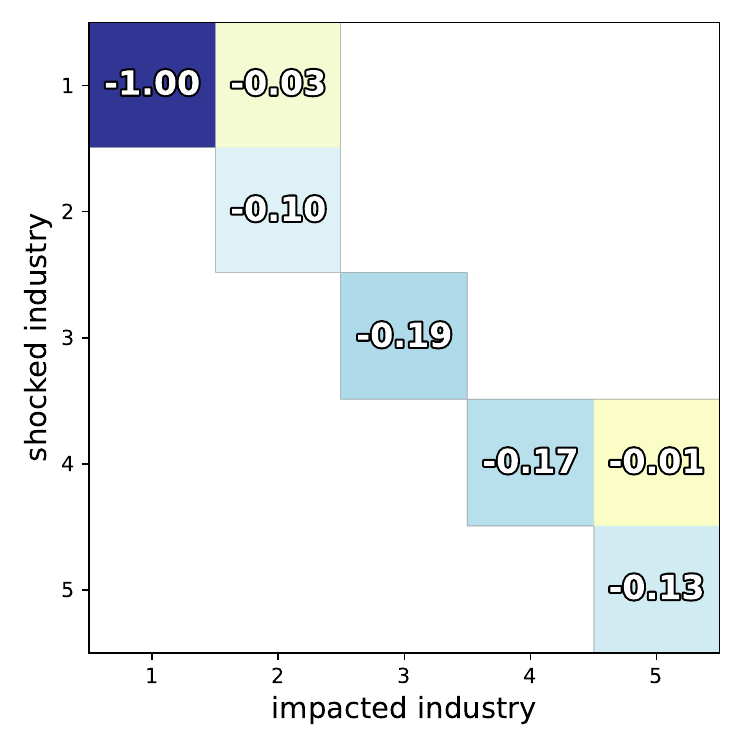}
    \end{minipage}
    \begin{minipage}{0.32\textwidth}
        \subcaption{Profits}\label{fig:shock_demand_matrix_ela.profts}
        \includegraphics[angle=0,width=1.\textwidth]{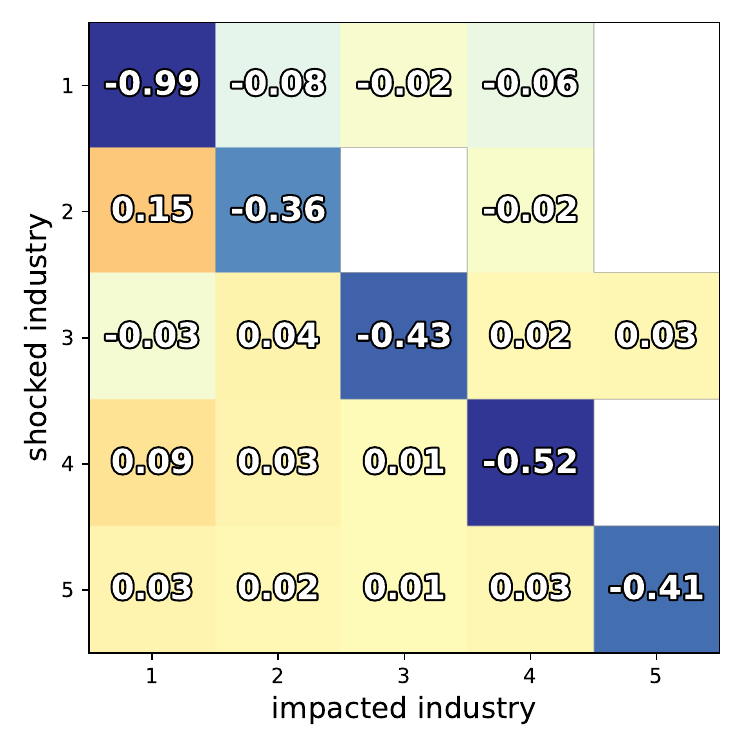}
    \end{minipage}
     \begin{minipage}{0.32\textwidth}
        \subcaption{Prices}\label{fig:shock_demand_matrix_inela.price}
        \includegraphics[angle=0,width=1.\textwidth]{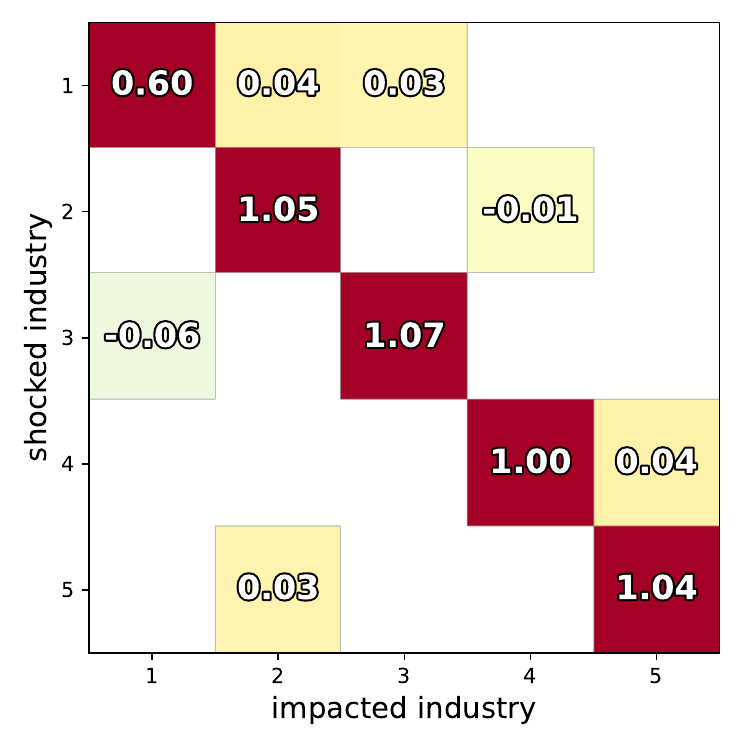}
    \end{minipage}
    \begin{minipage}{0.32\textwidth}
        \subcaption{Output volumes}\label{fig:shock_demand_matrix_inela.quantity}
        \includegraphics[angle=0,width=1.\textwidth]{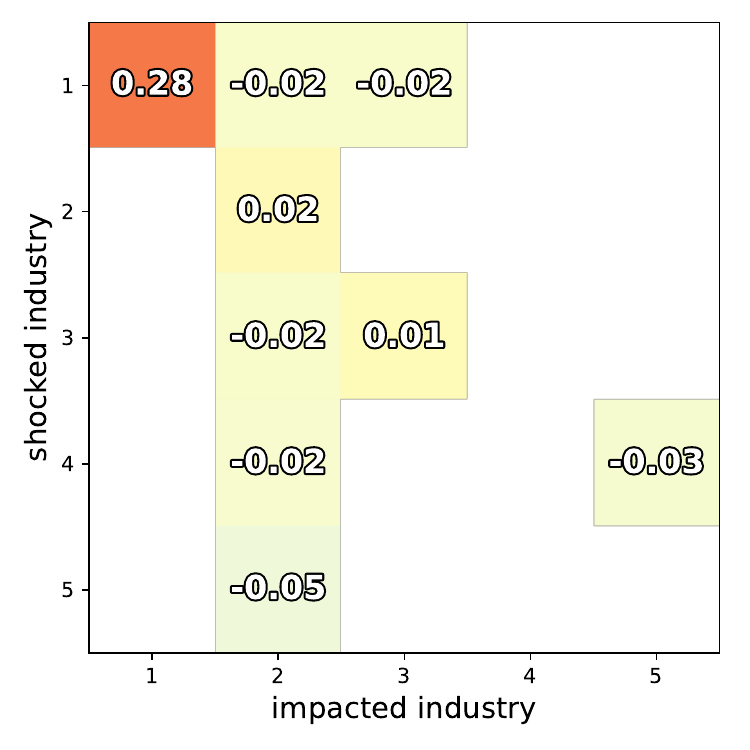}
    \end{minipage}
    \begin{minipage}{0.32\textwidth}
        \subcaption{Profits}\label{fig:shock_demand_matrix_inela.profts}
        \includegraphics[angle=0,width=1.\textwidth]{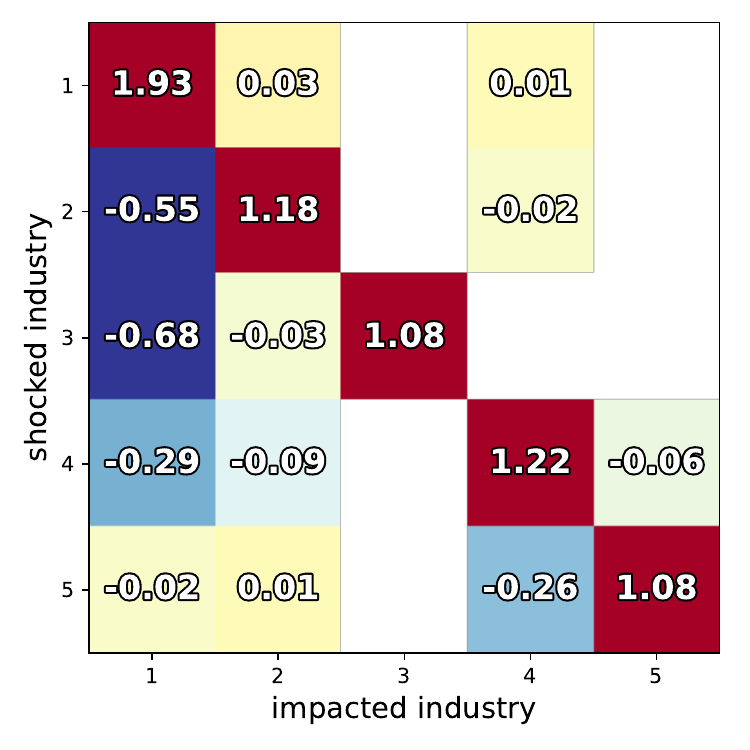}
    \end{minipage}
\caption*{\footnotesize 
\textit{\textbf{Notes}}: The entries in the matrices denote proportional changes in the steady-state outcomes after the shock.
They show changes that are greater than 0.01 (1\%) in absolute value.
Top panels correspond to an increase in demand elasticity through a 50\% change in the slope coefficient.
Bottom panels correspond to a decrease in demand elasticity through a 50\% change in the slope coefficient.\\}
\end{figure}

\begin{figure}[ht]
\centering
\caption{Inter-industry impacts from technological changes}\label{fig:shock_tech_matrix}
    \begin{minipage}{0.32\textwidth}
        \subcaption{Prices}\label{fig:shock_tech_matrix.price}
        \includegraphics[angle=0,width=1.\textwidth]{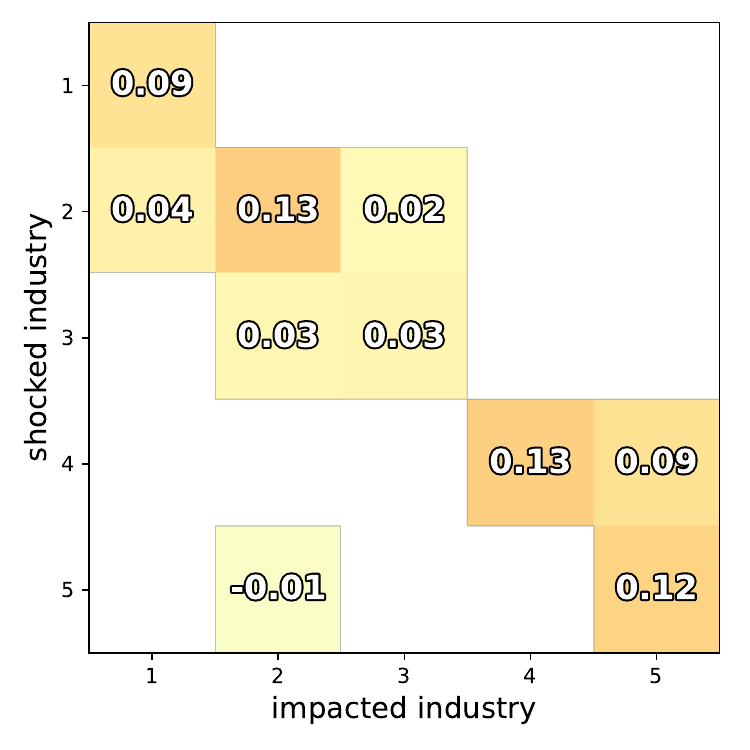}
    \end{minipage}
    \begin{minipage}{0.32\textwidth}
        \subcaption{Output volumes}\label{fig:shock_tech_matrix.quantity}
        \includegraphics[angle=0,width=1.\textwidth]{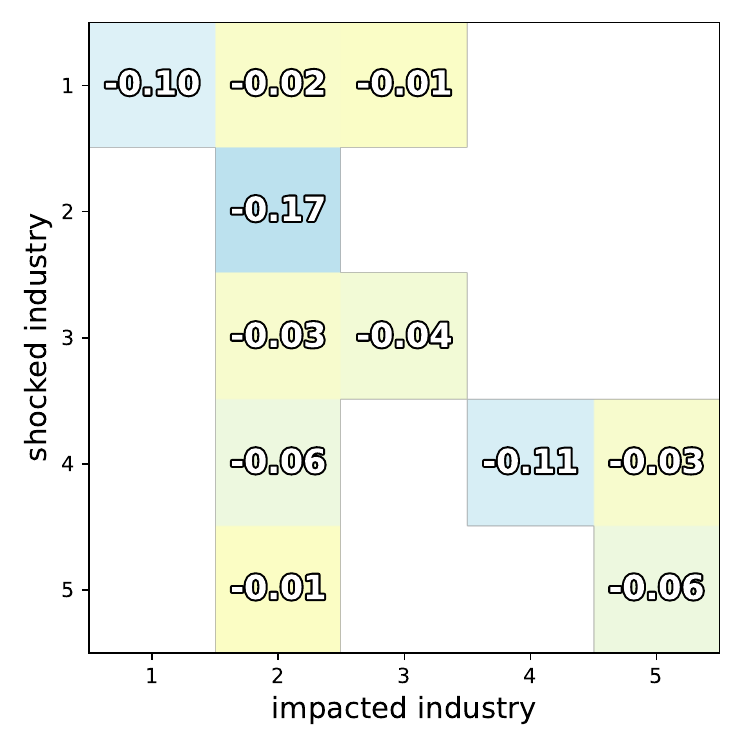}
    \end{minipage}
    \begin{minipage}{0.32\textwidth}
        \subcaption{Profits}\label{fig:shock_tech_matrix.profts}
        \includegraphics[angle=0,width=1.\textwidth]{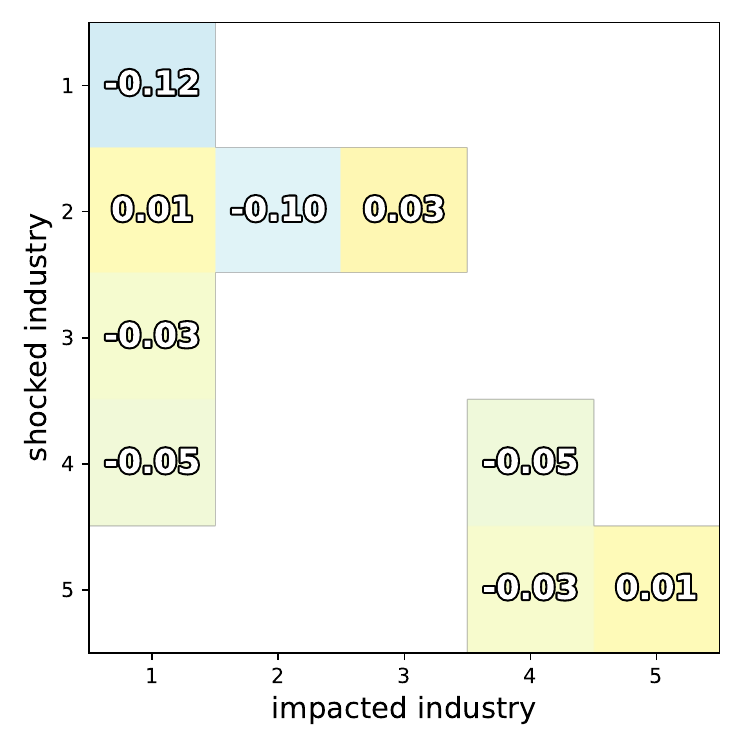}
    \end{minipage}
\caption*{\footnotesize 
\textit{\textbf{Notes}}: The entries in the matrices denote proportional changes in the steady-state outcomes after the shock.
They show changes that are greater than 1\% in absolute value.
The shock consists of changing the production function from a linear one to a Leontief one.\\}\end{figure}

\begin{figure}[ht]
\centering
\caption{Production network reorientation after recovery}\label{fig:shock_reorientation_matrix}
    \begin{minipage}{0.32\textwidth}
        \subcaption{Shock on industry 1}\label{fig:shock_reorientation_matrix.1}
        \includegraphics[angle=0,width=1.\textwidth]{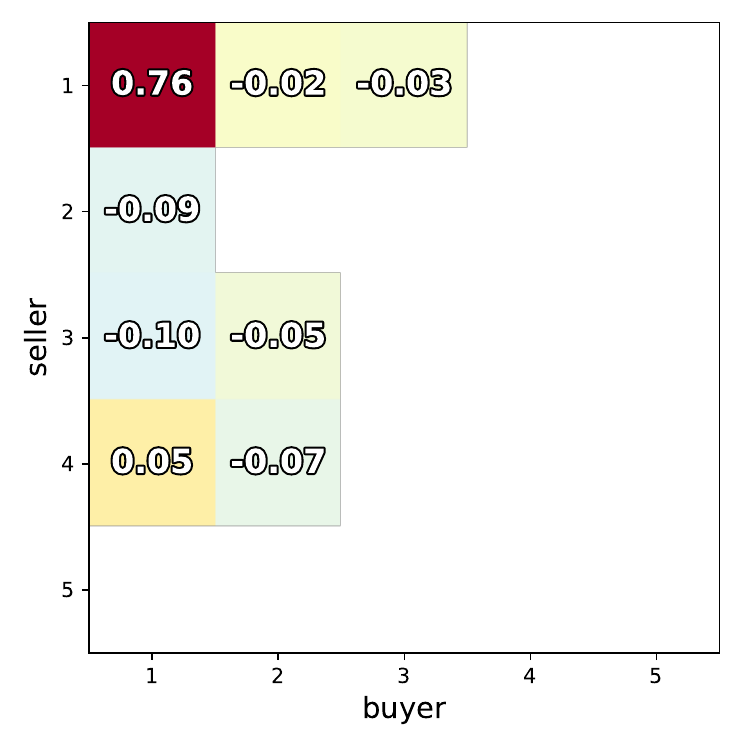}
    \end{minipage}
    \begin{minipage}{0.32\textwidth}
        \subcaption{Shock on industry 2}\label{fig:shock_reorientation_matrix.2}
        \includegraphics[angle=0,width=1.\textwidth]{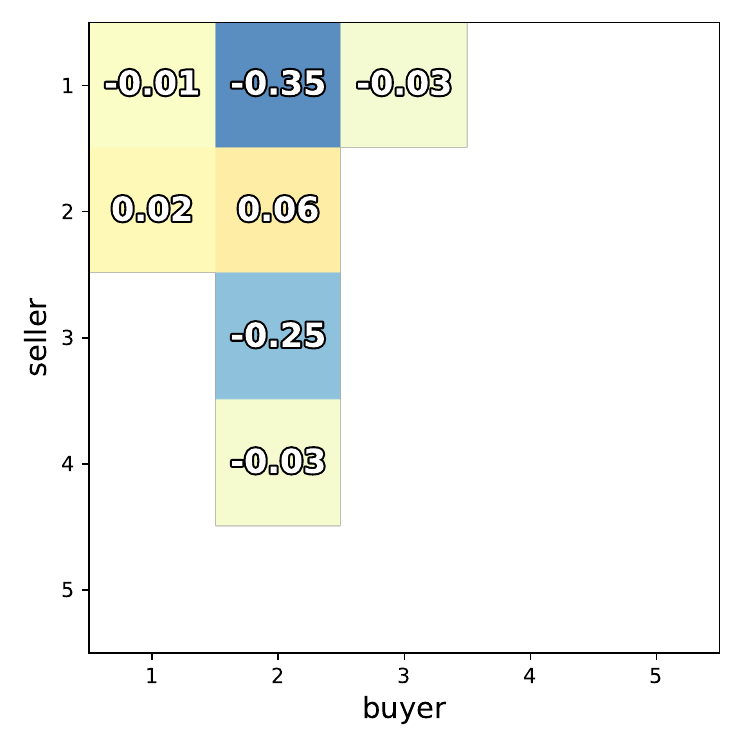}
    \end{minipage}
    \begin{minipage}{0.32\textwidth}
        \subcaption{Shock on industry 3}\label{fig:shock_reorientation_matrix.3}
        \includegraphics[angle=0,width=1.\textwidth]{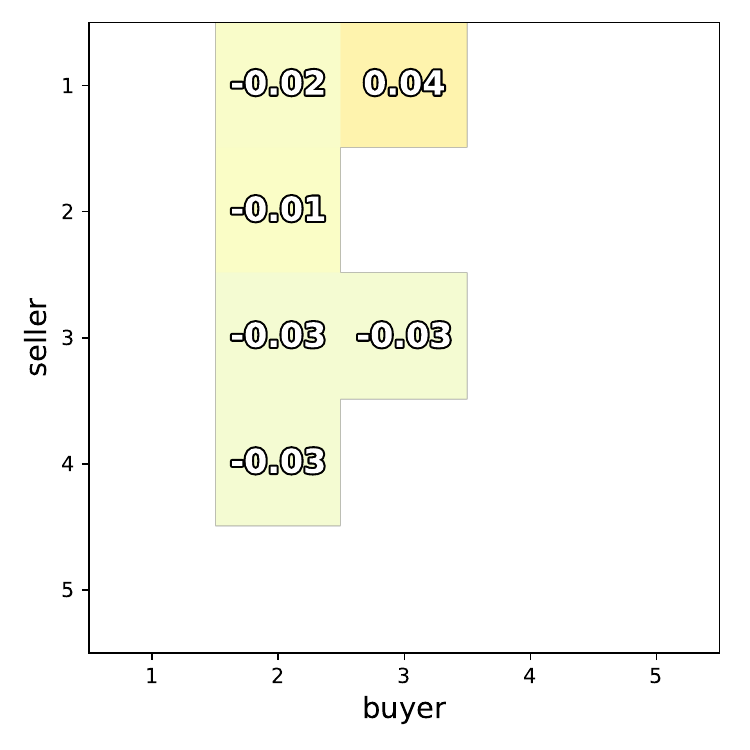}
    \end{minipage}
    \begin{minipage}{0.32\textwidth}
        \subcaption{Shock on industry 4}\label{fig:shock_reorientation_matrix.4}
        \includegraphics[angle=0,width=1.\textwidth]{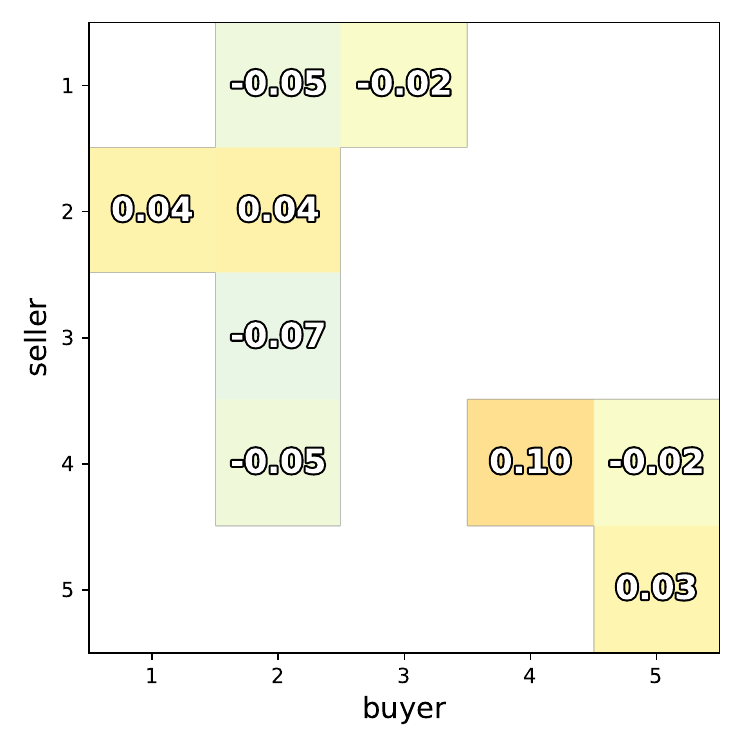}
    \end{minipage}
    \begin{minipage}{0.32\textwidth}
        \subcaption{Shock on industry 5}\label{fig:shock_reorientation_matrix.5}
        \includegraphics[angle=0,width=1.\textwidth]{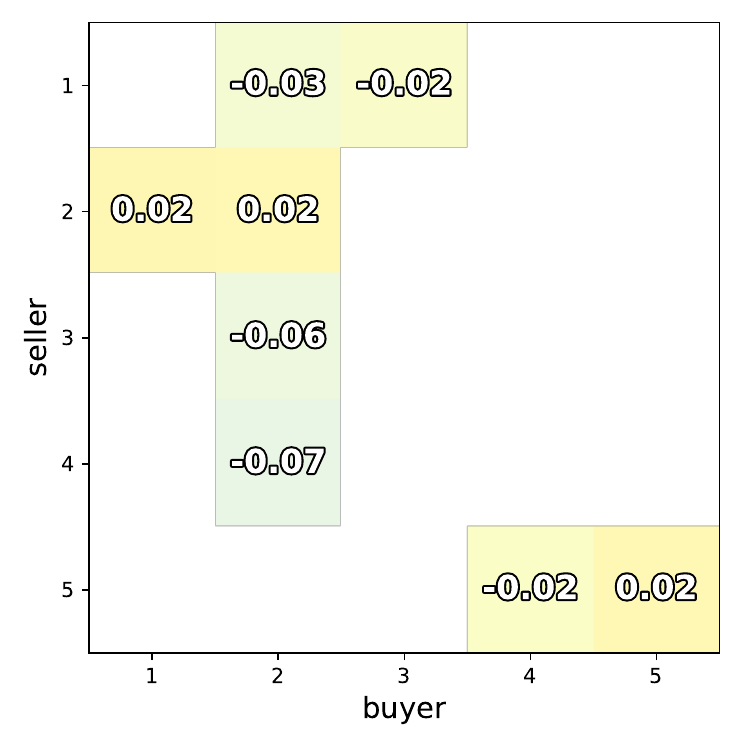}
    \end{minipage}
\caption*{\footnotesize 
\textit{\textbf{Notes}}: The entries in the matrices denote proportional changes in the value flows between industry pairs after the market demand recovers its original elasticity.
They show changes that are greater than 0.01 (1\%) in absolute value.\\}\end{figure}

\section{Appendix B: Scaling up}

\begin{figure}[ht]
\centering
\caption{Consistent learning in a larger economy (100 firms)}\label{fig:large_learning}
    \begin{minipage}{0.49\textwidth}
        \subcaption{Heterogeneous technologies}\label{fig:large_learning.learning.homo}
        \includegraphics[angle=0,width=1.\textwidth]{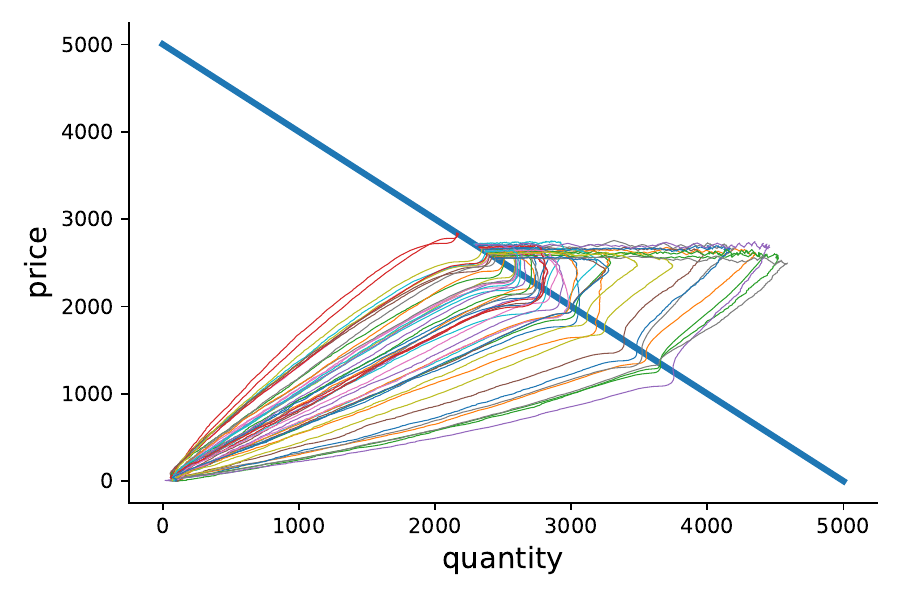}
    \end{minipage}
    \begin{minipage}{0.49\textwidth}
        \subcaption{Homogeneous technologies}\label{fig:large_learning.learning.heter}
        \includegraphics[angle=0,width=1.\textwidth]{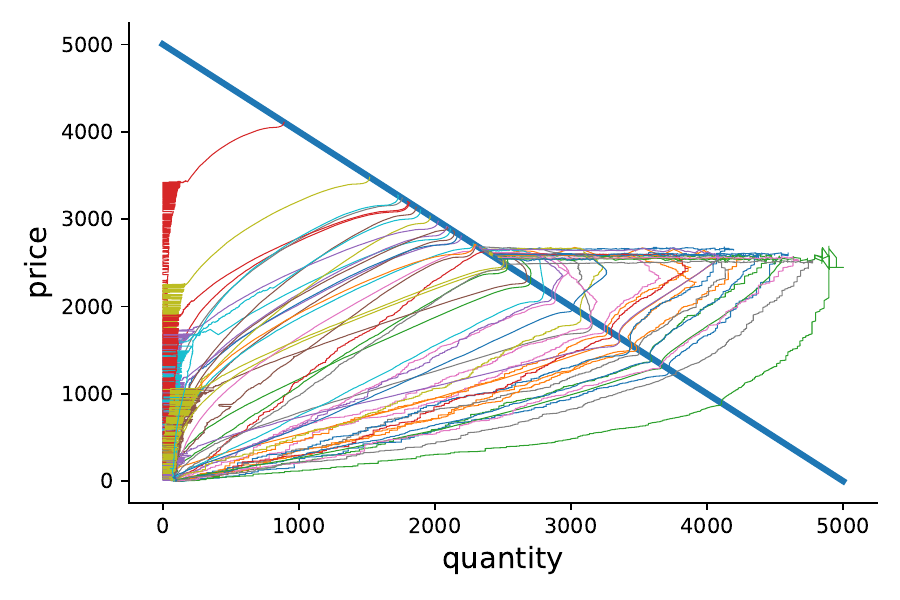}
    \end{minipage}
\end{figure}

\begin{figure}[ht]
\centering
\caption{Dynamics in a larger economy (100 firms)}\label{fig:large_dynamics}
    \begin{minipage}{0.32\textwidth}
        \subcaption{Prices}\label{fig:large_dynamics.price.homo}
        \includegraphics[angle=0,width=1.\textwidth]{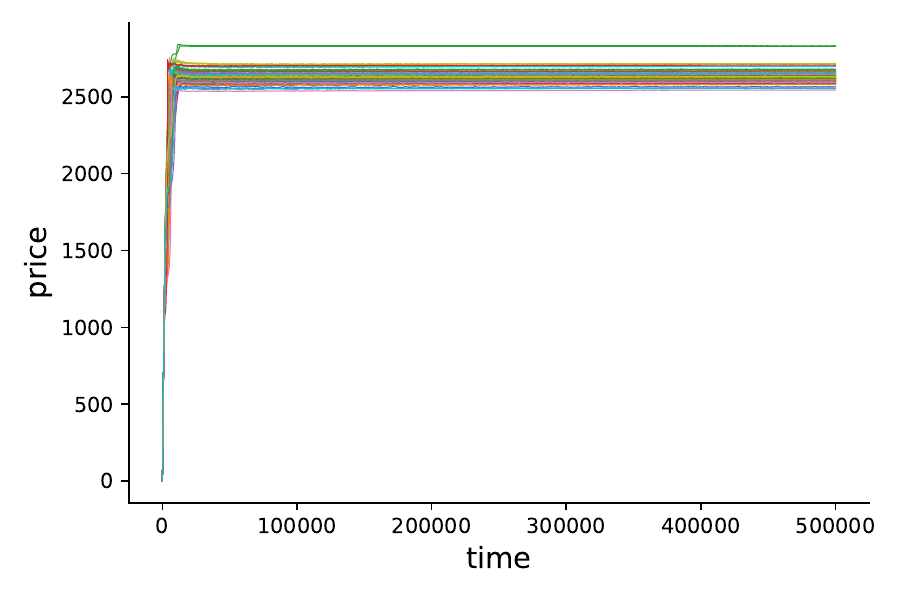}
    \end{minipage}
    \begin{minipage}{0.32\textwidth}
        \subcaption{Output volumes}\label{fig:large_dynamics.quantity.homo}
        \includegraphics[angle=0,width=1.\textwidth]{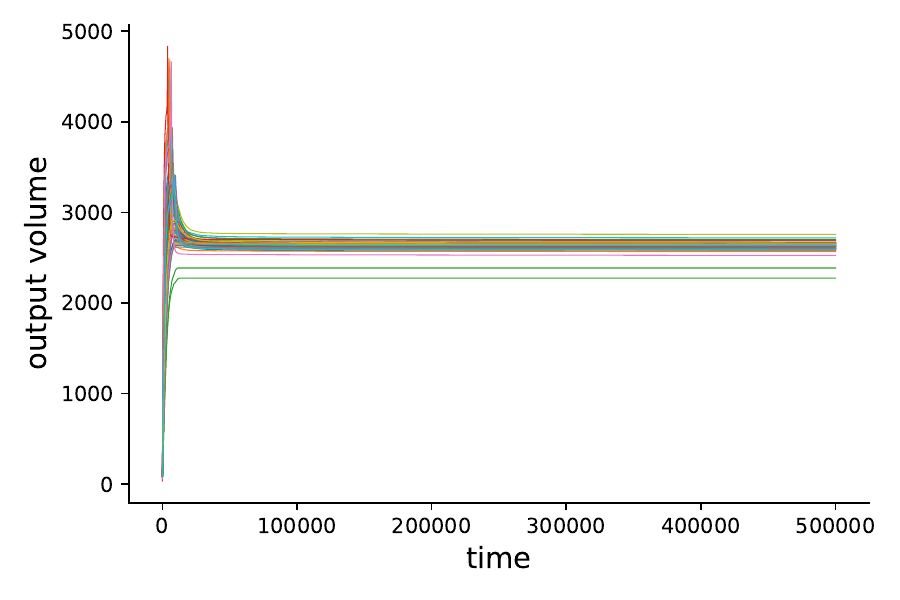}
    \end{minipage}
    \begin{minipage}{0.32\textwidth}
        \subcaption{Profits}\label{fig:large_dynamics.profit.homo}
        \includegraphics[angle=0,width=1.\textwidth]{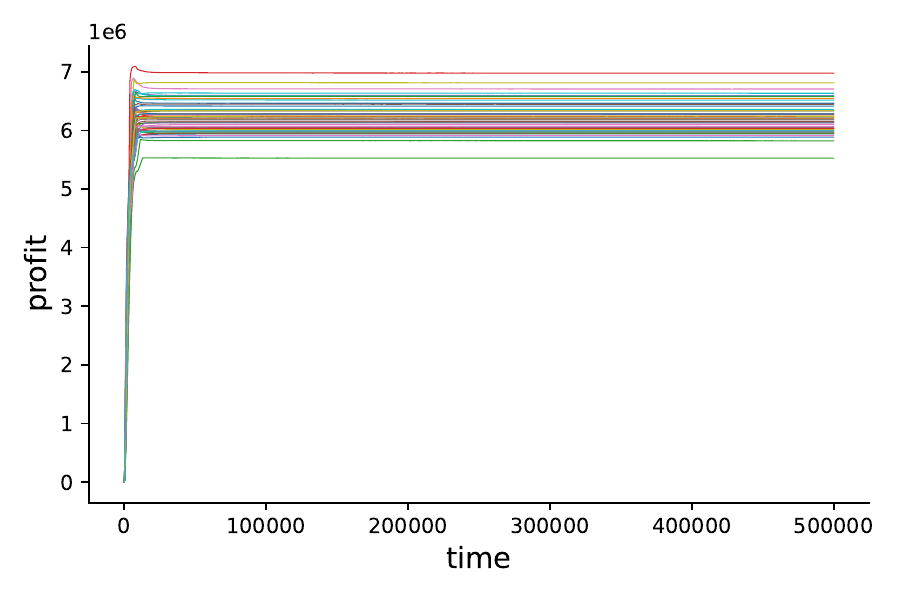}
    \end{minipage}

    \begin{minipage}{0.32\textwidth}
        \subcaption{Prices}\label{fig:large_dynamics.price.heter}
        \includegraphics[angle=0,width=1.\textwidth]{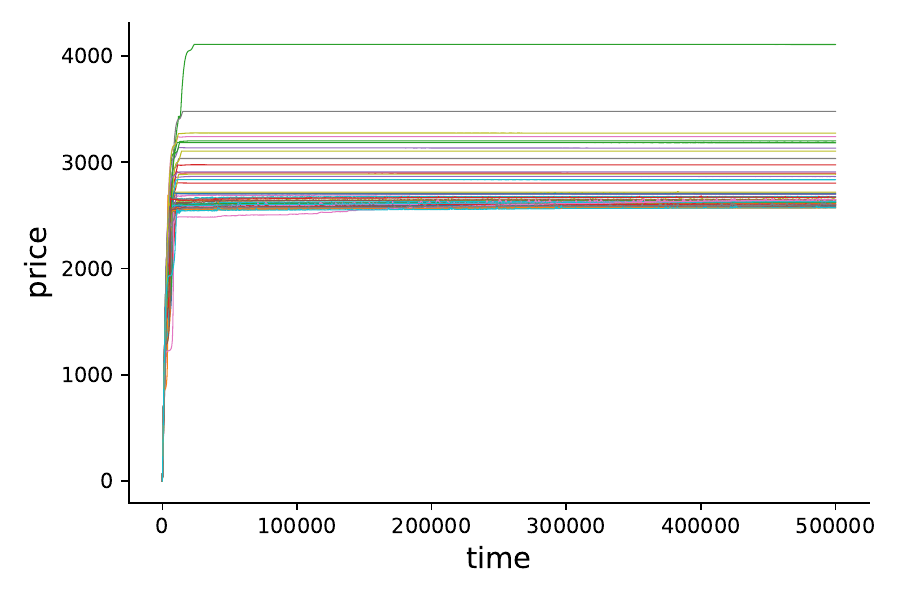}
    \end{minipage}
    \begin{minipage}{0.32\textwidth}
        \subcaption{Output volumes}\label{fig:large_dynamics.quantity.heter}
        \includegraphics[angle=0,width=1.\textwidth]{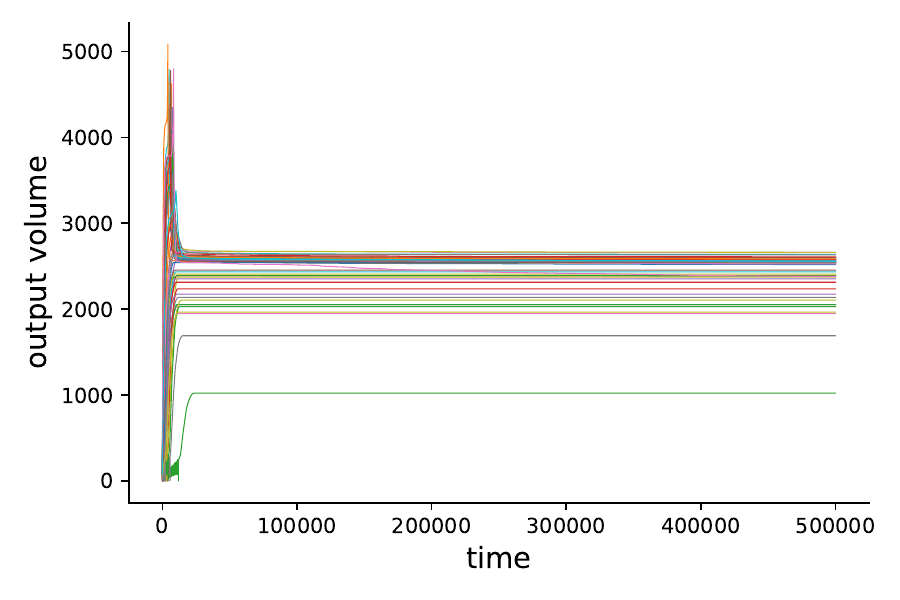}
    \end{minipage}
    \begin{minipage}{0.32\textwidth}
        \subcaption{Profits}\label{fig:large_dynamics.profit.heter}
        \includegraphics[angle=0,width=1.\textwidth]{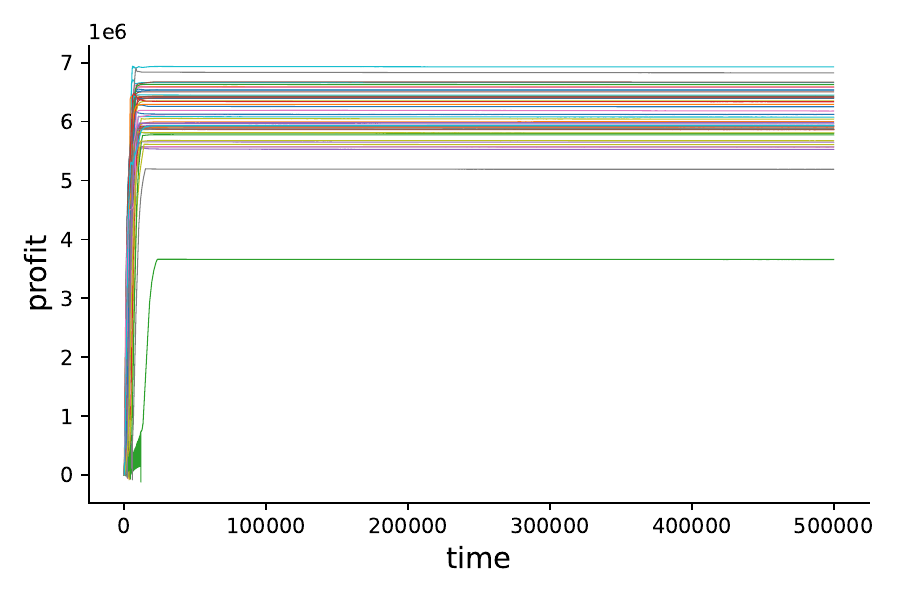}
    \end{minipage}
\end{figure}

\bibliography{main}

\end{document}